%% file: main.tex
\documentclass[11pt]{article}
\usepackage[
  letterpaper,
  top=1in,
  bottom=1in,
  left=1in,
  right=1in
]{geometry}
\usepackage{mathpazo}
\usepackage{amsmath,amssymb,amsthm,mathtools}
\usepackage{aliascnt}
\usepackage{microtype}
\usepackage{enumitem}
\usepackage[dvipsnames]{xcolor}
\usepackage{booktabs}
\usepackage{graphicx}
\usepackage{adjustbox}
\usepackage{float}
\usepackage{tikz}
\usetikzlibrary{arrows.meta,backgrounds,calc,fit}
\usepackage{fancyhdr}
\usepackage{appendix}
\usepackage[pagebackref]{hyperref}
\usepackage[nameinlink,capitalize]{cleveref}
\numberwithin{equation}{section}
\hypersetup{
  colorlinks=true,
  urlcolor=blue,
  linkcolor=RoyalBlue,
  citecolor=OliveGreen,
  linktocpage=true
}

\allowdisplaybreaks

\newtheorem{theorem}{Theorem}[section]

\newaliascnt{proposition}{theorem}
\newtheorem{proposition}[proposition]{Proposition}
\aliascntresetthe{proposition}

\newaliascnt{lemma}{theorem}
\newtheorem{lemma}[lemma]{Lemma}
\aliascntresetthe{lemma}

\newaliascnt{corollary}{theorem}
\newtheorem{corollary}[corollary]{Corollary}
\aliascntresetthe{corollary}

\newtheoremstyle{conjecturestyle}
  {\topsep}
  {0pt}
  {\itshape}
  {}
  {\bfseries}
  {.}
  {0.5em}
  {}
\theoremstyle{conjecturestyle}
\newtheorem*{conjecture}{Conjecture}

\theoremstyle{definition}
\newaliascnt{definition}{theorem}

\aliascntresetthe{definition}

\theoremstyle{remark}
\newaliascnt{remark}{theorem}
\newtheorem{remark}[remark]{Remark}
\aliascntresetthe{remark}

\theoremstyle{plain}
\newaliascnt{claim}{theorem}

\aliascntresetthe{claim}

\theoremstyle{definition}
\newaliascnt{example}{theorem}

\aliascntresetthe{example}
\theoremstyle{plain}

\newcommand{\spec}{\operatorname{Spec}}
\newcommand{\adv}{\operatorname{Adv}}
\newcommand{\rhosc}{\rho_{\mathrm{sc}}}
\newcommand{\E}{\mathbb E}
\newcommand{\Pp}{\mathbb P}
\newcommand{\R}{\mathbb R}
\newcommand{\C}{\mathbb C}
\newcommand{\Tr}{\operatorname{Tr}}
\newcommand{\Var}{\operatorname{Var}}
\newcommand{\dist}{\operatorname{dist}}
\newcommand{\op}{\mathrm{op}}
\newcommand{\SC}{\mathrm{SC}}
\newcommand{\supp}{\operatorname{supp}}
\newcommand{\diag}{\operatorname{diag}}
\newcommand{\pastur}{\Xi}
\title{Low-degree hardness rules out eigenvalue tests but not eigenvector tests}
\author{Xuan Chen\thanks{Peking University. \texttt{chenxuan@stu.pku.edu.cn}.}
\and
Shuangping Li\thanks{Yale University. \texttt{shuangping.li@yale.edu}.}}
\date{\today}

\begin{document}
\hypersetup{pageanchor=false}
\maketitle

\begin{abstract}
We study what low-degree hardness implies for spectral tests, with the Gaussian orthogonal ensemble (GOE) as the null distribution. We show that, once a fixed amount of Gaussian noise is added, it rules out every test based on eigenvalues, but not tests based on eigenvectors. For eigenvalues, if the low-degree advantage vanishes at logarithmic degree, then the joint law of the noisy eigenvalues converges to that of the GOE in total variation. For eigenvectors, we construct planted distributions with vanishing low-degree advantage whose eigenvalues follow the GOE law exactly at every noise level, yet whose noisy eigenvectors still distinguish them from the GOE. A single eigenvector can suffice. After a small amount of noise, a polynomial-time test that uses only the span of a fixed fraction of the top eigenvectors also succeeds, even when the low-degree advantage vanishes at a prescribed polylogarithmic degree.

To prove the eigenvalue result, we split the noise into three increments. The first uses the low-degree assumption to establish eigenvalue rigidity. The second couples the two spectra through Dyson Brownian motion and applies homogenization to obtain a transport estimate. The third uses Gaussian smoothing to convert this estimate into closeness in total variation.
\end{abstract}

\thispagestyle{empty}

\newpage

\tableofcontents

\thispagestyle{empty}

\newpage
\hypersetup{pageanchor=true}
\setcounter{page}{1}

\section{Introduction}
\label{section:intro}

In high-dimensional statistics and theoretical computer science, the low-degree method has become a central framework for predicting computational thresholds in average-case hypothesis testing \cite{HopkinsSteurer2017,HopkinsEtAl2017,Hopkins2018,KuniskyWeinBandeira2022,Wein2025Survey}. Originating from the Sum-of-Squares hierarchy and pseudocalibration \cite{BarakEtAl2019,HopkinsEtAl2017}, it treats multivariate polynomials of modest degree in the data, typically $O(\log n)$ for polynomial-time computation, as a proxy for efficient algorithms. Across many planted matrix and graph problems, low-degree bounds match the best known algorithms, and under mild conditions the framework is nearly equivalent to the statistical-query model \cite{BrennanEtAl2021}. Yet its role as evidence of hardness remains largely heuristic, raising the question of which richer classes of tests are genuinely ruled out when low-degree polynomials fail.

Concretely, the low-degree framework compares a planted distribution $P$ with a null distribution $Q$ through polynomial tests. This comparison is quantified by the degree-$D$ advantage, the largest absolute difference between the means of a degree-$\le D$ polynomial under $P$ and under $Q$, measured in units of its standard deviation under $Q$; see \eqref{eq:adv-def} below. The low-degree heuristic treats a vanishing advantage at degree somewhat larger than $\log n$ as evidence that no efficient algorithm can distinguish $P$ from $Q$.

A polynomial of degree $D$ sees only the low-degree part of the likelihood ratio, while the high-degree part can still carry information that some algorithm might exploit. Adding a small amount of noise bridges this gap. For Gaussian data, the natural noise operator is the Ornstein--Uhlenbeck (OU) operator, which replaces an observation $X$ by $e^{-t/2}X+\sqrt{1-e^{-t}}\,W$ for an independent null sample $W$. It preserves the null law and multiplies the degree-$d$ Hermite component of the likelihood ratio by $e^{-dt/2}$, damping precisely the high-degree components that low-degree tests cannot see. This heuristic is formalized by the following conjecture.

\begin{conjecture}[Low-degree conjecture, informal; {\cite[Conjecture~2.2.4]{Hopkins2018}}]
Let $Q_n$ be a product null distribution, and let $P_n$ be a planted distribution that is invariant under relabeling the indices. For any fixed noise level, if the low-degree advantage of $P_n$ over $Q_n$ stays bounded at degree $(\log n)^{1+\Omega(1)}$, then no polynomial-time algorithm can distinguish the noisy planted distribution from $Q_n$ with probability $1-o(1)$.
\end{conjecture}

Hopkins stated this conjecture for coordinate-resampling noise and, for standard Gaussian data, also conjectured it with the OU noise in place of resampling \cite[Section~2.2]{Hopkins2018}; Holmgren and Wein likewise proposed the OU noise for real-valued data \cite[Remark~6]{HolmgrenWein2021}; see also \cite[Conjecture~1.1]{HsiehKaneKothariLiMohantyTiegel2026} and \cite[Conjecture~1.1]{Mao2026}. Recent work shows that this prediction needs further conditions. Holmgren and Wein \cite{HolmgrenWein2021} gave counterexamples exploiting the resampling noise operator and showed that the symmetry assumption cannot be dropped; Buhai, Hsieh, Jain, and Kothari \cite{BuhaiHsiehJainKothari2025} refuted the quasipolynomial-time formulation; Mao \cite{Mao2026} refuted the polynomial-time formulation; and Jia and Vijayaraghavan \cite{JiaVijayaraghavan2026} gave a robust subspace recovery problem solved efficiently by anti-concentration despite low-degree hardness.

Because these counterexamples show that the low-degree conjecture fails in full generality, they motivate the positive question of which classes of tests are rigorously ruled out when the low-degree advantage vanishes. Hsieh, Kane, Kothari, Li, Mohanty, and Tiegel \cite{HsiehKaneKothariLiMohantyTiegel2026} initiated a systematic study of this question, showing in the Gaussian matrix setting that low-degree indistinguishability controls the noisy distribution of constant-size subgraph counts. At the same time, they highlighted ruling out spectral methods as a central problem left open by their approach \cite[Remark~1.3]{HsiehKaneKothariLiMohantyTiegel2026}. Indeed, constant-size subgraphs capture only bounded-degree local statistics, whereas even basic spectral quantities such as the largest eigenvalue depend on closed walks of length $O(\log n)$, and finer spectral statistics lie well beyond constant-size subgraph counts.

For matrix inference problems, spectral methods are among the most fundamental and widely used algorithms \cite{BaikBenArousPeche2005,FeralPeche2007,NagdaWein2026}. Low-degree polynomials and spectral methods are naturally linked through traces of powers: $\Tr X^k$ is a polynomial of degree $k$ that determines the operator norm up to a factor $n^{1/k}$ for even $k$, which is close to $1$ when $k$ is a large multiple of $\log n$. Using this observation, Kunisky, Wein, and Bandeira \cite{KuniskyWeinBandeira2022} showed that a low-degree bound rules out operator-norm tests with very small failure probability. That argument, however, does not reach finer spectral scales such as Tracy--Widom edge fluctuations, eigenvalue gaps, or short-interval eigenvalue counts.

In this paper, we prove that for the Gaussian orthogonal ensemble (GOE) null, vanishing low-degree advantage at degree $C\log n$ rules out every eigenvalue-based test after any fixed amount of OU noise, without any symmetry assumption on $P_n$. Specifically, the spectral law of the noisy matrix is close in total variation to that of the GOE (\cref{thm:main}). As a consequence, when $P_n$ is also invariant under orthogonal conjugation, the same total-variation closeness holds for the entire noisy matrix (\cref{cor:invariant}). By contrast, eigenvectors behave very differently, and the restriction to eigenvalues in \cref{thm:main} is essential. We construct planted distributions with vanishing low-degree advantage whose spectral law agrees exactly with that of the GOE at every noise level, yet whose noisy eigenvectors still distinguish them from the GOE, even in polynomial time after a small fixed amount of noise (\cref{ex:eigenvector}).

\subsection{Main results}
\label{subsec:main-result}

Let $Q_n$ be the GOE law on $n\times n$ real symmetric matrices, whose entries on and above the diagonal are independent centered Gaussians with variance $1/n$ off the diagonal and $2/n$ on the diagonal. Let $P_n$ be any distribution on $n\times n$ real symmetric matrices. For $t\ge0$ and independent $X\sim P_n$ and $W\sim Q_n$, let $\mathcal U_tP_n$ denote the law of
\begin{equation*}
H_t:=e^{-t/2}X+\sqrt{1-e^{-t}}\,W.
\end{equation*}
Thus $\mathcal U_tP_n$ adds independent GOE noise at level $t$, scaled so that the GOE null is unchanged ($\mathcal U_tQ_n=Q_n$); at a fixed time we also write $H$ for the noisy matrix. For a real symmetric matrix $B$, let $\spec(B)=(\lambda_1(B),\dots,\lambda_n(B))$ be its eigenvalues in nondecreasing order, and write $\spec(\mu)$ for the induced spectral law of a matrix law $\mu$. The low-degree advantage at degree $D_n$ is
\begin{equation}
\adv_{\le D_n}(Q_n,P_n) :=\sup_{\substack{q:\,\deg q\le D_n\\0<\Var_{Q_n} q<\infty}} \frac{|\E_{P_n} q-\E_{Q_n}q|}{\sqrt{\Var_{Q_n}q}},
\label{eq:adv-def}
\end{equation}
where $q$ ranges over real polynomials in the matrix entries on and above the diagonal (assuming $P_n$ has finite moments of order $D_n$).

We first show that, after any fixed amount of Gaussian noise, vanishing low-degree advantage implies that the spectral laws are close in total variation.

\begin{theorem}
\label{thm:main}
For every fixed $T>0$, there exists a constant $C>0$ such that the following holds. Let $\{P_n\}_{n\ge1}$ be any sequence of distributions on real symmetric matrices, and let $\{D_n\}_{n\ge1}$ be positive integers with $D_n\ge C\log n$ and $\adv_{\le D_n}(Q_n,P_n)=o(1)$ as $n\to\infty$. Then
\begin{equation*}
d_{\mathrm{TV}}\left(\spec (\mathcal U_T P_n),\spec (Q_n)\right)=o(1)
\end{equation*}
as $n\to\infty$.
\end{theorem}

In other words, no test depending only on the eigenvalues of the noisy matrix, regardless of its running time, can distinguish $\mathcal U_TP_n$ from $Q_n$ with nonvanishing advantage. In particular, this rules out tests based on the largest eigenvalue at any scale, the operator norm, the number of outliers, eigenvalue gaps, short-interval counts, and linear spectral statistics. Here a test based on the eigenvalues means a function of $\spec(H)$; a method that first transforms $H$ (for instance by a nonlinear entrywise map) and then inspects its eigenvalues is not covered by \cref{thm:main}.

We emphasize two strengths of \cref{thm:main}. First, it requires control only up to degree $C\log n$, well below $(\log n)^{1+\Omega(1)}$ in the low-degree conjecture. Second, it imposes no condition whatsoever on the full likelihood ratio. By contrast, a naive Hermite expansion of the OU noise operator gives
\begin{equation*}
d_{\mathrm{TV}}(\mathcal U_TP_n,Q_n)
\le\frac12\left(\adv_{\le D_n}(Q_n,P_n)^2+e^{-(D_n+1)T}\chi^2(P_n\,\|\,Q_n)\right)^{1/2},
\end{equation*}
which is vacuous whenever $\chi^2(P_n\,\|\,Q_n)\gtrsim e^{TD_n}$, including for low-rank or discrete planted distributions where $\chi^2(P_n\,\|\,Q_n)=\infty$. We overcome this barrier completely for spectral tests by combining low-degree information with the dynamics of the eigenvalue process.

Before turning to eigenvectors, we record two brief remarks on the hypotheses and scope of \cref{thm:main}.

\begin{remark}[Only spectral polynomials matter]
\label{rem:invariant-polynomials}
The assumption in \cref{thm:main} can be weakened so that \eqref{eq:adv-def} is taken only over spectral polynomials, namely linear combinations of products $\prod_j\Tr X^{k_j}$ with $\sum_jk_j\le D_n$. Indeed, letting $\mathsf SP_n$ be the law of $OXO^{\mathsf T}$ for $X\sim P_n$ and an independent Haar orthogonal matrix $O$, orthogonal invariance of the GOE gives $\spec(\mathcal U_T\mathsf SP_n)=\spec(\mathcal U_TP_n)$, while Haar averaging $\mathsf Aq(B)=\int q(OBO^{\mathsf T})\,dO$ and Jensen's inequality show that $\adv_{\le D_n}(Q_n,\mathsf SP_n)$ restricts \eqref{eq:adv-def} to orthogonally invariant polynomials. Applying \cref{thm:main} to $\mathsf SP_n$ gives the claim.
\end{remark}

\begin{remark}[Noise and the restriction to the spectrum]
\label{rem:necessity}
Both the added noise and the restriction to eigenvalues are essential in \cref{thm:main}. Fix $D_n=\lceil C\log n\rceil$, let $N_n$ be the dimension of mean-zero degree-$\le D_n$ polynomials under $Q_n$, and let $P_n$ be uniform on $M_n=nN_n$ independent GOE samples $A_1,\dots,A_{M_n}$. Expanding in an orthonormal basis gives $\E\,\adv_{\le D_n}(Q_n,P_n)^2=1/n$, so we may fix a realization with advantage at most $n^{-1/4}$ and $\|A_j\|_F^2=n+O(\log n)$. Without noise, $\spec(P_n)$ is discrete, so $d_{\mathrm{TV}}(\spec(P_n),\spec(Q_n))=1$. After noise, since $\log M_n=O((\log n)^2)\ll n$, Gaussian concentration and a union bound show that the brute-force test checking $\min_j\|H-e^{-T/2}A_j\|_F^2\le n$ distinguishes $\mathcal U_TP_n$ from $Q_n$ with probability $1-o(1)$, whereas \cref{thm:main} rules out every eigenvalue test. (This example is neither permutation-invariant nor efficient, so it is consistent with the low-degree conjecture.)
\end{remark}

We next show that the restriction to eigenvalues in \cref{thm:main} cannot be lifted to eigenvectors, even when the spectral law after noise  matches the GOE law exactly.

\begin{theorem}[Eigenvectors can carry the signal]
\label{ex:eigenvector}
For a planted distribution $P_n$ and noise level $T>0$, let $H\sim\mathcal U_TP_n$ be the noisy planted matrix, and let $v_1,\dots,v_n$ be unit eigenvectors of $H$ corresponding to its ordered eigenvalues $\lambda_1(H)\le\dots\le\lambda_n(H)$. In both parts below, the planted distribution $P_n$ has the exact GOE spectrum at every noise level, $\spec(\mathcal U_tP_n)=\spec(Q_n)$ for all $t\ge0$, so no test based on the eigenvalues has any power. Nevertheless, the eigenvectors of $H$ still distinguish the noisy planted matrix from the GOE.
\begin{enumerate}[label=(\alph*)]
\item \emph{One eigenvector.} Fix a noise level $T>0$ and a degree sequence $D_n=o(n/\log n)$. There exist planted distributions $P_n$ with vanishing low-degree advantage $\adv_{\le D_n}(Q_n,P_n)=o(1)$ such that, for every $i$, the line spanned by the single eigenvector $v_i$ distinguishes $\mathcal U_TP_n$ from $Q_n$ with probability $1-o(1)$. If we further require $P_n$ to be invariant under relabeling the indices, then the span of any $k$ eigenvectors with $C_T\log n\le k\le n-C_T\log n$ suffices.
\item \emph{Polynomial time.} Fix $\alpha\in(0,1)$, $q>0$, and a degree sequence $D_n=\lfloor(\log n)^q\rfloor$. There exist a constant $T_0>0$ and relabeling-invariant distributions $P_n$ with $\adv_{\le D_n}(Q_n,P_n)=o(1)$ such that, for every $T\in(0,T_0]$, a polynomial-time test using only the span of the top $k=\lfloor\alpha n\rfloor$ eigenvectors $v_{n-k+1},\dots,v_n$ distinguishes $\mathcal U_TP_n$ from $Q_n$ with probability $1-o(1)$.
\end{enumerate}
\end{theorem}

We prove \cref{ex:eigenvector} in \cref{sec:eigenvector-examples} and sketch both constructions in \cref{subsec:overview-eigenvectors}. In both parts, we form $X\sim P_n$ by conjugating a GOE matrix $Y\sim Q_n$ with a random orthogonal matrix (which may depend on $Y$), preserving the GOE spectral law at every noise level (\cref{lem:noisy-eigenbasis}). Part~(a) draws the eigenbasis from a fixed list of $n^{O(D_n)}$ random rotations and searches the list as in \cref{rem:necessity}, which is not efficient. To obtain a polynomial-time test in part~(b), we embed Mao's code \cite{Mao2026} into the signs of an off-diagonal block of the orthogonal projector onto the span of
the top eigenvectors (\cref{fig:coded-projector}) using the signed half-normal embedding of \cite{KlivansKothari2014}, extending Mao's Gaussian OU counterexample \cite[Sections~1.1 and~4]{Mao2026} so that the signal resides entirely in the eigenvectors.

\subsection{Consequences}
\label{subsec:consequences}

We record three consequences of \cref{thm:main}. First, consider orthogonally invariant planted distributions ($OXO^{\mathsf T}\stackrel{d}{=}X$ for every orthogonal $O$), whose eigenvectors are uniformly random and carry no information beyond the spectrum; examples include spherical or Gaussian spiked Wigner models, Wishart matrices $\frac{1}{m}GG^{\mathsf T}$, invariant ensembles $dP_n(X)\propto e^{-n\Tr V(X)}\,dX$, and randomly rotated spectra $U\diag(\mathbf d)U^{\mathsf T}$ with Haar orthogonal $U$ independent of $\mathbf d$. For such distributions, spectral indistinguishability immediately implies full matrix indistinguishability, confirming the information-theoretic form of the low-degree prediction at degree $C\log n$.

\begin{corollary}[Orthogonally invariant planted distributions]
\label{cor:invariant}
For every fixed $T>0$, there exists a constant $C>0$ such that the following holds. If $P_n$ is invariant under orthogonal conjugation ($OXO^{\mathsf T}\stackrel{d}{=}X$ for every orthogonal matrix $O$) and $\adv_{\le D_n}(Q_n,P_n)=o(1)$ for some $D_n\ge C\log n$, then $d_{\mathrm{TV}}(\mathcal U_TP_n,Q_n)=o(1)$.
\end{corollary}

\begin{proof}
Both $\mathcal U_TP_n$ and $Q_n$ are orthogonally invariant with densities, so each equals the law of $O\diag(\spec(H))O^{\mathsf T}$ for an independent Haar orthogonal matrix $O$. Data processing and \cref{thm:main} (taking its constant $C$) therefore give $d_{\mathrm{TV}}(\mathcal U_TP_n,Q_n)\le d_{\mathrm{TV}}(\spec(\mathcal U_TP_n),\spec(Q_n))=o(1)$.
\end{proof}

Second, consider additive Gaussian models $P_{n,\lambda}=\mathrm{Law}(\lambda S_n+W)$ for a random symmetric matrix $S_n$ and independent $W\sim Q_n$, covering structured spiked Wigner, sparse PCA, and planted submatrix models \cite{KuniskyWeinBandeira2022,DingKuniskyWeinBandeira2023,Wein2025Survey}. Here OU noise simply shrinks $\lambda$ to $e^{-T/2}\lambda$ without leaving the model family.

\begin{corollary}[Additive Gaussian models]
\label{cor:additive}
Let $P_{n,\lambda}$ be the law of $\lambda S_n+W$ with $W\sim Q_n$ independent of $S_n$. Fix $\lambda>0$, and suppose that $\adv_{\le D_n}(Q_n,P_{n,\lambda})=o(1)$ for some $D_n=\omega(\log n)$. Then, for every fixed $\lambda'\in[0,\lambda)$, $d_{\mathrm{TV}}(\spec(P_{n,\lambda'}),\spec(Q_n))=o(1)$.
\end{corollary}

\begin{proof}
For $\lambda'>0$, setting $T=2\log(\lambda/\lambda')$ gives $\mathcal U_TP_{n,\lambda}=P_{n,\lambda'}$; since $D_n\ge C\log n$ for all large $n$, \cref{thm:main} applies.
\end{proof}

Third, we say that the entries of $X\sim P_n$ are $D$-wise independent with GOE marginals if every $D$ entries on or above the diagonal are independent with GOE marginals.

\begin{corollary}[Bounded independence plus noise]
\label{cor:bounded-independence}
For every fixed $T>0$, there exists a constant $C>0$ such that the following holds. If the entries of $X\sim P_n$ are $D_n$-wise independent with GOE marginals for some $D_n\ge C\log n$, then $d_{\mathrm{TV}}(\spec(\mathcal U_TP_n),\spec(Q_n))=o(1)$.
\end{corollary}

\begin{proof}
Every monomial of degree at most $D_n$ involves at most $D_n$ upper-triangular entries, so $\adv_{\le D_n}(Q_n,P_n)=0$, and \cref{thm:main} applies.
\end{proof}

This gives a spectral analogue of bounded independence plus noise fooling structured tests \cite{HaramatyLeeViola2018,LeeViola2020,ForbesKelley2018}, and implies that every noisy spectral statistic shares its GOE limit law (for example, $n^{2/3}(\lambda_n(H)-2)$ converges to the Tracy--Widom law \cite{TracyWidom1996}).

\subsection{Related work}
\label{subsec:related-work}

\paragraph{Rigorous implications of low-degree bounds.}
Hsieh, Kane, Kothari, Li, Mohanty, and Tiegel \cite{HsiehKaneKothariLiMohantyTiegel2026} pioneered the systematic study of what vanishing low-degree advantage rigorously implies, proving a full total-variation result for permutation-invariant Boolean vectors and, for Gaussian matrices, controlling any fixed constant-size connected subgraph count at degree $\log n\log\log n$. Our results complement theirs and resolve the spectral problem highlighted in \cite[Remark~1.3]{HsiehKaneKothariLiMohantyTiegel2026}: rather than one constant-size local count at a time, we control the joint law of all $n$ eigenvalues down to microscopic scales with no symmetry assumption. In related settings, our companion paper \cite{ChenLi2026} controls the full noisy distribution for planted subgraphs in sparse random graphs under a treewidth condition, and Yu \cite{Yu2026} analyzes polynomial threshold functions and short cycle counts for noisy random lifts of regular graphs.

\paragraph{Spectral tests and other tests.}
The connection between low-degree polynomials and spectral methods has been studied from several angles. Algorithmically, low-degree matrix polynomials often match Sum-of-Squares relaxations \cite{HopkinsSchrammShiSteurer2016,HopkinsEtAl2017} and approximate message passing \cite{MontanariWein2025}. For spiked models, prior work characterized when PCA is optimal versus when entrywise pre-transformations help \cite{PerryWeinBandeiraMoitra2018}, studied linear spectral statistics below threshold \cite{NagdaWein2026}, and gave low-degree evidence for the optimality of spectral bounds in constrained PCA \cite{BandeiraKuniskyWein2020}. Directly comparing polynomials with spectral quantities, Kunisky, Wein, and Bandeira \cite{KuniskyWeinBandeira2022} ruled out operator-norm tests with very small failure probability, while Zhang \cite{Zhang2026} established sharp degree thresholds for approximating the top eigenvalue and eigenvector of the GOE. In the opposite direction, Buhai, Hsieh, Jain, and Kothari \cite{BuhaiHsiehJainKothari2025} gave a non-Gaussian low-rank construction where the largest eigenvalue beats low-degree polynomials, which \cref{thm:main} rules out for the GOE null with Gaussian noise. Conversely, spectral indistinguishability does not imply matrix indistinguishability: in the spirit of spectral planting \cite{BandeiraBanksKuniskyMooreWein2021}, \cref{rem:necessity} and \cref{ex:eigenvector} give noisy examples, and Forey, Fres\'an, Kowalski, and Wigderson \cite{ForeyFresanKowalskiWigderson2025} constructed explicit semicircle-spectrum graphs distinguished by a small subgraph count.

\paragraph{Dyson Brownian motion.}
Our proof builds on the dynamical approach to random matrix universality \cite{ERDOS20121435,BourgadeErdosYauYin2016,LandonYau2017LocalStatistics}, using deformed-Wigner local laws \cite{LeeSchnelli2015,LSSTY2016}, homogenization in the bulk \cite{Bourgade2022ExtremeGaps} and at the edges \cite{LandonXian2025}, the quantile comparison of \cite{AggarwalHuang2024}, and Gaussian entropy smoothing \cite{BoursierChafaiLabbe2023}. Unlike classical RMT settings with independent entries, here $X$ has no independence structure and enters only through degree-$O(\log n)$ polynomials.

\subsection{Organization and notation}
\label{subsec:organization}

\Cref{section:overview} gives a proof overview for both \cref{thm:main} and \cref{ex:eigenvector}, and \cref{sec:preliminaries} collects preliminaries. The proof of \cref{thm:main} occupies \cref{section:rigidity} (burn-in rigidity), \cref{sec:spectral-coupling} (DBM coupling and homogenization), and \cref{sec:spectral-smoothing} (Gaussian smoothing). Finally, \cref{sec:eigenvector-examples} proves \cref{ex:eigenvector}.

\paragraph{Notation.}
Throughout, $C,c,c_1,\dots$ denote positive constants that may change from line to line unless fixed explicitly (with dependencies on fixed parameters indicated explicitly). Asymptotic notation $O(\cdot)$, $o(1)$, $\lesssim$, and $\asymp$ is as $n\to\infty$. An event holds with overwhelming probability if its probability is at least $1-n^{-M}$ for every fixed $M>0$ and all sufficiently large $n$.

\section{Proof overview}
\label{section:overview}

We first outline the proof of \cref{thm:main} (Sections~\ref{subsec:overview-rigidity}--\ref{subsec:overview-smoothing}), and then sketch the eigenvector constructions for \cref{ex:eigenvector} (\cref{subsec:overview-eigenvectors}). For \cref{thm:main}, we fix a noise level $T>0$ and split the OU noise into three consecutive increments of durations $t_1,t_2,t_3>0$ with $t_1+t_2+t_3=T$, using the semigroup property $\mathcal U_T=\mathcal U_{t_3}\mathcal U_{t_2}\mathcal U_{t_1}$. The first increment acts as a short burn-in that brings every eigenvalue close to its semicircle quantile, into the rigidity tube \eqref{eq:rigidity-tube} illustrated in \cref{fig:rigidity}. During the second increment, we couple the noisy planted spectrum with a stationary GOE eigenvalue process and show that the two ordered spectra reach $2$-Wasserstein distance $o(n^{-1/2})$. The third increment then converts this $2$-Wasserstein bound into a total-variation bound. Throughout the proof, the assumption that the low-degree advantage vanishes enters only twice, first through degree-$O(\log n)$ traces and resolvent approximations in the burn-in step, and second through fixed-degree linear spectral statistics and degree-$O(\log n)$ moment bounds in the coupling step; every other estimate relies solely on the Gaussian noise.

\begin{figure}[H]
\centering
\begin{tikzpicture}[
  box/.style={draw, rounded corners=3pt, align=center, text width=3.1cm, minimum height=1.6cm, inner sep=3pt, font=\footnotesize},
  arr/.style={-{Stealth[length=2mm]}, thick},
  scale=0.92, transform shape]
\node[box, fill=gray!8] (a) at (0,0) {\textbf{Input}\\[1pt] $X\sim P_n$\\[1pt] $\adv_{\le C\log n}=o(1)$};
\node[box, fill=RoyalBlue!8] (b) at (4.1,0) {\textbf{Rigidity}\\[1pt] $\mathcal U_{t_1}P_n$, near $\gamma_i$\\ (\cref{lem:rigidity})};
\node[box, fill=OliveGreen!8] (c) at (8.2,0) {\textbf{Transport}\\[1pt] $\mathcal U_{t_1+t_2}P_n$, $nW_2^2=o(1)$\\ (\cref{lem:Homogenization})};
\node[box, fill=BrickRed!8] (d) at (12.3,0) {\textbf{Total variation}\\[1pt] $\mathcal U_TP_n$, $d_{\mathrm{TV}}=o(1)$\\ (\cref{thm:main})};
\draw[arr] (a) -- node[above, font=\scriptsize]{$t_1$} (b);
\draw[arr] (b) -- node[above, font=\scriptsize]{$t_2$} (c);
\draw[arr] (c) -- node[above, font=\scriptsize]{$t_3$} (d);
\node[font=\scriptsize, align=center, text width=3.2cm] at (4.1,-1.25) {degree-$O(\log n)$ resolvents; deformed GOE};
\node[font=\scriptsize, align=center, text width=3.2cm] at (8.2,-1.25) {homogenization; Chebyshev modes};
\node[font=\scriptsize, align=center, text width=3.2cm] at (12.3,-1.25) {Gaussian relative entropy and Pinsker};
\end{tikzpicture}
\end{figure}

\subsection{Rigidity after a short burn-in}
\label{subsec:overview-rigidity}

\begin{figure}[H]
\centering
\includegraphics[width=0.86\textwidth]{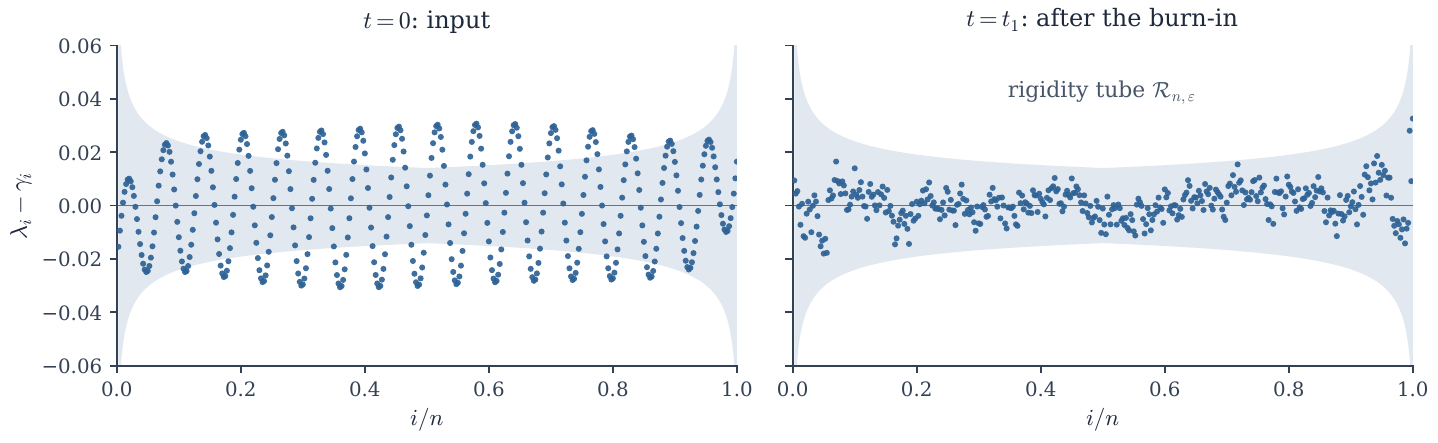}
\caption{Deviations $\lambda_i-\gamma_i$ for $n=400$. A rapidly oscillating input displacement perturbs each $\Tr X^k$ ($k\le6$) by less than a fifth of its GOE standard deviation while leaving the rigidity tube $\mathcal R_{n,\varepsilon}$ (shaded, $\varepsilon=1/4$). After a burn-in of time $t_1=0.3$, the eigenvalues lie inside the tube.}
\label{fig:rigidity}
\end{figure}

In the first step, we use a short burn-in of duration $t_1>0$ to bring every eigenvalue of the noisy planted matrix close to its semicircle classical location, supplying the microscopic regularity required by the coupling step. Let $\SC_1$ be the standard semicircle law with density $\rhosc(x)=\frac{1}{2\pi}\sqrt{(4-x^2)_+}$, and let $\gamma_i\in[-2,2]$ be its $i/n$-quantiles (the classical locations, i.e., the deterministic positions where the $n$ eigenvalues would sit if they matched $\SC_1$ exactly),
\begin{equation}
\int_{-2}^{\gamma_i}\rhosc(x)\,dx=\frac{i}{n}, \qquad 1\le i\le n,
\label{eq:classical-location}
\end{equation}
so $\gamma_n=2$. Write $\hat i:=\min\{i,n+1-i\}$ for the edge index and $\mathcal W_n:=\{\mathbf{x}\in\R^n:x_1\le\cdots\le x_n\}$ for ordered spectra. For $\varepsilon>0$, we define the rigidity tube (\cref{fig:rigidity}) as
\begin{equation}
\mathcal R_{n,\varepsilon}
:=\left\{\mathbf{x}\in\mathcal W_n:
|x_i-\gamma_i|\le\frac{n^\varepsilon}{n^{2/3}\hat i^{1/3}}
\ \text{for all }i\right\},
\label{eq:rigidity-tube}
\end{equation}
an envelope of width $O(n^{-1+\varepsilon})$ in the bulk and $O(n^{-2/3+\varepsilon})$ near the edges that contains a GOE spectrum with overwhelming probability \cite{ERDOS20121435}. While $X$ itself need not be rigid, low-degree polynomials control its global spectral shape, and a short burn-in turns this global information into rigidity (\cref{fig:rigidity}).

The following lemma, proved in \cref{section:rigidity}, formalizes the burn-in effect of \cref{fig:rigidity}.

\begin{lemma}%
  \label{lem:rigidity}
  Fix $t_1>0$, $\varepsilon>0$, and $M>0$. There exists a constant $C_1=C_1(t_1,\varepsilon,M)>0$ such that if $D_n\ge C_1\log n$ and $\adv_{\le D_n} (Q_n,P_n)=o(1)$, then for $H\sim \mathcal U_{t_1}P_n$ and all sufficiently large $n$,
\begin{equation}
\Pp\left\{\spec(H)\in\mathcal R_{n,\varepsilon}\right\}\ge 1-n^{-M}.
\label{eq:rigidity}
\end{equation}
\end{lemma}

We now outline the proof of \cref{lem:rigidity}. First, using the vanishing low-degree advantage, we establish two exterior estimates for the planted matrix $X$ that hold with probability at least $1-n^{-M}$, for any fixed $M$ once the degree constant is large enough. By controlling the trace powers $\Tr X^{2k}$ for $k=O(\log n)$, we bound the operator norm by $\|X\|_{\op}\le2+\delta$. Next, by approximating the resolvent kernels $x\mapsto(x-z)^{-j-1}$ with degree-$O(\log n)$ Chebyshev polynomials, transferring high moments from the GOE via \eqref{eq:adv-def}, and taking a union bound over a fine net, we show that the Stieltjes transform $m_X(z)=n^{-1}\Tr(X-z)^{-1}$ and its first three derivatives lie within $n^{-1+\eta}$ of those of the semicircle law $\SC_1$, uniformly on any fixed compact set away from $[-2,2]$.

Next, we turn these exterior estimates on $X$ into eigenvalue rigidity for the noisy matrix $H=e^{-t_1/2}X+\sqrt{1-e^{-t_1}}\,W$. Conditional on $X$, adding independent GOE noise smooths the empirical  spectral measure of $e^{-t_1/2}X$ into a deterministic limiting law $\mu_X$ (its free convolution with a semicircle law, defined in \cref{sec:preliminaries}). By deformed-Wigner rigidity \cite{LSSTY2016}, the eigenvalues of $H$ concentrate tightly around the quantiles of $\mu_X$, provided that $\mu_X$ is supported on a single interval with square-root edges. Our goal therefore reduces to showing that $\mu_X$ is close to the semicircle law $\SC_1$ (which arises when $X$ itself is GOE), with quantiles within $n^{-2/3+\varepsilon}\hat i^{-1/3}$ of $\gamma_i$. The key reason our exterior bounds on $X$ suffice is that adding a fixed amount of Gaussian noise washes out microscopic irregularities of $X$: analytically, the Stieltjes transform of $\mu_X$ depends on $m_X(z)$ only at complex points bounded away from $[-2,2]$ (\cref{fig:subordination}), precisely where degree-$O(\log n)$ polynomials control $m_X(z)$ to precision $n^{-1+\eta}$. Because the square-root behavior at the spectral edges makes pointwise density comparison unstable, we instead compare the cumulative distribution functions of $\mu_X$ and $\SC_1$ directly along a linear interpolation, which avoids the edge singularity and yields the required quantile bound. Applying \cite{LSSTY2016} conditionally on $X$ then establishes \cref{lem:rigidity}.

\subsection{Coupling the spectra through Dyson Brownian motion}
\label{subsec:overview-coupling}

\begin{figure}[H]
\centering
\includegraphics[width=0.86\textwidth]{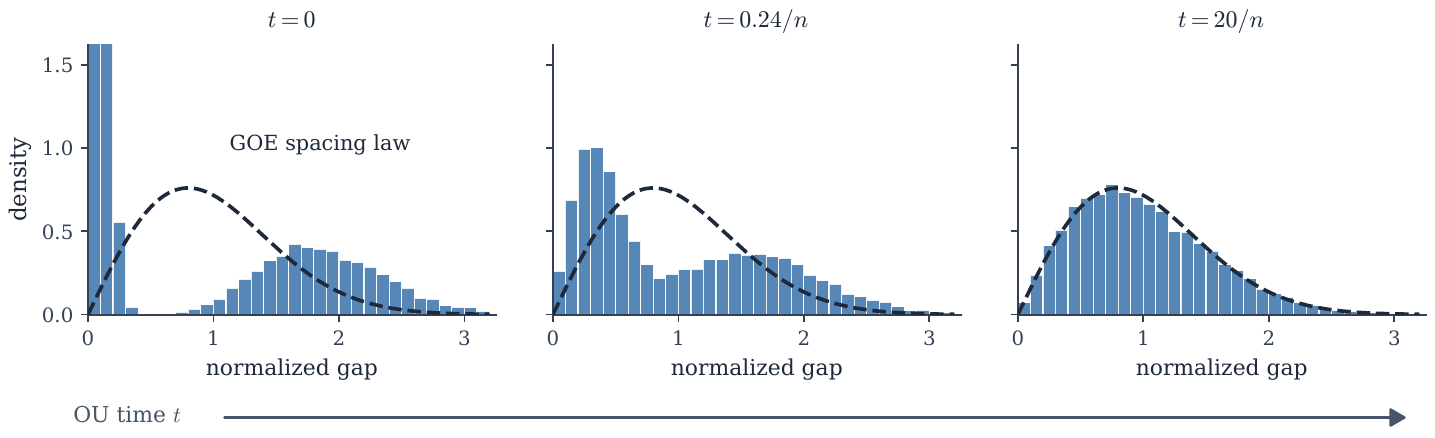}
\caption{Normalized neighboring eigenvalue gaps in $[-1.2,1.2]$ of $H_t=e^{-t/2}X+\sqrt{1-e^{-t}}\,W$ ($n=400$). Here $X$ is formed from a GOE matrix by moving each consecutive pair of eigenvalues toward its midpoint, which perturbs each $\Tr X^k$ by only $O(1/n)$. The dashed curve is the Wigner surmise $\frac{\pi s}{2}e^{-\pi s^2/4}$, a close approximation of the GOE spacing law; the artificial pairing disappears by time $O(1/n)$.}
\label{fig:simulation}
\end{figure}

Under OU noise, the spectrum evolves by the Dyson--Ornstein--Uhlenbeck equation \eqref{eq:Dyson-OU}, for which $\spec(Q_n)$ is stationary. We start one copy $\mathbf x(t)$ from $\spec(\mathcal U_{t_1}P_n)$ and another $\mathbf y(t)$ from $\spec(Q_n)$, drive both by the same Brownian motions, and choose their initial coupling below. \Cref{fig:simulation} illustrates how Gaussian noise rapidly erases local spectral structure invisible to low-degree polynomials.

Local relaxation alone does not suffice because slow global modes persist: for instance, if $X$ is a fixed rigid diagonal matrix with $\Tr X=0$, then $\Tr H_t\sim N(0,2(1-e^{-t}))$ stays at fixed total-variation distance from $\Tr W\sim N(0,2)$ at every fixed $t$. The vanishing low-degree advantage is what matches these global modes.

Our target is $n\E\|\mathbf x(t_2)-\mathbf y(t_2)\|_2^2=o(1)$. Rigidity alone gives only $n\|\mathbf x(0)-\mathbf y(0)\|_2^2\le cn^{2\varepsilon}$, and synchronous coupling contracts this merely by $e^{-t}$. We gain the missing factor from homogenization \cite{Bourgade2022ExtremeGaps,LandonXian2025}: after a fixed time $\tau$, the displacement $x_i(\tau)-y_i(\tau)$ is close, uniformly for all $i$ including the edges, to a deterministic heat kernel applied to the initial displacement (\cref{prop:regular-data-homogenization,lem:homogenization-transport-reduction}),
\begin{equation*}
x_i(\tau)-y_i(\tau)\approx\frac{e^{-\tau/2}}{n}\sum_{j=1}^np_\tau(\gamma_i,\gamma_j)\bigl(x_j(0)-y_j(0)\bigr),
\qquad
p_\tau(x,y)=\sum_{m\ge0}e^{-m\tau/2}\mathsf U_m(x)\mathsf U_m(y),
\end{equation*}
where $\mathsf U_m(x)=U_m(x/2)$ are rescaled Chebyshev polynomials of the second kind. Because $p_\tau$ is diagonal in this orthonormal basis, Parseval's identity splits the displacement energy into centered Chebyshev linear statistics $\Phi_{k,n}(\mathbf x)=\sum_i\psi_k(x_i)-\E\Tr\psi_k(W)$ with $\psi_k(x)=(2/k)T_k(x/2)$:
\begin{equation*}
n\sum_{i=1}^n\bigl(x_i(\tau)-y_i(\tau)\bigr)^2
\lesssim\sum_{k=1}^K e^{-k\tau}\bigl|\Phi_{k,n}(\mathbf{x}(0))-\Phi_{k,n}(\mathbf{y}(0))\bigr|^2
+c_\tau K^2e^{-K\tau}n^{2\varepsilon}+o(1)
\end{equation*}
on the rigidity event, for any $K=O(\log n)$ (\cref{lem:spectral-Parseval}). We handle the three mode ranges separately: for a fixed number $L$ of low modes, low-degree moment matching and the GOE CLT \cite{LytovaPastur2009} let us couple $(\Phi_{1,n},\dots,\Phi_{L,n})$ with mean squared error $o(1)$ (\cref{lem:Chebyshev-mode-coupling}); for middle modes $L<k\le K=\lfloor b\log n\rfloor$, degree-$2k$ moment transfer bounds $\E\Phi_{k,n}^2=O(k^2)$ and the weights $e^{-k\tau}$ make the sum small for large $L$; and tail modes beyond $K$ vanish once $b\tau>2\varepsilon$. Taking $n\to\infty$ then $L\to\infty$ yields the transport estimate (for the Euclidean $2$-Wasserstein metric $W_2$ on $\mathcal W_n$).

\begin{lemma}%
\label{lem:Homogenization}
Fix $t_1,t_2>0$. There exists a constant $C_2=C_2(t_1,t_2)>0$ such that if $D_n\ge C_2\log n$ and $\adv_{\le D_n}(Q_n,P_n)=o(1)$, then
\begin{equation}
nW_2^2\left(\spec(\mathcal U_{t_1+t_2}P_n),\spec(Q_n)\right)=o(1).
\label{eq:homogenization-transport}
\end{equation}
\end{lemma}

To prove the homogenization estimate \cref{prop:regular-data-homogenization}, we overcome two main technical obstacles. First, we establish simultaneous rigidity along all interpolating flows between $\mathbf x(0)$ and $\mathbf y(0)$ (\cref{subsec:interpolating-regularity}), combining \cite{AggarwalHuang2024} and L\'evy contraction away from the edges with a direct comparison to the GOE flow near the edges. Second, because our low-degree rigidity holds only up to a polynomial factor $n^\varepsilon$ rather than the subpolynomial factors assumed in \cite{LandonXian2025,Bourgade2022ExtremeGaps}, we adapt their homogenization argument by running a single induction step with control parameter $n^\delta$ (\cref{subsec:homogenization-proof}), which already yields error $n^{-1-\chi}$ per particle while keeping all degree requirements $O(\log n)$.

\subsection{From transport to total variation}
\label{subsec:overview-smoothing}

Finally, representing two ordered spectra as diagonal matrices and adding an independent OU increment of duration $t_3$ produces Gaussian matrix laws with common covariance $2(1-e^{-t_3})I/n$ in Frobenius orthonormal coordinates. By the Gaussian relative-entropy formula \cite[Lemma~A.5]{BoursierChafaiLabbe2023}, data processing under $\spec$, joint convexity over couplings, and Pinsker's inequality (\cref{lem:spectral-Gaussian-smoothing}), we obtain $d_{\mathrm{TV}}(\spec(\mathcal U_TP_n),\spec(Q_n))\le\bigl(\frac{ne^{-t_3}}{8(1-e^{-t_3})}\bigr)^{1/2}W_2(\spec(\mathcal U_{t_1+t_2}P_n),\spec(Q_n))$. Taking $t_1=t_2=t_3=T/3$ and combining with \cref{lem:Homogenization} proves \cref{thm:main}.

\subsection{Signals in the eigenvectors}
\label{subsec:overview-eigenvectors}

We now outline the two constructions in \cref{ex:eigenvector}, whose full proofs appear in \cref{sec:eigenvector-examples}. In both parts, we start from a GOE matrix $Y\sim Q_n$ and construct the planted matrix $X\sim P_n$ by keeping the eigenvalues of $Y$ intact while changing its eigenvectors, namely $X=OYO^{\mathsf T}$ for a random orthogonal matrix $O$ (which may depend on $Y$). By orthogonal invariance of the GOE, adding noise preserves the GOE spectral law, $\spec(\mathcal U_tP_n)=\spec(Q_n)$ for all $t\ge0$ (\cref{lem:noisy-eigenbasis}), so all distinguishing information must lie in the eigenvectors. Because eigenvectors are defined only up to signs (or basis rotations within a subspace), our tests observe a chosen subset $I\subset\{1,\dots,n\}$ of eigenvectors of the noisy matrix $H\sim\mathcal U_TP_n$ through the orthogonal projector $\Pi_I(H)$ onto their span.

\paragraph{Part~(a): a single noisy eigenvector.}
We keep the GOE eigenvalues of $Y$ and draw their eigenbasis uniformly from a fixed list of $n^{O(D_n)}$ random orthogonal matrices; averaging over Haar rotations shows that this makes $\adv_{\le D_n}(Q_n,P_n)=o(1)$. To detect $P_n$ from $H=rX+\sqrt s\,W$ ($r=e^{-T/2}$, $s=1-e^{-T}$), we form a deterministic candidate $A_j$ for each rotation in the list using the classical locations $\gamma_i$, and evaluate the normalized  residual $L(A,\Pi)/\sqrt{m_k}$ for $\Pi=\Pi_I(H)$ of rank $k$, where $L(A,\Pi):=\|(I_n-\Pi)A\Pi\|_F$ and $m_k:=k(n-k)/n$. This normalized residual vanishes for an exact eigenprojector of $A$ and concentrates near $1$ for an unrelated matrix, whereas under the planted law $L(X,\Pi)/\sqrt{m_k}\to\sqrt s<1$. GOE rigidity transfers this drop to the matching candidate $A_j$, while concentration and a union bound keep the null minimum near $1$. When $D_n=o(n/\log n)$, this works for $k=1$; random permutation gives relabeling invariance for $C_T\log n\le k\le n-C_T\log n$ (\cref{prop:one-eigenvector}).

\paragraph{Part~(b): a polynomial-time test from a constant fraction of eigenvectors.}
To replace list search by a polynomial-time test on $\Pi_k(H)$ ($k=\lfloor\alpha n\rfloor$), we plant Mao's binary code \cite{Mao2026} (whose small bit subsets are i.i.d.\ uniform, while polynomial lifts of its columns have a large rank deficit) into the signs of a small off-diagonal block $\Pi_{\mathcal A\mathcal B}$ of $\Pi_k(X)$ (partitioning rows into $\mathcal A$ and $\mathcal B$). Because $\Pi^2=\Pi$ constrains the Gram matrix to $\Pi_{\mathcal A\mathcal B}\Pi_{\mathcal A\mathcal B}^{\mathsf T}=\Pi_{\mathcal A\mathcal A}-\Pi_{\mathcal A\mathcal A}^2$, we cannot overwrite $\Pi_{\mathcal A\mathcal B}$ directly, but we can rotate its row space while preserving the spectrum. Specifically, we form $Z_{ij}=(-1)^{b_{ij}}|\zeta_{ij}|$ with independent standard Gaussians $\zeta_{ij}$ \cite{KlivansKothari2014}, orthonormalize its rows via $\operatorname{pol}(Z)=(ZZ^{\mathsf T})^{-1/2}Z$, and apply a block-orthogonal conjugation $X=(I_{|\mathcal A|}\oplus R)Y(I_{|\mathcal A|}\oplus R)^{\mathsf T}$ that maps the row space of $\Pi_{\mathcal A\mathcal B}(Y)$ onto $\operatorname{pol}(Z)$ (\cref{fig:coded-projector}), making the coded block equal to $\Sigma\operatorname{pol}(Z)$ with $\Sigma=(\Pi_{\mathcal A\mathcal A}-\Pi_{\mathcal A\mathcal A}^2)^{1/2}$.

Finally, we establish two main estimates to complete the proof of part~(b). First, local uniformity of the code matches low-degree moments of $Z$ to those of an i.i.d.\ Gaussian matrix (which reproduces the exact GOE); averaging over the random completion $R$ and Taylor-expanding $(ZZ^{\mathsf T})^{-1/2}$ reduces degree-$D$ moments of $X$ to degree-$O(D^3\log n)$ moments of $Z$, giving vanishing advantage at any polylogarithmic degree (\cref{lem:polar-moments}). Second, when $|\mathcal A|/n$ is small, the noiseless coded block is nearly a scalar multiple of $Z$; controlling the Frobenius perturbation of $\Pi_k(H)$ under small noise $T\le T_0$ (separating eigenvectors near and away from the cutoff) shows that only a tiny fraction of code signs flip (\cref{lem:coded-block-noise}). After random relabeling, a polynomial-time algorithm checks polynomially many sets of $O(\log n)$ rows using an $n^{-4}$-accurate projector approximation and tests the rank of the lifted columns, detecting the planted rank deficit while concentration keeps the rank
deficit below threshold for every row set under the null  (\cref{prop:poly-eigenvectors}).

\section{Preliminaries}
\label{sec:preliminaries}

In this section, we collect the basic tools used in the proof of \cref{thm:main}, namely polynomial
moment comparison from the vanishing low-degree advantage and Gaussian concentration for the GOE (\cref{subsec:prelim-ld-gauss}), spectral notation and Stieltjes transforms (\cref{subsec:prelim-spectral}), and free additive convolution with a semicircle law (\cref{subsec:prelim-fc}).

\subsection{Low-degree advantage and Gaussian concentration}
\label{subsec:prelim-ld-gauss}
A polynomial statistic is a real polynomial in the matrix coordinates, and its degree is its total degree in these coordinates. This degree is unchanged by an invertible linear change of coordinates, so it can equivalently be computed in the upper-triangular entries. For example, $B\mapsto\Tr B^k$ has degree $k$.

By \eqref{eq:adv-def} (see also \cite{KuniskyWeinBandeira2022} and \cite[Definition~1.1]{HsiehKaneKothariLiMohantyTiegel2026}), every real polynomial $q$ of degree at most $D_n$ satisfies
\begin{equation}
|\E_{P_n}q-\E_{Q_n}q| \le \adv_{\le D_n}(Q_n,P_n)\sqrt{\Var_{Q_n}q}.
\label{eq:low-degree-transfer}
\end{equation}
Applying \eqref{eq:low-degree-transfer} to powers of polynomial statistics gives moment comparisons, provided that the resulting total degree remains within $D_n$. For instance, if $q$ has degree $k$ and $p$ is a positive integer, then $q^{2p}$ is admissible whenever $2pk\le D_n$.

To bound the null variance and higher moments of trace polynomials under the GOE, we combine this transfer inequality with Gaussian concentration for Lipschitz functions \cite{Ledoux2001}, specialized to our GOE normalization as follows.

\begin{lemma}%
\label{lem:gaussian-concentration}
Let $W\sim Q_n$, and let $\Phi$ be a real-valued function on the space of $n\times n$ real symmetric matrices. Suppose that, for some $\ell>0$,
\begin{equation*}
|\Phi(B)-\Phi(B')|\le\ell\|B-B'\|_F
\end{equation*}
for all real symmetric matrices $B,B'$. Then, for every $u>0$,
\begin{equation*}
Q_n\{|\Phi(W)-\E_{Q_n}\Phi(W)|>u\}\le2\exp\left(-\frac{nu^2}{4\ell^2}\right).
\end{equation*}
Moreover, for every $p\ge1$,
\begin{equation*}
\bigl(\E_{Q_n}|\Phi(W)-\E_{Q_n}\Phi(W)|^p\bigr)^{1/p}\le c\ell\sqrt{\frac pn},
\end{equation*}
where $c>0$ is an absolute constant.
\end{lemma}

\begin{proof}
In a Frobenius orthonormal basis of the real symmetric matrices, the GOE has covariance $(2/n)I$ (the basis elements $E_{aa}$ and $(E_{ab}+E_{ba})/\sqrt2$ for $a<b$ have coordinates $W_{aa}$ and $\sqrt2\,W_{ab}$, each of variance $2/n$), so $W=\sqrt{2/n}\,\sum_ag_aV_a$ with $g$ standard Gaussian, and $g\mapsto\Phi(W(g))$ is $\ell\sqrt{2/n}$-Lipschitz. The tail bound is the Gaussian concentration inequality \cite{Ledoux2001}, and integrating it gives the moment bound.
\end{proof}

\subsection{Spectral notation}
\label{subsec:prelim-spectral}
We next fix the matrix norms, empirical spectral measures, and Stieltjes-transform conventions used throughout the proof. For a real symmetric matrix $B$, write $\|B\|_F=(\Tr B^2)^{1/2}$ for its Frobenius norm and $\|B\|_{\op}=\max_i|\lambda_i(B)|$ for its operator norm. For vectors, $\|\cdot\|_2$ and $\|\cdot\|_\infty$ denote the Euclidean and maximum norms, respectively.
Functions of a real symmetric matrix $B$ are defined through its spectral decomposition, and the empirical spectral measure of $B$ is $\widehat{\nu}_B:=n^{-1}\sum_{i=1}^n\delta_{\lambda_i(B)}$. Thus
\begin{equation}
\frac1n\Tr f(B)=\frac1n\sum_{i=1}^n f(\lambda_i(B))=\int_\R f\,d\widehat\nu_B
\label{eq:spectral-trace}
\end{equation}
for every function $f$ defined on the spectrum. The Hoffman--Wielandt inequality $\sum_{i}|\lambda_i(B)-\lambda_i(B')|^2\le\|B-B'\|_F^2$ \cite{HoffmanWielandt1953} (see also \cite[Lemma~2.1.19]{AGZ2010}) and the Cauchy--Schwarz inequality show that, if $f:\R\to\R$ is $\ell$-Lipschitz, then
\begin{equation}
|\Tr f(B)-\Tr f(B')|\le\ell\sqrt n\,\|B-B'\|_F.
\label{eq:Lipschitz-spectral-trace}
\end{equation}

To compare empirical spectral measures with the semicircle law in \cref{section:rigidity,sec:spectral-coupling}, we work with distribution functions and Stieltjes transforms \cite[Section~2.4]{AGZ2010}. For a probability measure $\mu$ on $\R$, write $F_\mu(E):=\mu(( -\infty,E])$ for its cumulative distribution function. In particular, $F_{\mathrm{sc}}:=F_{\SC_1}$.
For a finite Borel measure $\nu$ on $\R$, we use the Stieltjes-transform convention
\begin{equation*}
m_\nu(z):=\int_{\R}\frac{1}{x-z}\,\nu(dx), \qquad z\in\C\setminus\supp\nu.
\end{equation*}
With this convention the standard semicircle transform is
\begin{equation*}
m_{\mathrm{sc}}(z)=\frac{-z+\sqrt{z^2-4}}{2},
\end{equation*}
where the branch of the square root is chosen so that $\sqrt{z^2-4}\sim z$ as $|z|\to\infty$ and $\Im m_{\mathrm{sc}}(z)>0$ for $\Im z>0$.

For an $n\times n$ real symmetric matrix $B$, write
\begin{equation*}
m_B(z):=\frac1n\Tr(B-z)^{-1}=\frac1n\sum_{i=1}^n\frac{1}{\lambda_i(B)-z}, \qquad z\in\C\setminus\{\lambda_1(B),\dots,\lambda_n(B)\}.
\end{equation*}
Differentiating this finite sum gives, for every integer $j\ge0$,
\begin{equation}
m_B^{(j)}(z)=\frac{j!}{n}\Tr(B-z)^{-j-1}=\frac1n\sum_{i=1}^n\frac{j!}{(\lambda_i(B)-z)^{j+1}}.
\label{eq:resolvent-trace-derivatives}
\end{equation}

\subsection{Free convolution with a semicircle law}
\label{subsec:prelim-fc}
In \cref{section:rigidity}, conditional on the planted matrix $X$, the burn-in matrix $rX+\sqrt s\,W$ is a deformed GOE matrix whose spectrum is governed by the free additive convolution of $\widehat\nu_{rX}$ with a semicircle law. Here we record the analytic characterization of this operation. For $s>0$, let $\SC_s$ denote the centered semicircle law of variance $s$, with density
\begin{equation*}
\rho_{\SC_s}(x)=\frac{1}{2\pi s}\sqrt{(4s-x^2)_+}.
\end{equation*}
Thus $\supp\SC_s=[-2\sqrt s,2\sqrt s]$, and $\SC_1$ is the standard semicircle law from \cref{subsec:overview-rigidity}. The operation $\boxplus$ denotes free additive convolution \cite{Voiculescu1986,BercoviciVoiculescu1993}. For the convolution with a semicircle law used here, this operation is specified by the analytic characterization below.

Biane \cite[Sections~2--3]{Biane1997} describes this convolution through an analytic change of spectral parameter, called subordination. In our Stieltjes convention, $m(z)=m_{\nu\boxplus\SC_s}(z)$ is the solution of the Pastur equation \cite{Pastur1972}
\begin{equation}
m(z)=\int_\R\frac{\nu(dx)}{x-z-sm(z)}=m_\nu(z+sm(z)),\qquad z\in\C^+,
\label{eq:Pastur-equation}
\end{equation}
where $\C^+=\{z\in\C:\Im z>0\}$. We take the unique analytic solution satisfying $m:\C^+\to\C^+$ and $\lim_{y\to\infty}\mathrm{i}y\,m(\mathrm{i}y)=-1$. The resulting deterministic probability measure is denoted by $\nu\boxplus\SC_s$.

To track how $\nu\boxplus\SC_s$ responds to exterior perturbations of $m_\nu$, we isolate the subordination function and its inverse, the Pastur map, defined by
\begin{equation*}
\omega_\nu(z):=z+s\,m_{\nu\boxplus\SC_s}(z),\qquad
\pastur_\nu(w):=w-sm_\nu(w),\quad w\in\C\setminus\supp\nu.
\end{equation*}
Then
\begin{equation}
\pastur_\nu(\omega_\nu(z))=z,\qquad
m_{\nu\boxplus\SC_s}(z)=m_\nu(\omega_\nu(z)),\qquad z\in\C^+.
\label{eq:subordination-equations}
\end{equation}
Thus $\pastur_\nu$ maps $\omega_\nu(\C^+)$ conformally onto $\C^+$, with inverse $\omega_\nu$. The subordination function satisfies $\Im\omega_\nu(z)\ge\Im z$ and $\omega_\nu(\mathrm{i}y)/(\mathrm{i}y)\to1$. The map $\pastur_\nu$ takes the subordination parameter $w$ back to the spectral parameter $z$. The noise variance $s$ is fixed in each application and is suppressed in this notation.

The boundary regularity in \cite[Section~3]{Biane1997} gives a continuous density, denoted by $\rho_{\mathrm{fc},\nu}$, for $\nu\boxplus\SC_s$, with the uniform bound $\rho_{\mathrm{fc},\nu}\le1/(\pi\sqrt s)$. We write $g(E+\mathrm{i}0):=\lim_{\eta\downarrow0}g(E+\mathrm{i}\eta)$ for an upper-half-plane boundary value. Stieltjes inversion and the definition of $\omega_\nu$ give
\begin{equation*}
\rho_{\mathrm{fc},\nu}(E)
=\frac1\pi\Im m_{\nu\boxplus\SC_s}(E+\mathrm{i}0)
=\frac{1}{\pi s}\Im\omega_\nu(E+\mathrm{i}0).
\end{equation*}
For a compactly supported free-convolution law, we write $E_-(\nu)$ and $E_+(\nu)$ for its leftmost and rightmost support points. When the support is the single interval $[E_-(\nu),E_+(\nu)]$, its endpoints are the spectral edges, and a regular square-root edge means that the density is comparable to the square root of the distance to the edge from inside the support.

\section{Rigidity from vanishing low-degree advantage}
\label{section:rigidity}

In this section, we prove \cref{lem:rigidity} by adapting the deformed-Wigner rigidity estimates of \cite{LSSTY2016} to the burn-in matrix $H=rX+\sqrt s\,W\sim\mathcal U_{t_1}P_n$, where $t_1>0$ is fixed, $r=e^{-t_1/2}$, $s=1-r^2$, and $X\sim P_n$ is independent of $W\sim Q_n$. To verify the hypotheses of \cite{LSSTY2016} conditionally on $X$, we proceed in three steps. In \cref{subsec:norm-resolvent}, we use the low-degree assumption to prove high-probability operator-norm and exterior resolvent bounds for $X$. In \cref{subsec:pastur-stability}, we translate these exterior bounds via the Pastur equation into regularity of the free convolution $\widehat\nu_{rX}\boxplus\SC_s$ and a quantile comparison against the semicircle classical locations $\gamma_i$. Finally, in \cref{subsec:burnin-rigidity}, we apply deformed-Wigner rigidity conditionally on $X$ and integrate over $X$ to prove \cref{lem:rigidity}.
\subsection{Norm and exterior resolvent estimates}
\label{subsec:norm-resolvent}
In this subsection, we establish high-probability operator-norm and exterior resolvent bounds under $P_n$ by comparing moments of degree-$O(\log n)$ trace polynomials with their GOE counterparts. To bound the operator norm, we control the moments $\Tr X^{2k}$ for $k=O(\log n)$, which shows that $\|X\|_{\op}\le 2+\delta$ with high probability. To control the Stieltjes transform $m_X(z)$ and its first three derivatives at a fixed distance from $[-2,2]$, we approximate the analytic resolvent kernels $x\mapsto j!(x-z)^{-j-1}$ on $[-2-\delta,2+\delta]$ by degree-$O(\log n)$ Chebyshev polynomials up to error $n^{-3}$. In both arguments, we first bound high moments of these polynomial statistics under the GOE null $Q_n$, transfer those bounds to $P_n$ via \eqref{eq:low-degree-transfer}, and apply Markov's inequality (together with a union bound over a fine net for the resolvent) to obtain the desired high-probability estimates.

We begin by controlling the operator norm under both the GOE null and the planted distribution.

\begin{lemma}%
\label{lem:norm-confinement}
Let $W\sim Q_n$. For every fixed $M>0$ and all sufficiently large $n$,
\begin{equation}
Q_n\{\|W\|_{\op}>2+n^{-1/2}\}\le n^{-M}.
\label{eq:null-norm-shrinking}
\end{equation}
There are absolute constants $c_0$ and $n_0$ such that, for $n\ge n_0$,
\begin{equation}
\bigl(\E_{Q_n}\|W\|_{\op}^p\bigr)^{1/p}\le2+c_0\sqrt{\frac pn}, \qquad p\ge1.
\label{eq:null-norm-moments}
\end{equation}
Fix $\delta>0$ and $M>0$. There are $c=c(\delta,M)$ and $n_1=n_1(\delta,M)$ such that, if $D_n\ge c\log n$ and $\adv_{\le D_n}(Q_n,P_n)\le1$, then $P_n\{\|X\|_{\op}>2+\delta\}\le n^{-M}$ for all $n\ge n_1$.
\end{lemma}

\begin{proof} By the GOE extreme-eigenvalue tail bound of Ledoux and Rider \cite[Theorem~1]{LedouxRider2010} (together with the symmetry $W\stackrel{d}{=}-W$), we have $Q_n\{\|W\|_{\op}\ge 2(1+\varepsilon)\}\le C\exp(-cn\varepsilon^{3/2})$ for all $\varepsilon\in(0,1]$. Setting $\varepsilon=\frac12 n^{-1/2}$ yields $Q_n\{\|W\|_{\op}>2+n^{-1/2}\}\le C\exp(-c'n^{1/4})\le n^{-M}$ for all sufficiently large $n$, which proves \eqref{eq:null-norm-shrinking}. For \eqref{eq:null-norm-moments}, we apply \eqref{eq:null-norm-shrinking} with $M=2$ and use $\E_{Q_n}\|W\|_{\op}^2\le\E_{Q_n}\Tr W^2=n+1$. The Cauchy--Schwarz inequality gives $\E_{Q_n}\|W\|_{\op}\le2+n^{-1/2}+(n+1)^{1/2}n^{-1}\le2+3n^{-1/2}$. The operator norm is $1$-Lipschitz for the Frobenius norm, so the moment bound of \cref{lem:gaussian-concentration} and Minkowski's inequality give \eqref{eq:null-norm-moments}.

For the planted bound, let $k=\lceil c_1\log n\rceil$, with $c_1=c_1(\delta,M)$ chosen below, and take $c\ge2c_1+2$, so that $q(B)=\Tr B^{2k}$ has degree $2k\le D_n$. By the Cauchy--Schwarz inequality over the eigenvalues, $(\Tr W^{2k})^2\le n\Tr W^{4k}\le n^2\|W\|_{\op}^{4k}$. Hence \eqref{eq:null-norm-moments} with $p=4k$ gives, for large $n$,
\begin{equation*}
\|\Tr W^{2k}\|_{L^2(Q_n)}\le n\Bigl(2+c_0\sqrt{4k/n}\Bigr)^{2k}\le 2n\cdot 2^{2k},
\end{equation*}
since $k^{3/2}n^{-1/2}\to0$. By \eqref{eq:low-degree-transfer} and $\adv_{\le D_n}(Q_n,P_n)\le1$,
\begin{equation*}
\E_{P_n}\Tr X^{2k}\le\E_{Q_n}\Tr W^{2k}+\sqrt{\Var_{Q_n}\Tr W^{2k}}\le2\|\Tr W^{2k}\|_{L^2(Q_n)}\le4n\cdot2^{2k}.
\end{equation*}
Since $\|X\|_{\op}^{2k}\le\Tr X^{2k}$, Markov's inequality gives
\begin{equation*}
P_n\{\|X\|_{\op}>2+\delta\}\le4n\Bigl(\frac{2}{2+\delta}\Bigr)^{2k}\le n^{-M}
\end{equation*}
once $c_1$ is large in terms of $\delta$ and $M$. Every threshold for $n$ in this argument comes from a null estimate, so it does not depend on $P_n$.
\end{proof}

With the operator norm of $X$ confined to $[-2-\delta,2+\delta]$, we next show that degree-$O(\log n)$ polynomials also control the Stieltjes transform $m_X(z)$ and its first three derivatives uniformly on any compact set away from $[-2,2]$.

\begin{lemma}%
\label{lem:exterior-resolvent}
Let $X\sim P_n$ be the planted matrix, and write $m_X(z)=n^{-1}\Tr(X-z)^{-1}$. Fix a nonempty compact set $\mathcal K\subset\C\setminus[-2,2]$ and constants $M,\eta>0$. There exists $c=c(\mathcal K,M,\eta)>0$ such that, if
\begin{equation*}
D_n\ge c\log n, \qquad \adv_{\le D_n}(Q_n,P_n)\le1,
\end{equation*}
then, for all $n\ge n_0(\mathcal K,M,\eta)$, with $n_0$ independent of $P_n$, with probability at least $1-n^{-M}$ over $X\sim P_n$, the spectrum of $X$ is disjoint from $\mathcal K$ and
\begin{equation}
\sup_{z\in\mathcal K}\max_{0\le j\le3} \bigl|m_X^{(j)}(z)-m_{\mathrm{sc}}^{(j)}(z)\bigr| \le n^{-1+\eta}.
\label{eq:uniform-exterior-resolvent}
\end{equation}
Here $m^{(j)}$ denotes the $j$-th derivative with respect to $z$.
\end{lemma}

\begin{proof}
We approximate the kernels $x\mapsto j!(x-z)^{-j-1}$ by Chebyshev polynomials of degree $k=O(\log n)$. Under the null, GOE rigidity makes the resulting trace statistics accurate to $n^{-1+\eta/2}$, which bounds a moment of fixed order. We transfer this moment to $P_n$, apply Markov's inequality at each point of a polynomial net in $\mathcal K$, and extend the bound to all of $\mathcal K$ by continuity. We may assume $\eta\le1$.

\emph{Polynomial approximation.} Since $\mathcal K$ is compact and disjoint from $[-2,2]$, we can choose $\delta_0\in(0,1)$ and a closed region $\mathcal B$ bounded by a Bernstein ellipse with foci $\pm(2+\delta_0)$ such that $d_{\mathcal K}:=\dist(\mathcal K,\mathcal B)>0$. For $z\in\mathcal K$ and $0\le j\le3$, the function $f_{j,z}(x):=j!(x-z)^{-j-1}$ is analytic in a neighborhood of $\mathcal B$ and bounded there uniformly in $z$. Let $g_{j,z}$ be its Chebyshev truncation of degree $k=\lceil c_1\log n\rceil$ on $[-2-\delta_0,2+\delta_0]$. The Bernstein-ellipse estimates of \cite[Chapter~8]{Trefethen2013} bound the Chebyshev coefficients of $f_{j,z}$ by a constant $c_{\mathcal K}$, and give
\begin{equation}
\sup_{z\in\mathcal K}\max_{0\le j\le3}\sup_{|x|\le2+\delta_0}|f_{j,z}(x)-g_{j,z}(x)|\le n^{-3}
\label{eq:cheb-error}
\end{equation}
once $c_1=c_1(\mathcal K)$ is large. For real $y$, the Chebyshev polynomials satisfy $|T_m(y)|\le(1+2|y|)^m$ (applied to $y=x/(2+\delta_0)$, for which $1+2|y|\le1+|x|$), so the coefficient bound gives
\begin{equation}
|g_{j,z}(x)|\le c_{\mathcal K}(k+1)(1+|x|)^k,\qquad x\in\R.
\label{eq:cheb-growth}
\end{equation}
Define $F_{j,z}(B):=n^{-1}\Tr g_{j,z}(B)$. Its real and imaginary parts are real polynomials of degree at most $k$ in the entries of $B$. By \eqref{eq:spectral-trace}, \eqref{eq:resolvent-trace-derivatives} and \eqref{eq:cheb-error}, every real symmetric $B$ with $\|B\|_{\op}\le2+\delta_0$ satisfies
\begin{equation}
|F_{j,z}(B)-m_B^{(j)}(z)|\le n^{-3},\qquad z\in\mathcal K,\ 0\le j\le3.
\label{eq:poly-to-resolvent-X}
\end{equation}

\emph{Null moments.} Put $\zeta:=\eta/2$. We claim that, for every fixed integer $p\ge1$,
\begin{equation}
\sup_{z\in\mathcal K}\max_{0\le j\le3} \E_{Q_n}\bigl|F_{j,z}(W)-m_{\mathrm{sc}}^{(j)}(z)\bigr|^{4p} \le c_p n^{-4p(1-\zeta)}.
\label{eq:null-analytic-moment}
\end{equation}
Let $\mathcal E$ be the event that $\spec(W)\in\mathcal R_{n,\zeta/2}$. Applying the generalized Wigner rigidity estimate \cite[Theorem~2.2]{ERDOS20121435} to the unit-row-sum rescaling $W'=\sqrt{n/(n+1)}\,W$, together with $|\lambda_i(W)-\lambda_i(W')|\le\|W'\|_{\op}/n$, gives $Q_n(\mathcal E^{\mathrm c})\le n^{-M'}$ for every fixed $M'$ and large $n$. On $\mathcal E$ we have $\|W\|_{\op}\le2+\delta_0$, and we compare $m_W^{(j)}$ with $m_{\mathrm{sc}}^{(j)}$ directly. Put $\gamma_0:=-2$, so that each interval $[\gamma_{i-1},\gamma_i]$ carries semicircle mass $1/n$. Then
\begin{equation*}
m_W^{(j)}(z)-m_{\mathrm{sc}}^{(j)}(z)
=\sum_{i=1}^n\int_{\gamma_{i-1}}^{\gamma_i}\bigl(f_{j,z}(\lambda_i(W))-f_{j,z}(x)\bigr)\rhosc(x)\,dx.
\end{equation*}
For $x\in[\gamma_{i-1},\gamma_i]$, the segment between $x$ and $\lambda_i(W)$ lies in $[-2-\delta_0,2+\delta_0]$, at distance at least $d_{\mathcal K}$ from $z$. With the semicircle spacing $\gamma_i-\gamma_{i-1}\le cn^{-2/3}\hat i^{-1/3}$, this gives
\begin{equation*}
|f_{j,z}(\lambda_i(W))-f_{j,z}(x)|\le c_{\mathcal K}\bigl(|\lambda_i(W)-\gamma_i|+\gamma_i-\gamma_{i-1}\bigr)\le c_{\mathcal K}n^{\zeta/2}n^{-2/3}\hat i^{-1/3}.
\end{equation*}
Since $\sum_i\hat i^{-1/3}\le3n^{2/3}$, we obtain $|m_W^{(j)}(z)-m_{\mathrm{sc}}^{(j)}(z)|\le c_{\mathcal K}n^{-1+\zeta/2}$ on $\mathcal E$, and with \eqref{eq:poly-to-resolvent-X} also $|F_{j,z}(W)-m_{\mathrm{sc}}^{(j)}(z)|\le n^{-1+\zeta}$ for large $n$.
On $\mathcal E^{\mathrm c}$ we use \eqref{eq:cheb-growth} and \eqref{eq:null-norm-moments} with exponent $8pk=O(\log n)$.
Since $|F_{j,z}(W)|\le c_{\mathcal K}(k+1)(1+\|W\|_{\op})^k$ and $\|(1+\|W\|_{\op})^k\|_{L^{8p}(Q_n)}\le(3+c_0\sqrt{8pk/n})^k\le4^k$, they give $(\E_{Q_n}|F_{j,z}(W)-m_{\mathrm{sc}}^{(j)}(z)|^{8p})^{1/2}\le n^{c_p'}$, where $c_p'$ depends on $p$ and $c_1$. By the Cauchy--Schwarz inequality, with $M'$ large, the contribution of $\mathcal E^{\mathrm c}$ to \eqref{eq:null-analytic-moment} is at most $n^{-4p}$. This proves the claim.

\emph{Transfer to $P_n$.} For fixed $j$ and $z$, the function
\begin{equation*}
q_{j,z}(B):=|F_{j,z}(B)-m_{\mathrm{sc}}^{(j)}(z)|^{2p}=\Bigl(\bigl(\Re F_{j,z}(B)-\Re m_{\mathrm{sc}}^{(j)}(z)\bigr)^2+\bigl(\Im F_{j,z}(B)-\Im m_{\mathrm{sc}}^{(j)}(z)\bigr)^2\Bigr)^p
\end{equation*}
is a real polynomial of degree at most $2pk$, and $2pk\le D_n$ once $c\ge2p(c_1+1)$. By \eqref{eq:null-analytic-moment}, $\E_{Q_n}q_{j,z}+\sqrt{\Var_{Q_n}q_{j,z}}\le2(\E_{Q_n}q_{j,z}^2)^{1/2}\le2c_p^{1/2}n^{-2p(1-\zeta)}$, and \eqref{eq:low-degree-transfer} gives the same bound for $\E_{P_n}q_{j,z}$. Markov's inequality then yields
\begin{equation*}
P_n\Bigl\{|F_{j,z}(X)-m_{\mathrm{sc}}^{(j)}(z)|>\tfrac12n^{-1+\eta}\Bigr\}\le c_pn^{-p\eta}.
\end{equation*}

\emph{Net.} Choose $\mathcal Z_n\subset\mathcal K$ with $|\mathcal Z_n|\le c_{\mathcal K}n^6$ such that every point of $\mathcal K$ is within $n^{-3}$ of $\mathcal Z_n$. Choose the integer $p=p(M,\eta)$ with $p\eta\ge M+7$, and then $c$ as above and at least the constant of \cref{lem:norm-confinement} for $(\delta_0,M+1)$. A union bound over $\mathcal Z_n\times\{0,1,2,3\}$, together with \cref{lem:norm-confinement}, shows that with probability at least $1-n^{-M}$ we have $\|X\|_{\op}\le2+\delta_0$ and $|F_{j,z_0}(X)-m_{\mathrm{sc}}^{(j)}(z_0)|\le\frac12n^{-1+\eta}$ for all $z_0\in\mathcal Z_n$ and $j\le3$. On this event, the spectrum of $X$ is disjoint from $\mathcal K$, and $m_X^{(j+1)}$ and $m_{\mathrm{sc}}^{(j+1)}$ are bounded by $c_{\mathcal K}$ on the $d_{\mathcal K}/2$-neighborhood of $\mathcal K$. For $z\in\mathcal K$ and $z_0\in\mathcal Z_n$ with $|z-z_0|\le n^{-3}$, \eqref{eq:poly-to-resolvent-X} gives
\begin{equation*}
|m_X^{(j)}(z)-m_{\mathrm{sc}}^{(j)}(z)|\le\tfrac12n^{-1+\eta}+n^{-3}+c_{\mathcal K}n^{-3}\le n^{-1+\eta}.
\end{equation*}
All thresholds for $n$ come from null estimates, from \eqref{eq:cheb-error} and from the net, so they depend only on $\mathcal K$, $M$ and $\eta$.
\end{proof}

\subsection{Stability of the free convolution}
\label{subsec:pastur-stability}
In this subsection, we consider a deterministic probability measure $\nu$ supported near $[-2r,2r]$ whose Stieltjes transform $m_\nu$ and its first three derivatives lie within $\xi$ of those of $\SC_{r^2}$ away from the support. We show that its free convolution $\nu\boxplus\SC_s$ stays close to $\SC_1=\SC_{r^2}\boxplus\SC_s$ in support, density regularity, distribution function, and quantiles. Specifically, after setting up the reference subordination geometry for $\SC_{r^2}\boxplus\SC_s$, we prove that $\nu\boxplus\SC_s$ is supported on a single interval with edges within $O(\xi)$ of $\pm2$ (\cref{lem:pastur-one-cut}). We then interpolate linearly between $\SC_{r^2}$ and $\nu$ to control the subordination boundary curves (\cref{lem:pastur-boundary}), deduce bulk and square-root edge regularity for the density (\cref{lem:pastur-regularity}), and bound the difference of distribution functions by $O(\xi)$ uniformly on $\R$ (\cref{lem:pastur-cdf}). Finally, we convert these edge and distribution-function bounds into quantile comparisons (\cref{lem:quantile-comparison}).

To set up the reference subordination geometry for the burn-in time $t_1>0$, we recall that $r=e^{-t_1/2}$ and $s=1-r^2$, and take $\SC_{r^2}$ as our reference deformation because $\SC_{r^2}\boxplus\SC_s=\SC_1$. Its Stieltjes transform is
\begin{equation}
m_r(w):=m_{\SC_{r^2}}(w)=\frac1r\,m_{\mathrm{sc}}(w/r),
\label{eq:mr-def}
\end{equation}
and we write $\pastur_r:=\pastur_{\SC_{r^2}}$ for its Pastur map. By \eqref{eq:subordination-equations}, the inverse branch of $\pastur_r$ is $\omega_0(z)=z+s\,m_{\mathrm{sc}}(z)$, with boundary values
\begin{equation}
\omega_0(E)=\frac{u_0}{2}E+\mathrm{i}\frac s2\sqrt{4-E^2},\qquad -2\le E\le2,\qquad u_0:=1+r^2.
\label{eq:omega0-def}
\end{equation}
The curve $\Gamma_0:=\omega_0([-2,2])$ is the upper half of the ellipse with foci $\pm2r$ and real vertices $\pm u_0$. Equivalently, $\Gamma_0=\{u+\mathrm{i}h_0(u):|u|\le u_0\}$ with $h_0(u):=s(1-u^2/u_0^2)^{1/2}$. Its distance from $[-2r,2r]$ is $u_0-2r=(1-r)^2$, attained at the real vertices (by the focal property of the ellipse, every point on $\Gamma_0$ lies at distance at least $u_0-2r$ from the focal segment $[-2r,2r]$). In the Joukowski variable $w=r(\zeta+\zeta^{-1})$, $|\zeta|>1$, we have $m_r(w)=-1/(r\zeta)$ and hence $\pastur_r(w)=r\zeta+(r\zeta)^{-1}$ (whose derivative in $\zeta$ vanishes only at $\zeta=\pm r^{-1}$, corresponding to $w=\pm u_0$). Therefore the only zeros of $\pastur_r'$ in $\C\setminus[-2r,2r]$ are $\pm u_0$, and
\begin{equation}
\pastur_r(\pm u_0)=\pm2,\qquad \pastur_r'(\pm u_0)=0,\qquad \pastur_r''(\pm u_0)=\pm\frac2{s^2},\qquad m_r'(\pm u_0)=\frac1s.
\label{eq:reference-critical}
\end{equation}

\begin{figure}[H]
\centering
\definecolor{figblue}{HTML}{2E6295}%
\definecolor{figband}{HTML}{DCE7F3}%
\definecolor{figplum}{HTML}{8E244D}%
\definecolor{figslate}{HTML}{475569}%
\begin{tikzpicture}[
axis/.style={-{Stealth[length=1.9mm]}, figslate!70, thin},
arr/.style={-{Stealth[length=2.1mm]}, thick, figslate},
scale=0.95]
\begin{scope}[shift={(0,0)}]
\draw[figband, line width=0.52cm, line cap=round]
(2.65,0) arc[start angle=0, end angle=180, x radius=2.65, y radius=2.15];
\draw[axis] (-3.65,0) -- (3.80,0) node[right, font=\footnotesize, text=figslate] {$\Re w$};
\draw[axis] (0,-0.45) -- (0,2.95) node[above, font=\footnotesize, text=figslate] {$\Im w$};
\draw[figslate!60, thick] (-1.85,0.08) -- (-1.85,-0.08) (1.85,0.08) -- (1.85,-0.08);
\draw[figslate!45, line width=2.3pt] (-1.85,0) -- (1.85,0);
\draw[black!85, line width=2.8pt] (-1.55,0) -- (1.55,0);
\node[font=\scriptsize, text=black!85] at (0,-0.32) {$[-2r,2r]\subset I_r$};
\draw[figblue, very thick]
(2.65,0) arc[start angle=0, end angle=180, x radius=2.65, y radius=2.15];
\draw[figplum, thick, densely dashed]
(-2.79,0) .. controls (-2.64,1.58) and (-1.15,2.28) .. (0,2.00)
.. controls (1.15,1.72) and (2.38,1.42) .. (2.51,0);
\fill[figblue] (-2.65,0) circle (1.6pt) (2.65,0) circle (1.6pt);
\fill[figplum] (-2.79,0) circle (1.5pt) (2.51,0) circle (1.5pt);
\node[font=\scriptsize, text=figblue] at (-2.42,-0.33) {$-u_0$};
\node[font=\scriptsize, text=figblue] at (2.95,-0.33) {$u_0$};
\draw[thin, figplum!75] (-2.79,0.06) -- (-3.15,0.52);
\node[font=\scriptsize, text=figplum, above] at (-3.22,0.50) {$u_-(\theta)$};
\draw[thin, figplum!75] (2.51,0.06) -- (2.10,0.54);
\node[font=\scriptsize, text=figplum, left] at (2.12,0.56) {$u_+(\theta)$};
\node[font=\footnotesize, text=figblue] at (-2.12,2.10) {$\Gamma_0$};
\node[font=\footnotesize, text=figplum] at (1.02,1.38) {$\Gamma_\theta$};
\draw[thin, figblue!70] (2.28,1.42) -- (2.78,1.88);
\node[font=\footnotesize, text=figblue!90!black, right] at (2.76,1.90) {$\mathcal N$};
\node[font=\footnotesize, text=figslate] at (-0.82,2.65) {$\Omega_\theta$};
\draw[thin, figslate!45, densely dotted] (1.85,-0.10) -- (1.85,-0.66);
\draw[thin, figslate!45, densely dotted] (2.65,-0.10) -- (2.65,-0.66);
\draw[{Stealth[length=1.4mm]}-{Stealth[length=1.4mm]}, figslate, thin]
(1.86,-0.62) -- (2.64,-0.62);
\node[font=\scriptsize, text=figslate, below] at (2.25,-0.65) {$\tfrac78(1-r)^2$};
\node[font=\footnotesize\bfseries, text=black!85] at (0,-1.55) {Subordination plane ($w$)};
\end{scope}
\draw[arr] (4.20,1.55) to[bend left=18] node[above, font=\footnotesize, text=black!85] {$\pastur_\theta$} (5.65,1.55);
\draw[arr] (5.65,0.92) to[bend left=18] node[below, font=\footnotesize, text=black!85] {$\omega_\theta$} (4.20,0.92);
\begin{scope}[shift={(8.95,0)}]
\draw[axis] (-2.95,0) -- (3.05,0) node[right, font=\footnotesize, text=figslate] {$\Re z$};
\draw[axis] (0,-0.45) -- (0,2.95) node[above, font=\footnotesize, text=figslate] {$\Im z$};
\fill[figband!85] (-1.95,0) arc[start angle=180, end angle=0, x radius=1.95, y radius=1.95] -- cycle;
\draw[figblue!85, thick] (-1.95,0) arc[start angle=180, end angle=0, x radius=1.95, y radius=1.95];
\draw[figplum, thick, densely dashed]
(-2.12,0) .. controls (-2.12,1.32) and (-0.98,2.10) .. (0,1.86)
.. controls (0.98,1.64) and (2.10,1.25) .. (2.10,0);
\draw[figblue, very thick] (-1.95,0) -- (1.95,0);
\fill[figblue] (-1.95,0) circle (1.6pt) (1.95,0) circle (1.6pt);
\fill[figplum] (-2.12,0) circle (1.5pt) (2.10,0) circle (1.5pt);
\node[font=\scriptsize, text=figblue] at (-1.78,-0.32) {$-2$};
\node[font=\scriptsize, text=figblue] at (1.78,-0.32) {$2$};
\draw[thin, figplum!75] (-2.12,0.06) -- (-2.42,0.52);
\node[font=\scriptsize, text=figplum, above] at (-2.48,0.50) {$E_-(\theta)$};
\draw[thin, figplum!75] (2.10,0.06) -- (2.40,0.52);
\node[font=\scriptsize, text=figplum, above] at (2.46,0.50) {$E_+(\theta)$};
\node[font=\footnotesize, text=figblue] at (-1.52,2.08) {$\SC_1$};
\node[font=\footnotesize, text=figplum] at (1.88,2.08) {$\mu_\theta=\nu_\theta\boxplus\SC_s$};
\node[font=\footnotesize, text=figslate] at (2.45,2.60) {$\C^+$};
\node[font=\footnotesize\bfseries, text=black!85] at (0,-1.55) {Spectral plane ($z$)};
\end{scope}
\end{tikzpicture}
\caption{Subordination geometry for the free convolution $\mu_\theta=\nu_\theta\boxplus\SC_s$ in \cref{subsec:pastur-stability}. Because the reference curve $\Gamma_0$ stays at distance at least $\frac78(1-r)^2$ from $I_r$, the exterior resolvent bound \eqref{eq:C3-stieltjes-close} on the shaded neighborhood $\mathcal N$ keeps every interpolating curve $\Gamma_\theta$ inside $\mathcal N$ (\cref{lem:pastur-boundary}). The Pastur map $\pastur_\theta$ sends $\Omega_\theta$ conformally onto $\C^+$ and maps $\Gamma_\theta$ onto the single-interval support $[E_-(\theta),E_+(\theta)]$, allowing exterior polynomial information to control the edges and distribution function of $\mu_\theta$ down to scale $O(\xi)$.}
\label{fig:subordination}
\end{figure}
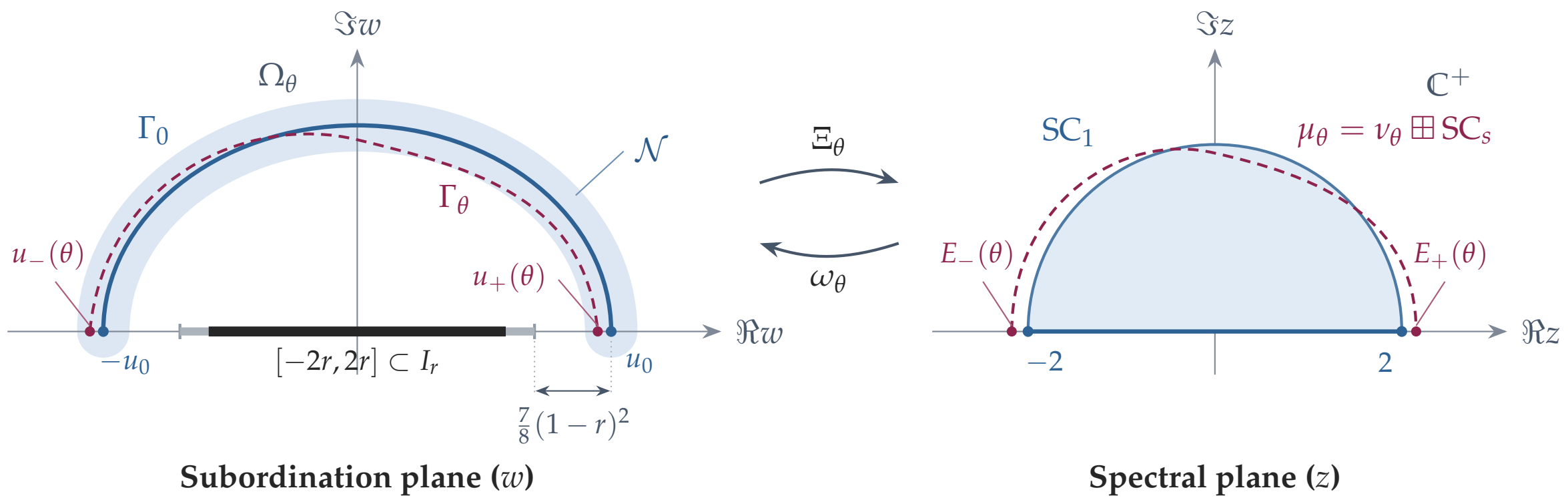

Next, we enlarge $[-2r,2r]$ slightly to contain the spectrum of $rX$ and fix a neighborhood $\mathcal N$ of $\Gamma_0$ that stays bounded away from it. Specifically, set $\delta:=(1-r)^2/(8r)$, $\ell_r:=r(2+\delta)$, and $I_r:=[-\ell_r,\ell_r]$. Then $u_0-\ell_r=\frac78(1-r)^2$ and $\dist(\Gamma_0,I_r)\ge\frac78(1-r)^2$. Let
\begin{equation}
\mathcal N:=\{w\in\C:\dist(w,\Gamma_0)<\rho_r\},\qquad \rho_r:=\tfrac14(1-r)^2.
\label{eq:N-gap}
\end{equation}
This is a bounded open set with $\dist(\overline{\mathcal N},I_r)\ge\frac12(1-r)^2$ (\cref{fig:subordination}). \Cref{lem:exterior-resolvent} needs a compact set, and we will apply it to the compact set $r^{-1}\overline{\mathcal N}$. All analytic arguments below take place in $\mathcal N$, mostly in disks of radius $\rho_r/2$ around points of $\Gamma_0$, whose closures lie in $\mathcal N$.

Throughout this subsection, $\nu$ is a probability measure with $\supp\nu\subset I_r$, and for some $\xi>0$,
\begin{equation}
\sup_{w\in\mathcal N}\max_{0\le j\le3}
|m_\nu^{(j)}(w)-m_r^{(j)}(w)|\le\xi.
\label{eq:C3-stieltjes-close}
\end{equation}
We write $\Delta m:=m_\nu-m_r$, $\mu_\nu:=\nu\boxplus\SC_s$, $\rho_\nu:=\rho_{\mathrm{fc},\nu}$ for its density, and $\pastur_\nu(w)=w-s\,m_\nu(w)$ for the Pastur map of \cref{sec:preliminaries}. Constants $c,C$ and smallness thresholds $\xi_0$ depend only on $r$, unless stated otherwise, and all estimates are uniform over the measures $\nu$ considered. Two consequences of the assumptions are used repeatedly. First, if $\dist(w,I_r)\ge d$, then $|m_\nu^{(j)}(w)|\le j!\,d^{-j-1}$, and the same holds for $m_r$. Second, $\Delta m$ is analytic in $\C\setminus I_r$, tends to zero at infinity, and satisfies $\Delta m(\bar w)=\overline{\Delta m(w)}$. Write ${}^*$ for reflection in the real axis. The full ellipse $\Gamma_0\cup\Gamma_0^*$ surrounds $I_r$ and lies in $\mathcal N\cup\mathcal N^*$. So the maximum principle, applied to each $\Delta m^{(j)}$ on the closed exterior $\mathcal O$ of this ellipse, gives
\begin{equation}
|\Delta m^{(j)}(w)|\le\xi,\qquad w\in\mathcal N\cup\mathcal N^*\cup\mathcal O,\ 0\le j\le3.
\label{eq:C3-extended}
\end{equation}
Since $\pastur_\nu-\pastur_r=-s\Delta m$, the derivatives of $\pastur_\nu-\pastur_r$ satisfy the same bound with $s\xi$.

To characterize the support of $\mu_\nu$, we introduce the inverse-square integral $\Phi_\nu(u):=\int_\R(u-x)^{-2}\,\nu(dx)\in(0,\infty]$ for $u\in\R$. By Biane's support characterization \cite[Sections~3--4]{Biane1997}, in the form of \cite[Theorem~2.1]{CapitaineDonatiMartinFeralFevrier2011}, there is an increasing homeomorphism $\Psi_\nu$ of $\R$ such that $\supp\mu_\nu$ is the closure of $\Psi_\nu(\{\Phi_\nu>s^{-1}\})$, the density is positive on $\Psi_\nu(\{\Phi_\nu>s^{-1}\})$, and $\Psi_\nu(u)=\pastur_\nu(u)$ at each boundary point $u\notin\supp\nu$ of $\{\Phi_\nu>s^{-1}\}$.
We first show that $\mu_\nu$ is supported on a single interval, locate the two real critical points of $\pastur_\nu$ outside $I_r$, and prove that the resulting spectral edges lie within $O(\xi)$ of $\pm2$.
\begin{lemma}%
\label{lem:pastur-one-cut}
There are $\xi_0>0$ and $\alpha_*>0$ such that the following holds if $\xi\le\xi_0$.
\begin{enumerate}[label=(\alph*)]
\item For every $u\in I_r$,
\begin{equation}
\Phi_\nu(u)\ge\frac{1+\alpha_*}{s}.
\label{eq:one-cut-resolvent-margin}
\end{equation}
\item $\{u\in\R:\Phi_\nu(u)>s^{-1}\}=(u_-,u_+)$ with $u_-<-\ell_r$ and $u_+>\ell_r$. The points $u_\pm$ are the only real critical points of $\pastur_\nu$ outside $I_r$, and $|u_\pm\mp u_0|\le C\xi$.
\item $\supp\mu_\nu=[E_-,E_+]$ with $E_\pm=\pastur_\nu(u_\pm)$, the density $\rho_\nu$ is positive on $(E_-,E_+)$, and
\begin{equation}
|E_+-2|+|E_-+2|\le C\xi.
\label{eq:edge-location-stability}
\end{equation}
\end{enumerate}
\end{lemma}

\begin{proof}
The reference identity for $\SC_{r^2}$ holds at the positive height $h_0(u)$ above each $u\in I_r$, and \eqref{eq:C3-stieltjes-close} transfers it to $\nu$ up to $O(\xi)$. Removing the height gains a fixed factor because $\nu$ is supported on the bounded interval $I_r$, and this gives the margin in (a). Monotonicity of $\Phi_\nu$ outside $I_r$ then gives (b) and (c).

For (a), fix $u\in I_r$ and write $w_0(u):=u+\mathrm{i}h_0(u)\in\Gamma_0$. The height is at least $h_*:=s(1-\ell_r^2/u_0^2)^{1/2}>0$. Since $\pastur_r$ is real on $\Gamma_0$, the imaginary part of $\pastur_r(w_0(u))$ vanishes, which is the reference identity
\begin{equation*}
\int_\R\frac{\SC_{r^2}(dx)}{(u-x)^2+h_0(u)^2}=\frac{\Im m_r(w_0(u))}{h_0(u)}=\frac1s.
\end{equation*}
By \eqref{eq:C3-stieltjes-close} at $w_0(u)$, the same integral for $\nu$ is at least $s^{-1}-\xi/h_*$. Now we remove the height. For $x,u\in I_r$ we have $(u-x)^2\le4\ell_r^2$, so with $\alpha:=h_*^2/(4\ell_r^2)$,
\begin{equation*}
\frac1{(u-x)^2}=\Bigl(1+\frac{h_0(u)^2}{(u-x)^2}\Bigr)\frac1{(u-x)^2+h_0(u)^2}\ge\frac{1+\alpha}{(u-x)^2+h_0(u)^2}.
\end{equation*}
Integrating against $\nu$ gives $\Phi_\nu(u)\ge(1+\alpha)(s^{-1}-\xi/h_*)\ge(1+\alpha/2)s^{-1}$ once $\xi\le\alpha h_*/(2s(1+\alpha))$. This is (a) with $\alpha_*=\alpha/2$.

For (b), $\Phi_\nu$ is finite and strictly decreasing on $(\ell_r,\infty)$, since $\Phi_\nu'(u)=-2\int(u-x)^{-3}\nu(dx)<0$ there, and it tends to zero at infinity. By monotone convergence and (a), $\Phi_\nu(u)\to\Phi_\nu(\ell_r)>s^{-1}$ as $u\downarrow\ell_r$. So there is a unique $u_+>\ell_r$ with $\Phi_\nu(u_+)=s^{-1}$, and in the same way a unique $u_-<-\ell_r$. Together with (a), this identifies the set. Outside $I_r$ we have $\pastur_\nu'(u)=1-s\Phi_\nu(u)$, so $u_\pm$ are the only real critical points there. To locate $u_+$, note that the interval $J:=[u_0-\rho_r/2,u_0+\rho_r/2]$ lies in $\mathcal N$, to the right of $I_r$. On $J$, $\Phi_\nu=m_\nu'$ is within $\xi$ of $m_r'$, and $m_r'$ is strictly decreasing, with $m_r''\le-c<0$ and $m_r'(u_0)=s^{-1}$ by \eqref{eq:reference-critical}. For small $\xi$, $\Phi_\nu-s^{-1}$ changes sign on $J$, so $u_+\in J$ and $c|u_+-u_0|\le|m_r'(u_+)-\Phi_\nu(u_+)|\le\xi$. The point $u_-$ is treated in the same way.

For (c), the support characterization gives $\supp\mu_\nu=[\pastur_\nu(u_-),\pastur_\nu(u_+)]$, with positive density inside. By \eqref{eq:reference-critical} and \eqref{eq:C3-stieltjes-close},
\begin{equation*}
|E_+-2|\le|\pastur_\nu(u_+)-\pastur_r(u_+)|+|\pastur_r(u_+)-\pastur_r(u_0)|\le s\xi+C|u_+-u_0|^2\le C\xi,
\end{equation*}
and the left edge is treated in the same way.
\end{proof}

With the support identified, we linearly interpolate between $\SC_{r^2}$ and $\nu$ so that we can track the subordination boundary curves and compare the full distribution functions.
For $\theta\in[0,1]$ put
\begin{equation*}
\nu_\theta:=(1-\theta)\SC_{r^2}+\theta\nu,\qquad m_\theta:=m_{\nu_\theta}=m_r+\theta\Delta m,\qquad \mu_\theta:=\nu_\theta\boxplus\SC_s,\qquad M_\theta:=m_{\mu_\theta},
\end{equation*}
and let $\omega_\theta(z):=z+sM_\theta(z)$ and $\pastur_\theta(w):=w-sm_\theta(w)=\pastur_r(w)-s\theta\Delta m(w)$. Then $\mu_0=\SC_1$ and $\mu_1=\mu_\nu$. Every $\nu_\theta$ satisfies the standing assumptions with the same $\xi$, so \cref{lem:pastur-one-cut} applies to it, and we write $u_\pm(\theta)$ and $E_\pm(\theta)$ for the corresponding points.

We use Biane's description of the subordination domain, as recalled in \cite[Proposition~2.1 and Theorem~2.1]{CapitaineDonatiMartinFeralFevrier2011}. Let
\begin{equation*}
J_\theta(u,q):=\int_\R\frac{\nu_\theta(dx)}{(u-x)^2+q},\qquad h_\theta(u):=\inf\{h\ge0:J_\theta(u,h^2)\le s^{-1}\}.
\end{equation*}
For $h>0$ we have $\Im\pastur_\theta(u+\mathrm{i}h)=h(1-sJ_\theta(u,h^2))$, and $q\mapsto J_\theta(u,q)$ is strictly decreasing. Hence $\Omega_\theta:=\{w\in\C^+:\Im\pastur_\theta(w)>0\}=\{u+\mathrm{i}h:h>h_\theta(u)\}$, and $\pastur_\theta$ maps $\Omega_\theta$ conformally onto $\C^+$, with inverse $\omega_\theta$. By \cref{lem:pastur-one-cut}, $h_\theta(u)>0$ exactly when $u_-(\theta)<u<u_+(\theta)$. Let $\Gamma_\theta:=\{u+\mathrm{i}h_\theta(u):u_-(\theta)\le u\le u_+(\theta)\}$, which for $\theta=0$ is the curve $\Gamma_0$ above. The map $\omega_\theta$ extends continuously to $\C^+\cup\R$. For $E\in\supp\mu_\theta$ the boundary value $\omega_\theta(E+\mathrm{i}0)$ lies on $\Gamma_\theta$, and otherwise it is real and lies outside $(u_-(\theta),u_+(\theta))$. The density of $\mu_\theta$ is $\Im\omega_\theta(E+\mathrm{i}0)/(\pi s)$.

We next show that every interpolating boundary curve $\Gamma_\theta$ stays within $O(\sqrt\xi)$ of $\Gamma_0$, and hence inside $\mathcal N$, allowing us to apply the exterior bound \eqref{eq:C3-stieltjes-close} along the entire interpolation.

\begin{lemma}%
\label{lem:pastur-boundary}
There is $\xi_0>0$ such that, if $\xi\le\xi_0$, then for all $\theta\in[0,1]$,
\begin{equation*}
|h_\theta(u)^2-h_0(u)^2|\le C\xi\quad\text{for }|u|\le u_0,\qquad h_\theta(u)^2\le C\xi\quad\text{for }u_0<|u|,\ u_-(\theta)\le u\le u_+(\theta).
\end{equation*}
Consequently,
\begin{equation}
\sup_{0\le\theta\le1}\sup_{w\in\Gamma_\theta}
\dist(w,\Gamma_0)\le C\sqrt{\xi},
\label{eq:Pastur-boundary-comparison}
\end{equation}
and every $\Gamma_\theta$ lies within distance $\rho_r/2$ of $\Gamma_0$.
\end{lemma}

\begin{proof}
We compare squared heights through the equation $J_\theta(u,h^2)=s^{-1}$, which is uniformly monotone in $h^2$ even where the height vanishes. This lets the argument reach the edges, where $h_0$ has a square-root singularity.

First, $|J_\theta(u,h_0(u)^2)-s^{-1}|\le C\xi$ for $|u|\le u_0$. Indeed, $J_0(u,h_0(u)^2)=s^{-1}$ by the reference identity in the proof of \cref{lem:pastur-one-cut}, and if $h_0(u)>0$,
\begin{equation*}
J_\theta(u,h_0(u)^2)-J_0(u,h_0(u)^2)=\theta\,\frac{\Im\Delta m(u+\mathrm{i}h_0(u))}{h_0(u)}.
\end{equation*}
Fix $\tau>0$ small in terms of $r$. If $|u|\le u_0-\tau$, then $h_0(u)$ is bounded below and the right side is $O(\xi)$ by \eqref{eq:C3-stieltjes-close}. If $u_0-\tau<|u|\le u_0$, the vertical segment from $u$ to $u+\mathrm{i}h_0(u)$ lies in $\mathcal N$ and $u\notin I_r$. Then $\Delta m(u)$ is real, and the bound on $\Delta m'$ gives $|\Im\Delta m(u+\mathrm{i}h_0(u))|\le\xi h_0(u)$. At $u=\pm u_0$ we use $J_\theta(u,0)-J_0(u,0)=\theta\Delta m'(u)$.

Second, for $|u|\le u_0+1$ and $0\le q_1<q_2\le s$,
\begin{equation}
J_\theta(u,q_1)-J_\theta(u,q_2)=(q_2-q_1)\int_\R\frac{\nu_\theta(dx)}{((u-x)^2+q_1)((u-x)^2+q_2)}\ge c_h(q_2-q_1),
\label{eq:Pastur-height-monotonicity}
\end{equation}
with $c_h:=((u_0+1+\ell_r)^2+s)^{-2}$. Both $h_\theta(u)^2$ and $h_0(u)^2$ lie in $[0,s]$, since $J_\theta(u,s)\le s^{-1}$.

Now let $|u|\le u_0$. If $h_\theta(u)>0$, then $J_\theta(u,h_\theta(u)^2)=s^{-1}$, and the two steps give $|h_\theta(u)^2-h_0(u)^2|\le C\xi/c_h$. If $h_\theta(u)=0$, then $J_\theta(u,0)\le s^{-1}\le J_\theta(u,h_0(u)^2)+C\xi$, and \eqref{eq:Pastur-height-monotonicity} gives $h_0(u)^2\le C\xi/c_h$. If instead $u_0<u\le u_+(\theta)$, the exterior monotonicity of $J_\theta(\cdot,0)=\Phi_{\nu_\theta}$ gives $s^{-1}\le J_\theta(u,0)\le J_\theta(u_0,0)=s^{-1}+\theta\Delta m'(u_0)\le s^{-1}+\xi$, so $h_\theta(u)^2\le\xi/c_h$. The left side is symmetric. Finally, $|h_\theta-h_0|\le|h_\theta^2-h_0^2|^{1/2}$ on $[-u_0,u_0]$, and $|u_\pm(\theta)\mp u_0|\le C\xi$ by \cref{lem:pastur-one-cut}. This proves \eqref{eq:Pastur-boundary-comparison}.
\end{proof}
Using this control on the boundary curves, we next verify that the density $\rho_\nu$ is bounded below in the bulk and exhibits square-root vanishing at both edges, together with the matching bounds on $\pastur_\nu'$.
\begin{lemma}%
\label{lem:pastur-regularity}
There are $\xi_0,d_0,\rho_1>0$ and $C\ge1$ such that the following holds if $\xi\le\xi_0$.
\begin{enumerate}[label=(\alph*)]
\item $\rho_\nu\le(\pi\sqrt s)^{-1}$. For every $\kappa\in(0,2)$, $\rho_\nu\ge\sqrt\kappa/(4\pi)$ on $[-2+\kappa,2-\kappa]$ once $\xi\le\xi_0(r,\kappa)$.
\item For $z\in\C^+$ with $|z-E_\pm|<d_0$, the point $\omega_\nu(z)$ lies in the disk $B(u_\pm,\rho_1)$, and
\begin{equation*}
C^{-1}|z-E_\pm|^{1/2}\le|\pastur_\nu'(\omega_\nu(z))|\le C|z-E_\pm|^{1/2}.
\end{equation*}
\item There are $c_\pm\in[C^{-1},C]$ such that, for $0<d<d_0$,
\begin{equation*}
|\rho_\nu(E_+-d)-c_+\sqrt d|+|\rho_\nu(E_-+d)-c_-\sqrt d|\le Cd.
\end{equation*}
\end{enumerate}
\end{lemma}

\begin{proof}
In the bulk, since the density $\rho_\nu$ is proportional to the height of the boundary curve $\Gamma_1$, we apply \cref{lem:pastur-boundary} to bound it from below. Near each edge, we invert the quadratic expansion of $\pastur_\nu$ around $u_\pm$ to obtain the square-root behavior, and we use the injectivity of $\pastur_\nu$ on the subordination domain to identify this local inverse with $\omega_\nu$.

The upper bound in (a) is recalled in \cref{sec:preliminaries}. For the lower bound, let $E\in[-2+\kappa,2-\kappa]$, which lies inside the support for small $\xi$ by \eqref{eq:edge-location-stability}. The boundary value $w:=\omega_\nu(E+\mathrm{i}0)$ lies on $\Gamma_1$, and $\pastur_\nu(w)=E$. By \cref{lem:pastur-boundary}, there is $E'\in[-2,2]$ with $|w-\omega_0(E')|\le C\sqrt\xi$. Since $\pastur_r(\omega_0(E'))=E'$, and $\pastur_r$ is Lipschitz on disks of radius $\rho_r/2$ around points of $\Gamma_0$, we get
\begin{equation*}
|E-E'|\le|\pastur_\nu(w)-\pastur_r(w)|+|\pastur_r(w)-\pastur_r(\omega_0(E'))|\le C\sqrt\xi.
\end{equation*}
So $|E'|\le2-\kappa/2$ for small $\xi$, and \eqref{eq:omega0-def} gives $\pi s\rho_\nu(E)=\Im w\ge\frac s2\sqrt\kappa-C\sqrt\xi\ge\frac s4\sqrt\kappa$.

For (b) and (c), we treat the right edge. By \cref{lem:pastur-one-cut}, $u_+$ is real, $|u_+-u_0|\le C\xi$, and $\pastur_\nu'(u_+)=0$. On the disk $B(u_+,\rho_r/2)\subset\mathcal N$, the function $\pastur_\nu$ is analytic and real on the real axis, its third derivative is bounded, and $\pastur_\nu''(u_+)=2s^{-2}+O(\xi)$ by \eqref{eq:reference-critical} and \eqref{eq:C3-stieltjes-close}. Hence there are $\rho_1\in(0,\rho_r/2)$ and an analytic function $g$ on $B(u_+,\rho_1)$, real on the real axis, with
\begin{equation*}
\pastur_\nu(w)-E_+=g(w)^2,\qquad g(w)=\Bigl(\tfrac12\pastur_\nu''(u_+)\Bigr)^{1/2}(w-u_+)\bigl(1+O(|w-u_+|)\bigr),
\end{equation*}
and $\Re g'\ge\frac12g'(u_+)>0$ on $B(u_+,\rho_1)$. Since $\Im g(u+\mathrm{i}v)=\int_0^v\Re g'(u+\mathrm{i}t)\,dt$ (by the Cauchy--Riemann equation $\partial_v\Im g=\Re g'$ and $\Im g(u)=0$), the map $g$ sends the upper half of $B(u_+,\rho_1)$ into $\C^+$ and the lower half into $\C^-$. By the inverse function theorem there is $d_0>0$ such that $g^{-1}$ is defined and analytic on $\{|\zeta|^2<d_0\}$, with values in $B(u_+,\rho_1)$.

For $z\in\C^+$ with $|z-E_+|<d_0$, let $\sqrt{z-E_+}$ be the square root in $\C^+$, and put $w(z):=g^{-1}(\sqrt{z-E_+})$. Then $w(z)\in\C^+$ (as $g$ maps the lower half and real diameter of $B(u_+,\rho_1)$ into $\C^-\cup\R$) and $\pastur_\nu(w(z))=z\in\C^+$, so $w(z)\in\Omega_1$. Since $\pastur_\nu$ is injective on $\Omega_1$, we get $w(z)=\omega_\nu(z)$. This proves the first claim in (b), and the second follows from $\pastur_\nu'=2gg'$, $|g(\omega_\nu(z))|=|z-E_+|^{1/2}$ and $|g'|\asymp1$. For (c), let $z\to E_+-d$. Then $\sqrt{z-E_+}\to\mathrm{i}\sqrt d$, and
\begin{equation*}
\omega_\nu(E_+-d+\mathrm{i}0)=g^{-1}(\mathrm{i}\sqrt d)=u_++\frac{\mathrm{i}\sqrt d}{g'(u_+)}+O(d).
\end{equation*}
So $\rho_\nu(E_+-d)=c_+\sqrt d+O(d)$ with $c_+=(\pi s\,g'(u_+))^{-1}=(\pi s)^{-1}(2/\pastur_\nu''(u_+))^{1/2}$. At the left edge $\pastur_\nu''(u_-)=-2s^{-2}+O(\xi)$, and the same argument applies to $E_--\pastur_\nu=g^2$, with the square root of $E_--z$ in $\C^+$.
\end{proof}

We next bound the difference of distribution functions uniformly on $\R$ to precision $O(\xi)$, even near the edges where the boundary curves move by $O(\sqrt\xi)$.
\begin{lemma}%
\label{lem:pastur-cdf}
There is $\xi_0>0$ such that, if $\xi\le\xi_0$, then
\begin{equation}
\sup_{E\in\R}|F_{\mu_\nu}(E)-F_{\mathrm{sc}}(E)|\le C\xi.
\label{eq:Pastur-cdf-stability}
\end{equation}
\end{lemma}

\begin{proof}
Along the interpolation, the $\theta$-derivative of the Stieltjes transform of $\mu_\theta$ is the $z$-derivative of $\mathcal A(\omega_\theta(z))$, where $\mathcal A$ is a primitive of $\Delta m$. Integrating in the spectral variable therefore removes the derivative of the inverse map $\omega_\theta$, which is unbounded at the edges, and expresses the difference of the distribution functions through $\Im\mathcal A$ on the boundary curves $\Gamma_\theta$. \Cref{lem:pastur-boundary} is used only to keep these curves inside $\mathcal N$, where the increments of $\mathcal A$ along paths of bounded length are $O(\xi)$.

Let $\sigma:=\nu-\SC_{r^2}$, a signed measure on $I_r$ with $\sigma(\R)=0$ and $|\sigma|(\R)\le2$, and define
\begin{equation*}
\mathcal A(w):=-\int_\R\operatorname{Log}(w-x)\,\sigma(dx),\qquad w\in\C^+,
\end{equation*}
with the branch of the logarithm whose argument lies in $(0,\pi)$. Then $\mathcal A'=\Delta m$, $\mathcal A(w)=O(|w|^{-1})$ at infinity (writing $\operatorname{Log}(w-x)=\operatorname{Log} w+\operatorname{Log}(1-x/w)$, the leading term integrates to zero since $\sigma(\R)=0$, and similarly $\Delta m(w)=O(|w|^{-2})$), and $|\Im\mathcal A|\le\pi|\sigma|(\R)\le2\pi$. The function $\mathcal A$ extends continuously to $\R\setminus I_r$ with real values. Indeed, to the right of $I_r$ all arguments vanish, and to the left they all equal $\pi$, which cancels because $\sigma(\R)=0$.

Differentiating $\pastur_\theta(\omega_\theta(z))=z$ in $\theta$ and in $z$ gives $\pastur_\theta'(\omega_\theta)\,\partial_\theta\omega_\theta=s\Delta m(\omega_\theta)$ and $\pastur_\theta'(\omega_\theta)\,\partial_z\omega_\theta=1$, where $\pastur_\theta'(\omega_\theta(z))\ne0$ because $\omega_\theta$ is conformal. Here $\theta\mapsto\omega_\theta(z)$ is differentiable by the implicit function theorem, since $\pastur_\theta(w)$ is affine in $\theta$ and $\pastur_\theta'(\omega_\theta(z))\ne0$. Since $M_\theta=(\omega_\theta-z)/s$, it follows that
\begin{equation}
\partial_\theta M_\theta(z)=\Delta m(\omega_\theta(z))\,\partial_z\omega_\theta(z)=\partial_z\bigl[\mathcal A(\omega_\theta(z))\bigr],\qquad z\in\C^+.
\label{eq:Pastur-interpolation-derivative}
\end{equation}
The factor $\partial_z\omega_\theta$ is unbounded near the edges, but it disappears once we integrate in $z$.

Fix $y>0$, and let $G_{\theta,y}(E):=\pi^{-1}\int_{-\infty}^E\Im M_\theta(x+\mathrm{i}y)\,dx$ be the distribution function of $\mu_\theta$ convolved with the Cauchy law of scale $y$. Since $\Delta m(w)=O(|w|^{-2})$ and $|\partial_z\omega_\theta|\le1+sy^{-2}$ on the line $\Im z=y$, we may differentiate under the integral. As $\omega_\theta(x+\mathrm{i}y)\to\infty$ when $x\to-\infty$, and $\mathcal A$ vanishes at infinity, \eqref{eq:Pastur-interpolation-derivative} gives
\begin{equation*}
G_{1,y}(E)-G_{0,y}(E)=\frac1\pi\int_0^1\Im\mathcal A(\omega_\theta(E+\mathrm{i}y))\,d\theta.
\end{equation*}
Let $y\downarrow0$. The measures $\mu_\theta$ have continuous densities, so the left side tends to $F_{\mu_\nu}(E)-F_{\mathrm{sc}}(E)$. The boundary value $\omega_\theta(E+\mathrm{i}0)$ lies on $\Gamma_\theta$ or on $\R\setminus I_r$, where $\mathcal A$ is continuous, and $|\Im\mathcal A|\le2\pi$. Dominated convergence therefore gives
\begin{equation}
F_{\mu_\nu}(E)-F_{\mathrm{sc}}(E)=\frac1\pi\int_0^1\Im\mathcal A(\omega_\theta(E+\mathrm{i}0))\,d\theta.
\label{eq:Pastur-cdf-interpolation}
\end{equation}

It remains to bound $\Im\mathcal A$ on $\Gamma_\theta$, since it vanishes on $\R\setminus I_r$. For $w\in\Gamma_\theta$, choose $w_0\in\Gamma_0$ with $|w-w_0|\le C\sqrt\xi\le\rho_r/2$ by \eqref{eq:Pastur-boundary-comparison}. The path that follows $\Gamma_0$ from $-u_0$ to $w_0$ and then the segment from $w_0$ to $w$ lies in $\mathcal N$ and in the closed upper half-plane, and its length is at most the length of $\Gamma_0$ plus one. Since $\mathcal A'=\Delta m$ and $|\Delta m|\le\xi$ on $\mathcal N$, we get $|\mathcal A(w)-\mathcal A(-u_0)|\le C\xi$. As $\Im\mathcal A(-u_0)=0$, this gives $|\Im\mathcal A(w)|\le C\xi$, and \eqref{eq:Pastur-cdf-interpolation} proves \eqref{eq:Pastur-cdf-stability}.
\end{proof}

Finally, we convert these edge and distribution-function bounds into quantile comparisons covering both the $i/n$ and midpoint conventions.
For a probability measure $\mu$ and $p\in(0,1]$, let $q_\mu(p):=\inf\{E:F_\mu(E)\ge p\}$. Write $q_{\mathrm{sc}}:=q_{\SC_1}$, extended by $q_{\mathrm{sc}}(p)=-2$ for $p\le0$ and $q_{\mathrm{sc}}(p)=2$ for $p\ge1$, so that $\gamma_i=q_{\mathrm{sc}}(i/n)$.

\begin{lemma}%
\label{lem:quantile-comparison}
There is an absolute constant $C$ such that the following holds. Let $\mu$ be a probability measure on $\R$ and $0<\Delta\le1/8$, and suppose that $\supp\mu\subset[-2-\Delta,2+\Delta]$ and $\sup_E|F_\mu(E)-F_{\mathrm{sc}}(E)|\le\Delta$. Then, for $p\in(0,1]$ and $p'\in[0,1]$ with $|p-p'|\le\Delta$,
\begin{equation}
|q_\mu(p)-q_{\mathrm{sc}}(p')|\le C\Delta\bigl(\min\{p',1-p'\}+\Delta\bigr)^{-1/3}.
\label{eq:quantile-comparison}
\end{equation}
In particular, if $\Delta\ge1/n$, then for $1\le i\le n$ and $p\in\{i/n,(i-1/2)/n\}$,
\begin{equation}
|q_\mu(p)-\gamma_i|\le C\Delta\,(n/\hat i)^{1/3}.
\label{eq:quantile-comparison-index}
\end{equation}
\end{lemma}

\begin{proof}

If $p-\Delta>0$ and $E<q_{\mathrm{sc}}(p-\Delta)$, then $F_\mu(E)\le F_{\mathrm{sc}}(E)+\Delta<p$, so $q_\mu(p)\ge q_{\mathrm{sc}}(p-\Delta)$. If $p+\Delta<1$, then $F_\mu(q_{\mathrm{sc}}(p+\Delta))\ge p$, so $q_\mu(p)\le q_{\mathrm{sc}}(p+\Delta)$. In the remaining cases we use $-2-\Delta\le q_\mu(p)\le2+\Delta$. Hence
\begin{equation*}
q_{\mathrm{sc}}(p-\Delta)-\Delta\le q_\mu(p)\le q_{\mathrm{sc}}(p+\Delta)+\Delta.
\end{equation*}
Put $h:=2\Delta$ and $m:=\min\{p',1-p'\}$, so that $p\pm\Delta$ lies within $h$ of $p'$. The square-root vanishing of $\rhosc$ at $\pm2$ gives $\rhosc(q_{\mathrm{sc}}(v))\ge c\min\{v,1-v\}^{1/3}$, so $q_{\mathrm{sc}}'(v)\le C\min\{v,1-v\}^{-1/3}$ on $(0,1)$. If $h\le m/2$, then $\min\{v,1-v\}\ge m/2$ whenever $|v-p'|\le h$, and $|q_{\mathrm{sc}}(v)-q_{\mathrm{sc}}(p')|\le Chm^{-1/3}$. If $h>m/2$, integrating the derivative bound over an interval of length $h$ gives $|q_{\mathrm{sc}}(v)-q_{\mathrm{sc}}(p')|\le Ch^{2/3}$. In both cases the bound is at most $Ch(m+h)^{-1/3}\le2C\Delta(m+\Delta)^{-1/3}$, and $\Delta\le\Delta(m+\Delta)^{-1/3}$ since $m+\Delta\le1$. This proves \eqref{eq:quantile-comparison}. For \eqref{eq:quantile-comparison-index}, take $p'=i/n$. Then $|p-p'|\le1/(2n)\le\Delta$ and $\min\{p',1-p'\}+\Delta\ge\hat i/n$.
\end{proof}

\subsection{Rigidity after Gaussian noise}
\label{subsec:burnin-rigidity}

In this subsection, we assemble the results of \cref{subsec:norm-resolvent,subsec:pastur-stability} to complete the proof of \cref{lem:rigidity}. Conditional on $X$, orthogonal invariance of the GOE implies that $H=rX+\sqrt s\,W$ has the same spectral law as $A+\sqrt s\,W$ with $A=r\diag(\lambda_1(X),\dots,\lambda_n(X))$.

We first record a uniform version of the deformed-Wigner rigidity estimate of \cite[Section~3]{LSSTY2016} (\cref{lem:conditional-deformed-wigner}), and then use \cref{subsec:pastur-stability} to verify its hypotheses whenever $\widehat\nu_A$ satisfies our exterior $C^3$ Stieltjes-transform bound \eqref{eq:C3-stieltjes-close} (\cref{lem:deformation-class}). Combining this with the high-probability bounds of \cref{subsec:norm-resolvent} and the quantile comparison \cref{lem:quantile-comparison}, we integrate over $X$ to obtain uniform burn-in rigidity (\cref{prop:burnin-rigidity}), from which \cref{lem:rigidity} follows immediately.

To state the conditional rigidity estimate of \cite{LSSTY2016} cleanly, we first fix notation for a deterministic diagonal deformation $A=\diag(a_1,\dots,a_n)$ with $\supp\widehat\nu_A\subset I_r$. Write $\mu_A:=\widehat\nu_A\boxplus\SC_s$, with Stieltjes transform $m_{\mathrm{fc},A}$, density $\rho_{\mathrm{fc},A}$ and support edges $E_\pm(A)$. Let $\omega_A(z):=z+s\,m_{\mathrm{fc},A}(z)$, and define
\begin{equation}
R_j(z):=\frac1n\sum_{i=1}^n\bigl(a_i-\omega_A(z)\bigr)^{-j},\qquad\kappa_A(E):=\min\{|E-E_-(A)|,|E-E_+(A)|\},
\label{eq:deformed-coefficient-definitions}
\end{equation}
and $\mathcal D:=\{E+\mathrm{i}y:|E|\le5,\ 0<y\le3\sqrt s\}$. Let $\bar\gamma_i^A$ be the $(i-1/2)/n$-quantile of $\mu_A$.

\begin{lemma}[{Adapted from \cite[Section~3]{LSSTY2016}}]%
\label{lem:conditional-deformed-wigner}
Fix $C_*\ge1$, and let $A$ be as above. Assume the following conditions.
\begin{enumerate}[label=(\roman*),ref=(\roman*)]
\item\label{ass:deformed-regularity} $\supp\mu_A=[E_-(A),E_+(A)]\subset[-3,3]$, and $C_*^{-1}\sqrt{\kappa_A(E)}\le\rho_{\mathrm{fc},A}(E)\le C_*\sqrt{\kappa_A(E)}$ on this interval.
\item\label{ass:deformed-inverse-square} $\int(a-x)^{-2}\,\widehat\nu_A(da)\ge(1+C_*^{-1})s^{-1}$ for every $x\in I_r$.
\item\label{ass:deformed-stability} For $z=E+\mathrm{i}y\in\mathcal D$ and $x\in I_r$,
\begin{equation}
\begin{gathered}
C_*^{-1}\le|x-\omega_A(z)|\le C_*,\qquad
C_*^{-1}\sqrt{\kappa_A(E)+y}\le|1-sR_2(z)|\le C_*\sqrt{\kappa_A(E)+y},\\
|R_3(z)|\le C_*,\qquad |R_3(z)|\ge C_*^{-1}\ \text{if }\kappa_A(E)+y\le C_*^{-1}.
\end{gathered}
\label{eq:deformed-stability-bounds}
\end{equation}
\item\label{ass:deformed-stieltjes} For $z=E+\mathrm{i}y\in\mathcal D$, the ratio of $\Im m_{\mathrm{fc},A}(z)$ to $\sqrt{\kappa_A(E)+y}$ if $E\in[E_-(A),E_+(A)]$, and to $y/\sqrt{\kappa_A(E)+y}$ otherwise, lies in $[C_*^{-1},C_*]$.
\end{enumerate}
Then, for every fixed $\varepsilon,M>0$, there is $n_0=n_0(r,C_*,\varepsilon,M)$ such that, for all $n\ge n_0$ and all such $A$, the matrix $H=A+\sqrt s\,W$ with $W\sim Q_n$ satisfies
\begin{equation}
\Pp\Bigl\{|\lambda_i(H)-\bar\gamma_i^A|\le\frac{n^\varepsilon}{n^{2/3}\hat i^{1/3}}\ \text{for all }1\le i\le n\Bigr\}\ge1-n^{-M}.
\label{eq:conditional-deformed-midpoint-rigidity}
\end{equation}
\end{lemma}

\begin{proof}
We first match our normalization and conditions with those of \cite{LSSTY2016}. Then we apply their rigidity estimate to a single sequence of deformations, and finally we obtain uniformity over the class by a subsequence argument.

Put $V:=s^{-1/2}A$ and $\widetilde H:=s^{-1/2}H=V+W$. The GOE matrix $W$ has the variances of \cite[Definition~2.1]{LSSTY2016}, and we use the results of \cite[Section~3]{LSSTY2016} with $\vartheta=1$. For $\widetilde z=s^{-1/2}z$, the free-convolution Stieltjes transform $\widehat m_{\mathrm{fc}}$ of \cite[(3.6)]{LSSTY2016} for $V$ is $\widehat m_{\mathrm{fc}}(\widetilde z)=\sqrt s\,m_{\mathrm{fc},A}(z)$. Hence $\widetilde z+\widehat m_{\mathrm{fc}}(\widetilde z)=s^{-1/2}\omega_A(z)$, and for $v=s^{-1/2}x$,
\begin{equation*}
v-\widetilde z-\widehat m_{\mathrm{fc}}(\widetilde z)=\frac{x-\omega_A(z)}{\sqrt s},\qquad
\int\frac{\widehat\nu_V(dv)}{(v-\widetilde z-\widehat m_{\mathrm{fc}})^{2}}=sR_2(z),\qquad
\int\frac{\widehat\nu_V(dv)}{(v-\widetilde z-\widehat m_{\mathrm{fc}})^{3}}=s^{3/2}R_3(z).
\end{equation*}
Under this change of variables, \ref{ass:deformed-regularity} is \cite[(3.18)]{LSSTY2016}, \ref{ass:deformed-inverse-square} is the inverse-square condition \cite[(2.13)]{LSSTY2016}, \ref{ass:deformed-stability} gives \cite[(3.21) and (A.5)--(A.7)]{LSSTY2016}, and \ref{ass:deformed-stieltjes} is \cite[(3.20)]{LSSTY2016}, all written for $\widehat\nu_V$ and with constants depending only on $r$ and $C_*$. They hold on the domain $\mathcal D'$ of \cite[(3.17)]{LSSTY2016} with $E_0=5/\sqrt s$, which is at least $1+\max_\pm|E_\pm(A)|/\sqrt s$.

The proof of \cite[Theorem~3.3 and Corollary~3.4]{LSSTY2016}, in \cite[Section~3.2]{LSSTY2016}, compares $m_{\widetilde H}$ with the free-convolution Stieltjes transform $\widehat m_{\mathrm{fc}}$. It uses the deformation only through these estimates for $\widehat m_{\mathrm{fc}}$, which \cite[Lemma~3.6]{LSSTY2016} and its proof in the appendix of \cite{LSSTY2016} provide, and through the bound on $\|V\|$, which holds here because $\supp\widehat\nu_V\subset s^{-1/2}I_r$. The limiting law of \cite[Lemma~3.5]{LSSTY2016} and the estimate \cite[(3.3)]{LSSTY2016} enter that proof only through the proof of \cite[Lemma~3.6]{LSSTY2016}, where they produce \cite[(3.18), (3.20), (3.21) and (A.5)--(A.7)]{LSSTY2016} for $\widehat m_{\mathrm{fc}}$. There they are derived from \cite[Assumption~2.2]{LSSTY2016}, which asks for a limiting measure $\nu$ and for the transform bound \cite[(2.6)]{LSSTY2016} on every compact subset of $\C^+$ at positive distance from $\supp\nu$. In our setting this assumption is not available in that form: the degree constant in \cref{lem:exterior-resolvent} grows as the compact set approaches $[-2,2]$, so a fixed multiple of $\log n$ controls the transform only at a fixed distance from the support. Since the assumption enters the proof only through the estimates above, we state them directly for $\widehat\nu_V$ as conditions \ref{ass:deformed-regularity}--\ref{ass:deformed-stieltjes}. With this input, \cite[(3.15)]{LSSTY2016} gives, for a fixed sequence $(A_n)$ satisfying \ref{ass:deformed-regularity}--\ref{ass:deformed-stieltjes},
\begin{equation*}
|\lambda_i(\widetilde H)-\widehat\gamma_i|\le e^{O((\log\log n)^2)}\,n^{-2/3}\hat i^{-1/3},\qquad 1\le i\le n,
\end{equation*}
with failure probability smaller than any fixed power of $n$ for large $n$. Here $\widehat\gamma_i=s^{-1/2}\bar\gamma_i^A$ are the midpoint quantiles  of \cite[(3.14)]{LSSTY2016}. Multiplying by $\sqrt s$ gives \eqref{eq:conditional-deformed-midpoint-rigidity} for all large $n$ along the sequence.

The constants in \cite[(3.15)]{LSSTY2016} depend on the deformation, so uniformity over the class needs one more step. Suppose that it fails for some $\varepsilon$ and $M$. Then there are infinitely many $n$ and, for each of them, a matrix $A_n$ satisfying \ref{ass:deformed-regularity}--\ref{ass:deformed-stieltjes} with the constant $C_*$ for which \eqref{eq:conditional-deformed-midpoint-rigidity} fails. Extending $(A_n)$ to all $n$ by placing the quantiles of $\SC_{r^2}$ on the remaining indices gives a full sequence satisfying \ref{ass:deformed-regularity}--\ref{ass:deformed-stieltjes} with fixed constants, so \eqref{eq:conditional-deformed-midpoint-rigidity} holds for all sufficiently large $n$. This contradiction proves the lemma.
\end{proof}

We next verify that every deformation $A$ satisfying the exterior Stieltjes-transform bound \eqref{eq:C3-stieltjes-close} belongs to the admissible class of \cref{lem:conditional-deformed-wigner}, with constants depending only on $r$.

\begin{lemma}%
\label{lem:deformation-class}
There are $\xi_0>0$ and $C_*\ge1$, depending only on $r$, such that the following holds. Let $A$ be a deterministic diagonal matrix with $\supp\widehat\nu_A\subset I_r$ such that $\nu=\widehat\nu_A$ satisfies \eqref{eq:C3-stieltjes-close} with $\xi\le\xi_0$. Then $A$ satisfies conditions \ref{ass:deformed-regularity}--\ref{ass:deformed-stieltjes} of \cref{lem:conditional-deformed-wigner} with the constant $C_*$.
\end{lemma}

\begin{proof}
Conditions \ref{ass:deformed-regularity}, \ref{ass:deformed-inverse-square} and \ref{ass:deformed-stieltjes} follow quickly from \cref{lem:pastur-one-cut,lem:pastur-regularity}. For \ref{ass:deformed-stability}, the main points are that the subordination domain stays a fixed distance away from $I_r$, and that $\pastur_\nu'$ is bounded below on it except near the critical points $u_\pm$, where \cref{lem:pastur-regularity}(b) gives the square-root behavior. Write $\nu=\widehat\nu_A$, so that $\mu_A=\mu_\nu$ and $\omega_A=\omega_\nu$.

\emph{Conditions \ref{ass:deformed-regularity} and \ref{ass:deformed-inverse-square}.} By \cref{lem:pastur-one-cut}, the support is $[E_-,E_+]\subset[-3,3]$. By \cref{lem:pastur-regularity}(c), there is $d_1\in(0,d_0)$ such that $\rho_\nu\asymp\sqrt{\kappa_A}$ where $\kappa_A<d_1$. Where $\kappa_A\ge d_1$, we have $E\in[-2+d_1/2,2-d_1/2]$ for small $\xi$, so $\rho_\nu$ is bounded below by \cref{lem:pastur-regularity}(a), and above by $(\pi\sqrt s)^{-1}$. Since $\kappa_A\le3$, this gives \ref{ass:deformed-regularity}. Condition \ref{ass:deformed-inverse-square} is \eqref{eq:one-cut-resolvent-margin}.

\emph{The subordination domain.} By \cref{lem:pastur-boundary} with $\theta=1$, the lower boundary of $\Omega_1=\omega_\nu(\C^+)$ is the graph of $h_1$, which lies within $C\sqrt\xi$ of $\Gamma_0\cup(\R\setminus(-u_0,u_0))$. A point $u+\mathrm{i}h$ with $h\ge h_1(u)$ is at least as far from $I_r$ as the graph point $u+\mathrm{i}h_1(u)$. Hence every point of $\overline{\Omega_1}$ is at distance at least $\frac12(1-r)^2$ from $I_r$. Moreover, such a point lies either in $\mathcal O$, or inside the ellipse and then within $C\sqrt\xi$ of $\Gamma_0$, hence in $\mathcal N$.

\emph{Condition \ref{ass:deformed-stability}.} Let $z\in\mathcal D$ and $x\in I_r$. The previous step gives $|x-\omega_\nu(z)|\ge\frac12(1-r)^2$ and $|R_3(z)|\le8(1-r)^{-6}$. Taking imaginary parts in the Pastur equation \eqref{eq:Pastur-equation} gives $\Im m_{\mathrm{fc},A}(z)=\Im\omega_\nu(z)\int|a-\omega_\nu(z)|^{-2}\nu(da)$, and since $\Im\omega_\nu(z)=y+s\Im m_{\mathrm{fc},A}(z)>s\Im m_{\mathrm{fc},A}(z)$, it follows that $s\int|a-\omega_\nu(z)|^{-2}\nu(da)\le1$. Cauchy--Schwarz then gives $|m_{\mathrm{fc},A}(z)|\le s^{-1/2}$, so $|\omega_\nu(z)|\le|z|+s|m_{\mathrm{fc},A}(z)|\le\sqrt{25+9s}+\sqrt s\le7$, and $|x-\omega_\nu(z)|\le\ell_r+7<10$ because $\ell_r=2r+(1-r)^2/8<2$. Next, $1-sR_2(z)=\pastur_\nu'(\omega_\nu(z))$ and $2sR_3(z)=-\pastur_\nu''(\omega_\nu(z))$, and the separation bounds $|1-sR_2|$ from above (giving the upper bound in \eqref{eq:deformed-stability-bounds} whenever $\kappa_A(E)+y\ge d_0$). If $|z-E_\pm|<d_0$ for one sign, \cref{lem:pastur-regularity}(b) gives $|\pastur_\nu'(\omega_\nu(z))|\asymp|z-E_\pm|^{1/2}\asymp\sqrt{\kappa_A(E)+y}$ and $\omega_\nu(z)\in B(u_\pm,\rho_1)$, where $|\pastur_\nu''|\ge s^{-2}$ once $\rho_1$ in \cref{lem:pastur-regularity} is chosen small enough, by \eqref{eq:reference-critical} and $|\pastur_\nu''-\pastur_r''|\le s\xi$. Since $\kappa_A(E)+y<d_0$ forces $|z-E_\pm|<d_0$ for the nearer edge, this also gives the lower bound on $|R_3|$. If $|z-E_\pm|\ge d_0$ for both signs, then $\kappa_A(E)+y\ge d_0$, and it remains to bound $|\pastur_\nu'(\omega_\nu(z))|$ from below. Since $|\pastur_\nu(w)-E_\pm|\le C|w-u_\pm|^2$ on $B(u_\pm,\rho_1)$, there is $\rho_2>0$ such that $\omega_\nu(z)\notin B(u_\pm,\rho_2)$, and hence $\omega_\nu(z)\notin B(\pm u_0,\rho_2/2)$ for small $\xi$. The set of $w\in\overline{\mathcal N}\cup\mathcal O$ with $|w|\le7$ and $|w\mp u_0|\ge\rho_2/2$ is compact, lies at positive distance from $[-2r,2r]$, and avoids $\pm u_0$. By \eqref{eq:reference-critical}, $|\pastur_r'|\ge c$ on it, and \eqref{eq:C3-extended} gives $|\pastur_\nu'(\omega_\nu(z))|\ge c-s\xi\ge c/2$.

\emph{Condition \ref{ass:deformed-stieltjes}.} It follows from \ref{ass:deformed-regularity} and the representation $\Im m_{\mathrm{fc},A}(E+\mathrm{i}y)=\int y\rho_{\mathrm{fc},A}(x)((x-E)^2+y^2)^{-1}dx$ by an elementary computation, as for the semicircle law.
\end{proof}

Combining these deterministic estimates with the high-probability norm and exterior resolvent bounds of \cref{subsec:norm-resolvent}, we now establish burn-in rigidity uniformly over all planted laws with bounded low-degree advantage.

\begin{proposition}%
\label{prop:burnin-rigidity}
Fix $t_1>0$, $\varepsilon>0$, and $M>0$. There exist $c=c(t_1,\varepsilon,M)<\infty$ and $n_0=n_0(t_1,\varepsilon,M)$ such that, if
\begin{equation*}
D_n\ge c\log n,\qquad \adv_{\le D_n}(Q_n,P_n)\le1,
\end{equation*}
then, for independent $X\sim P_n$ and $W\sim Q_n$, the matrix
\begin{equation*}
H=rX+\sqrt s\,W\sim \mathcal U_{t_1}P_n,\qquad
r=e^{-t_1/2},\quad s=1-r^2,
\end{equation*}
satisfies, for all $n\ge n_0$,
\begin{equation}
\Pp\!\left(
|\lambda_i(H)-\gamma_i|
\le\frac{n^\varepsilon}{n^{2/3}\hat i^{1/3}}
\text{ for every }1\le i\le n
\right)\ge1-n^{-M}.
\label{eq:burnin-final-rigidity}
\end{equation}
In particular, $c$ and $n_0$ are uniform over all planted laws $P_n$ with $\adv_{\le D_n}(Q_n,P_n)\le1$.
\end{proposition}

\begin{proof}
We may assume $\varepsilon<1$, and we put $\eta:=\varepsilon/4$. We first find an event for $X$ on which $\widehat\nu_{rX}$ satisfies the assumptions of \cref{subsec:pastur-stability}. On this event we apply \cref{lem:conditional-deformed-wigner,lem:deformation-class} conditionally on $X$, and then compare the midpoint quantiles of the free convolution with $\gamma_i$ by \cref{lem:quantile-comparison}.

Let $\mathcal K:=r^{-1}\overline{\mathcal N}$. It is compact, and since $[-2r,2r]\subset I_r$, its distance from $[-2,2]$ is at least $r^{-1}\dist(\overline{\mathcal N},I_r)>0$. Apply \cref{lem:norm-confinement} with $(\delta,M+1)$, where $\delta=(1-r)^2/(8r)$ as in \cref{subsec:pastur-stability}, and \cref{lem:exterior-resolvent} with $(\mathcal K,M+1,\eta/2)$. Take $c$ at least their degree constants, and let $\mathcal E_n$ be the intersection of their good events, so that $P_n(\mathcal E_n^{\mathrm c})\le2n^{-M-1}$ for large $n$. On $\mathcal E_n$ we have $\supp\widehat\nu_{rX}\subset I_r$. The scaling identities $m_{\widehat\nu_{rX}}^{(j)}(w)=r^{-j-1}m_X^{(j)}(w/r)$ and $m_r^{(j)}(w)=r^{-j-1}m_{\mathrm{sc}}^{(j)}(w/r)$ show that $\nu=\widehat\nu_{rX}$ satisfies \eqref{eq:C3-stieltjes-close} with $\xi=r^{-4}n^{-1+\eta/2}\le n^{-1+\eta}$ for large $n$.

Fix $X\in\mathcal E_n$. Conditional on $X$, the orthogonal matrix that diagonalizes $X$ is fixed and $W$ keeps its law under conjugation, so $\spec(H)$ has the law of $\spec(A+\sqrt s\,W)$ with $A:=r\diag(\lambda_1(X),\dots,\lambda_n(X))$. For large $n$ we have $\xi\le\xi_0$, so \cref{lem:deformation-class} verifies the conditions of \cref{lem:conditional-deformed-wigner} with a constant $C_*$ depending only on $r$. That lemma, applied with $(\varepsilon/2,M+1)$, gives, with conditional probability at least $1-n^{-M-1}$,
\begin{equation}
|\lambda_i(H)-\bar\gamma_i^A|\le\frac{n^{\varepsilon/2}}{n^{2/3}\hat i^{1/3}},\qquad1\le i\le n.
\label{eq:conditional-rigidity}
\end{equation}
By \eqref{eq:edge-location-stability} and \cref{lem:pastur-cdf}, the measure $\mu_A$ satisfies the hypotheses of \cref{lem:quantile-comparison} with $\Delta:=Cn^{-1+\eta}$. Then \eqref{eq:quantile-comparison-index} with $p=(i-1/2)/n$ gives $|\bar\gamma_i^A-\gamma_i|\le Cn^{\eta}n^{-2/3}\hat i^{-1/3}$. Since $\eta<\varepsilon/2$, the triangle inequality and \eqref{eq:conditional-rigidity} give the bound in \eqref{eq:burnin-final-rigidity} for large $n$. Integrating over $X$, we obtain
\begin{equation*}
\Pp\Bigl(\exists i:|\lambda_i(H)-\gamma_i|>\frac{n^\varepsilon}{n^{2/3}\hat i^{1/3}}\Bigr)\le P_n(\mathcal E_n^{\mathrm c})+\sup_{X\in\mathcal E_n}\Pp\bigl(\eqref{eq:conditional-rigidity}\text{ fails}\bigm|X\bigr)\le3n^{-M-1}\le n^{-M}.
\end{equation*}
Every threshold for $n$ in this argument comes from \cref{lem:norm-confinement,lem:exterior-resolvent,lem:conditional-deformed-wigner,lem:deformation-class,lem:quantile-comparison} and from the fixed choices above, so it depends only on $(t_1,\varepsilon,M)$ and not on $P_n$.
\end{proof}

To conclude this subsection, we complete the proof of \cref{lem:rigidity}.

\begin{proof}[Proof of \cref{lem:rigidity}]
For all sufficiently large $n$, the assumption
$\adv_{\le D_n}(Q_n,P_n)=o(1)$ implies $\adv_{\le D_n}(Q_n,P_n)\le1$.
Choosing the fixed parameter $C_1=C_1(t_1,\varepsilon,M)$ at least as large as the threshold in \cref{prop:burnin-rigidity} gives \eqref{eq:rigidity}.
\end{proof}

\section{Coupling the spectra by homogenization}
\label{sec:spectral-coupling}
\label{subsec:spectral-semigroup}
In this section, we prove \cref{lem:Homogenization}. Fix $t_1,t_2>0$ and write $\sigma_n:=\spec(\mathcal U_{t_1}P_n)$ and $\pi_n:=\spec(Q_n)$. Starting from a coupled pair $\mathbf x(0)\sim\sigma_n$ and $\mathbf y(0)\sim\pi_n$, we run two copies $\mathbf x(t)$ and $\mathbf y(t)$ of the Dyson--Ornstein--Uhlenbeck flow driven by the same Brownian motions, aiming to show that $nW_2^2(\spec(\mathcal U_{t_1+t_2}P_n),\pi_n)=o(1)$. Because the rigidity bound of \cref{section:rigidity} only controls $n\|\mathbf x(0)-\mathbf y(0)\|_2^2$ up to $n^{2\varepsilon}$, synchronous contraction alone does not suffice. Instead, we combine a fixed-time homogenization estimate at $\tau=\min\{t_2,1/2\}$ (\cref{subsec:homogenization-input}) with a Chebyshev mode decomposition (\cref{subsec:spectral-Parseval}) and a low-mode initial coupling (\cref{subsec:Chebyshev-coupling}) to prove \cref{lem:Homogenization} in \cref{subsec:microscopic-transportation}. We supply the homogenization input in \cref{subsec:interpolating-regularity,subsec:homogenization-proof}.

For a constant $C>0$, we say that hypothesis (LD$_C$) holds if
\begin{equation}
D_n\ge C\log n\qquad\text{and}\qquad \adv_{\le D_n}(Q_n,P_n)\le1
\label{eq:standing-hypothesis}
\end{equation}
for all sufficiently large $n$. When a statement below says that there is $C=C(\dots)$ such that a conclusion holds under (LD$_C$) for all sufficiently large $n$, the constant $C$ and the threshold for $n$ depend only on the listed fixed parameters. They do not depend on $P_n$ or on the coupling of the initial spectra, and all error terms are uniform in the same sense. We write $\ell_i:=n^{-2/3}\hat i^{-1/3}$ for the rigidity scale at index $i$, so that $\mathcal R_{n,\varepsilon}$ is the set of $\mathbf x\in\mathcal W_n$ with $|x_i-\gamma_i|\le n^\varepsilon\ell_i$ for all $i$. Semicircle quantile spacing gives $\gamma_i-\gamma_{i-1}\le c\ell_i$, $\sum_i\ell_i\le c$, and $\sum_i\ell_i^2\le c/n$.

We now introduce the Markov semigroup and stochastic differential equation governing the evolution of the eigenvalues under OU noise. For $s\ge0$, define the spectral transition kernel on $\mathcal W_n$ by
\begin{equation}
K_s^{\mathrm{spec}}(\mathbf{x},\cdot)
:=\spec\left(\mathcal U_s\delta_{\diag(\mathbf{x})}\right).
\label{eq:spectral-kernel}
\end{equation}
Here $\delta_{\diag(\mathbf{x})}$ is the point mass at the diagonal matrix with entries $x_1,\ldots,x_n$. By orthogonal invariance of the Gaussian increment, the spectral law after applying the matrix OU channel depends on the initial matrix only through its spectrum. The matrix OU semigroup therefore induces a Markov semigroup on $\mathcal W_n$. In particular, for every matrix law $\mu$,
\begin{equation}
\spec(\mathcal U_s\mu)=\spec(\mu)K_s^{\mathrm{spec}},
\qquad
\pi_nK_s^{\mathrm{spec}}=\pi_n.
\label{eq:spectral-semigroup}
\end{equation}
The second identity follows from stationarity of $Q_n$.

For probability laws $\mu,\nu$ on $\mathcal W_n$ with finite second moments, we use the $2$-Wasserstein distance associated with the Euclidean norm:
\begin{equation*}
W_2^2(\mu,\nu)
:=\inf_{\Gamma\in\Pi(\mu,\nu)}
\int_{\mathcal W_n\times\mathcal W_n}\|\mathbf{x}-\mathbf{y}\|_2^2\,\Gamma(d\mathbf{x},d\mathbf{y}),
\end{equation*}
where $\Pi(\mu,\nu)$ is the set of couplings of $\mu$ and $\nu$.

By Dyson's classical calculation \cite{Dyson1962} (see also \cite[Chapter~4]{AGZ2010}), the spectral kernel in \eqref{eq:spectral-kernel} is the transition kernel of the following Dyson--Ornstein--Uhlenbeck system under our GOE normalization:
\begin{equation}
dx_i(t)=\sqrt{\frac2n}\,dB_i(t)
+\left(\frac1n\sum_{j\ne i}\frac1{x_i(t)-x_j(t)}
-\frac{x_i(t)}2\right)dt,
\qquad 1\le i\le n,
\label{eq:Dyson-OU}
\end{equation}
where $B_1,\ldots,B_n$ are independent standard Brownian motions. Starting from any coupling of the almost surely strictly ordered spectra $\mathbf{x}(0)\sim\sigma_n$ and $\mathbf{y}(0)\sim\pi_n$, drive the two copies of \eqref{eq:Dyson-OU} by the same Brownian motions, independently of the initial pair. Let $\mathcal F_t$ be the $\sigma$-algebra generated by the initial pair and by $B_i(s)$, $s\le t$, $1\le i\le n$. Since the initial pair is independent of the Brownian motions, $(B_i)$ is an $(\mathcal F_t)$-Brownian motion and the initial pair is $\mathcal F_0$-measurable. If
$d_i(t):=x_i(t)-y_i(t)$, subtraction gives
\begin{equation}
\frac{d}{dt}\left(e^{t/2}d_i(t)\right)
=\sum_{j\ne i}
\frac{e^{t/2}d_j(t)-e^{t/2}d_i(t)}
{n(x_i(t)-x_j(t))(y_i(t)-y_j(t))}.
\label{eq:synchronous-difference}
\end{equation}
Here, the coefficients $[n(x_i-x_j)(y_i-y_j)]^{-1}$ are positive and symmetric in $i,j$.
Multiplying \eqref{eq:synchronous-difference} by $e^{t/2}d_i(t)$ and summing over $i$ (and symmetrizing over $(i,j)$, which gives $\frac12\frac{d}{dt}(e^t\|\mathbf d(t)\|_2^2)=-\frac{e^t}{2n}\sum_{i\ne j}\frac{(d_i-d_j)^2}{(x_i-x_j)(y_i-y_j)}\le0$) gives
\begin{equation}
\|\mathbf{x}(t)-\mathbf{y}(t)\|_2^2\le e^{-t}\|\mathbf{x}(0)-\mathbf{y}(0)\|_2^2.
\label{eq:synchronous-contraction}
\end{equation}
In particular,
$W_2(\mu K_t^{\mathrm{spec}},\nu K_t^{\mathrm{spec}})
\le e^{-t/2}W_2(\mu,\nu)$.

Following \cite[(1.7)--(1.10)]{Bourgade2022ExtremeGaps} and \cite[Section~2.2]{LandonXian2025}, we also run the flow from convex combinations of the initial spectra. For any coupling of $\mathbf{x}(0)\sim\sigma_n$ and $\mathbf{y}(0)\sim\pi_n$ and each $\theta\in[0,1]$, let $\mathbf{x}^{(\theta)}$ solve \eqref{eq:Dyson-OU} with
\begin{equation}
\mathbf{x}^{(\theta)}(0)=\theta \mathbf{x}(0)+(1-\theta)\mathbf{y}(0),
\qquad \mathbf{x}^{(1)}=\mathbf{x},\quad \mathbf{x}^{(0)}=\mathbf{y},
\label{eq:interpolating-DBM}
\end{equation}
using the same Brownian motions for all $\theta$, independently of the initial pair. The initial spectra and their convex interpolations are strictly ordered almost surely. The maximum principle applied to \eqref{eq:synchronous-difference} gives
\begin{equation}
\sup_{0\le t\le1}\|\mathbf{x}^{(\theta)}(t)-\mathbf{x}^{(\theta')}(t)\|_\infty
\le |\theta-\theta'|\,\|\mathbf{x}(0)-\mathbf{y}(0)\|_\infty.
\label{eq:interpolation-Lipschitz}
\end{equation}
One can first construct the flows for a countable dense set of interpolation parameters and then use \eqref{eq:interpolation-Lipschitz} to extend the flows continuously to every $\theta\in[0,1]$. If both endpoint initial spectra lie in $\mathcal R_{n,\varepsilon}$, so does every interpolated initial spectrum. In \cref{subsec:homogenization-proof}, the derivative of $\mathbf x^{(\theta)}$ in $\theta$ solves a linear parabolic equation, homogenization estimates this derivative, and integration over $\theta$ recovers $\mathbf x-\mathbf y$. This requires rigidity along all the flows $\mathbf x^{(\theta)}$, which we prove in \cref{subsec:interpolating-regularity}.

Finally, we record the Chebyshev heat kernel that governs the homogenized displacement after a fixed time, defined by
\begin{equation}
p_t(x,y)=\sum_{m=0}^{\infty}e^{-mt/2}\mathsf{U}_m(x)\mathsf{U}_m(y),
\qquad t>0,
\label{eq:Dyson-heat-kernel}
\end{equation}
where $\mathsf U_m$ are the rescaled Chebyshev polynomials defined in \eqref{eq:Chebyshev-coordinates} below. For each fixed $t>0$, this series and its first derivatives converge uniformly on $[-2,2]^2$.

\subsection{Homogenization at a fixed time}
\label{subsec:homogenization-input}

In this subsection, we state our fixed-time homogenization estimate and use it to reduce the normalized transport cost $nW_2^2$ to a deterministic quadratic form in the initial displacement. We first record this estimate in \cref{prop:regular-data-homogenization}: after a fixed time $\tau\in(0,1/2]$, every particle displacement $x_i(\tau)-y_i(\tau)$ is approximated with high probability by $e^{-\tau/2}$ times the heat kernel $p_\tau$ applied to the initial displacement, up to a particlewise error $n^{-1-\chi}$ for a small constant $\chi>0$. We then show that the OU channel contracts the low-degree advantage and establish a fourth-moment bound to control the contribution of the exceptional event. Combining these estimates, we reduce the transport problem to bounding the expectation of a deterministic quadratic form in the initial displacement.

\begin{proposition}
\label{prop:regular-data-homogenization}
Let $\chi:=10^{-6}$. Fix $\tau\in(0,1/2]$, $\varepsilon\in(0,1/10)$, and $M>0$. There is $C=C(t_1,\tau,\varepsilon,M)$ such that the following holds under (LD$_C$) for all sufficiently large $n$. For every coupling of $\mathbf x(0)\sim\sigma_n$ and $\mathbf y(0)\sim\pi_n$, with probability at least $1-n^{-M}$, we have $\mathbf x(0),\mathbf y(0)\in\mathcal R_{n,\varepsilon}$ and
\begin{equation}
\left|x_i(\tau)-y_i(\tau)
-\frac{e^{-\tau/2}}n\sum_{j=1}^n
p_\tau(\gamma_i,\gamma_j)\bigl(x_j(0)-y_j(0)\bigr)\right|
\le n^{-1-\chi},
\qquad 1\le i\le n.
\label{eq:regular-data-homogenization}
\end{equation}
\end{proposition}

The exponent $\chi=10^{-6}$ comes from the contraction factor $1-10^{-5}$ in a single induction step of \cite[(4.63)]{LandonXian2025}. Although weaker than the iterated bound of \cite[Theorem~1.2]{LandonXian2025}, it already suffices for our transport bound because the squared particlewise errors contribute at most $n\cdot n\cdot n^{-2-2\chi}=n^{-2\chi}=o(1)$ to $nW_2^2$. Intuitively, in the parabolic equation \eqref{eq:synchronous-difference} for $e^{t/2}(\mathbf x(t)-\mathbf y(t))$, the short-range interactions average out local fluctuations of the displacement (localized by finite speed of propagation \cite{Bourgade2022ExtremeGaps,LandonXian2025}), leaving the long-range interactions to govern the evolution. By rigidity along the flow, the long-range coefficients are close to the deterministic weights $[n(\gamma_i-\gamma_j)^2]^{-1}$, which correspond to the continuum operator $\mathcal L$ of \cref{subsec:spectral-Parseval} with heat kernel $p_t$. We prove \cref{prop:regular-data-homogenization} in \cref{subsec:interpolating-regularity,subsec:homogenization-proof} by establishing simultaneous rigidity for all interpolating flows $\mathbf x^{(\theta)}$ and applying one induction step of \cite{LandonXian2025} (together with the bulk argument of \cite{Bourgade2022ExtremeGaps}) to their $\theta$-derivative before integrating in $\theta$.

To pass from the high-probability bound of \cref{prop:regular-data-homogenization} to an expectation bound for $nW_2^2$, we next record two elementary consequences of the low-degree assumption. First, we verify that the matrix OU semigroup  does not increase the low-degree advantage.

Let $\mathcal T_s$ denote the Markov operator of the matrix OU semigroup  on test functions:
\begin{equation*}
(\mathcal T_s q)(B)
:=\E_{W\sim Q_n}q\left(e^{-s/2}B+\sqrt{1-e^{-s}}W\right).
\end{equation*}
Then $\E_{\mathcal U_sP_n}q=\E_{P_n}[\mathcal T_s q]$, and $\mathcal T_s q$ has degree at most $D_n$ whenever $q$ does. Stationarity of $Q_n$ and conditional Jensen's inequality give
\begin{equation*}
\E_{Q_n}[\mathcal T_s q]=\E_{Q_n}q,
\qquad
\Var_{Q_n}(\mathcal T_s q)\le\Var_{Q_n}q.
\end{equation*}
Applying \eqref{eq:low-degree-transfer} to $\mathcal T_s q$ and taking the supremum over $q$ yields
\begin{equation}
\adv_{\le D_n}(Q_n,\mathcal U_sP_n)
\le \adv_{\le D_n}(Q_n,P_n).
\label{eq:OU-low-degree-contraction}
\end{equation}

Second, we establish a fourth-moment bound to control the exceptional event on which rigidity or homogenization fails. For $q(B)=(\Tr B^2)^2$, \cref{lem:norm-confinement} and $\Tr W^2\le n\|W\|_{\op}^2$ give $\E_{Q_n}q=O(n^2)$ and $\Var_{Q_n}q\le\E_{Q_n}q^2=O(n^4)$. Since $q$ has degree four, applying \eqref{eq:low-degree-transfer} together with \eqref{eq:OU-low-degree-contraction} yields, whenever $D_n\ge4$ and $\adv_{\le D_n}(Q_n,P_n)\le1$, uniformly in $s\ge0$,
\begin{equation}
\E_{H\sim \mathcal U_sP_n}(\Tr H^2)^2=O(n^2).
\label{eq:spectral-fourth-moment}
\end{equation}

Combining \cref{prop:regular-data-homogenization} with the fourth-moment bound \eqref{eq:spectral-fourth-moment} to control the exceptional event, we now reduce $nW_2^2$ to a deterministic quadratic form in the initial displacement.

\begin{lemma}%
\label{lem:homogenization-transport-reduction}

Fix $\tau\in(0,1/2]$ and $\varepsilon\in(0,1/10)$. There is $C=C(t_1,\tau,\varepsilon)$ such that the following holds under (LD$_C$) for all sufficiently large $n$. For every coupling of $\mathbf{x}(0)\sim\sigma_n$ and $\mathbf{y}(0)\sim\pi_n$, run \eqref{eq:Dyson-OU} with the same Brownian motions, independently of the initial pair. There is an event $\Omega_n$ with $\Pp(\Omega_n^c)\le n^{-10}$, on which $\mathbf{x}(0),\mathbf{y}(0)\in\mathcal R_{n,\varepsilon}$, such that the following inequality between expectations holds:

\begin{equation}
\begin{aligned}
&nW_2^2\bigl(\spec(\mathcal U_{t_1+\tau}P_n),\pi_n\bigr)\\
&\quad\le2\E\left[\mathbf1_{\Omega_n}\,
n\sum_{i=1}^n\left|
\frac{e^{-\tau/2}}n\sum_{j=1}^n
p_\tau(\gamma_i,\gamma_j)(x_j(0)-y_j(0))
\right|^2\right]+o(1).
\end{aligned}
\label{eq:homogenization-transport-reduction}
\end{equation}
The $o(1)$ is uniform over the initial couplings.
\end{lemma}

\begin{proof}
The time-$\tau$ marginals of the synchronous coupling are $\spec(\mathcal U_{t_1+\tau}P_n)$ and $\pi_n$. The definition of $W_2$ gives
\begin{equation*}
nW_2^2\bigl(\spec(\mathcal U_{t_1+\tau}P_n),\pi_n\bigr)
\le\E\bigl[n\|\mathbf{x}(\tau)-\mathbf{y}(\tau)\|_2^2\bigr].
\end{equation*}

Let $C$ be the constant of \cref{prop:regular-data-homogenization} with $M=10$, and let $\Omega_n$ be the event of that proposition. On $\Omega_n$, write $r_i$ for the remainder in \eqref{eq:regular-data-homogenization}. Then
\begin{equation*}
n\sum_i r_i^2\le n\cdot n\cdot n^{-2-2\chi}=n^{-2\chi}=o(1).
\end{equation*}
Using $|u+v|^2\le2|u|^2+2|v|^2$ on $\Omega_n$ bounds $\E[n\|\mathbf x(\tau)-\mathbf y(\tau)\|_2^2\mathbf1_{\Omega_n}]$ by the right-hand side of \eqref{eq:homogenization-transport-reduction}.

On the complementary event, put $Z_n:=\sum_i x_i(0)^2+\sum_i y_i(0)^2$. Equation~\eqref{eq:spectral-fourth-moment}, also applied with $P_n=Q_n$, gives $\E Z_n^2=O(n^2)$ uniformly over the initial couplings. By \eqref{eq:synchronous-contraction} and Cauchy--Schwarz,
\begin{equation*}
\E\bigl[n\|\mathbf{x}(\tau)-\mathbf{y}(\tau)\|_2^2\mathbf1_{\Omega_n^c}\bigr]
\le2n(\E Z_n^2)^{1/2}\Pp(\Omega_n^c)^{1/2}
=O(n^{-3}).
\end{equation*}
Adding the two contributions proves \eqref{eq:homogenization-transport-reduction}. The error bounds are uniform over the initial couplings.
\end{proof}

\subsection{Chebyshev decomposition of the displacement}
\label{subsec:spectral-Parseval}

In this subsection, we rewrite the quadratic form in \cref{lem:homogenization-transport-reduction} as an exponentially weighted sum of squared differences of polynomial linear statistics, up to small errors and a high-mode tail. Because the integral operator with kernel $p_\tau$ is diagonal in the orthonormal basis $\{\mathsf U_m\}$ of $L^2(\SC_1)$ with eigenvalues $e^{-m\tau/2}$, we use Parseval's identity to expand the quadratic form in Chebyshev modes. Moreover, each basis function $\mathsf U_{k-1}$ is the derivative of a degree-$k$ Chebyshev polynomial $\psi_k$, so we use a first-order Taylor expansion at the classical locations $\gamma_i$ to identify the $k$th displacement coefficient with the difference of linear statistics $\sum_i\psi_k(x_i)-\sum_i\psi_k(y_i)$. We then control the quadrature error, the Taylor remainder, and the tail of modes above a logarithmic cutoff $K\le b\log n$ to obtain the mode decomposition in \cref{lem:spectral-Parseval}.

Let $T_k$ and $U_k$ denote the Chebyshev polynomials of the first and second kind, respectively. They are characterized by the trigonometric identities
\begin{equation*}
T_k(\cos\theta)=\cos(k\theta),
\qquad
U_k(\cos\theta)=\frac{\sin((k+1)\theta)}{\sin\theta},
\qquad 0<\theta<\pi,
\end{equation*}
for $k\ge0$; see \cite[Section~18.5(i)]{DLMF}. Both are polynomials of degree $k$, defined on all of $\R$, with their values at $\pm1$ obtained by continuity. In particular, $T_0=U_0=1$, $T_1(x)=x$, and $U_1(x)=2x$. Differentiating the trigonometric identities gives $T_k'=kU_{k-1}$ for $k\ge1$.

For the semicircle support $[-2,2]$, we use the rescaled polynomials
\begin{equation}
\psi_k(x):=\frac{2}{k}T_k(x/2),
\qquad
\mathsf{U}_{k-1}(x):=U_{k-1}(x/2)=\psi_k'(x),
\qquad k\ge1.
\label{eq:Chebyshev-coordinates}
\end{equation}
Thus $\mathsf{U}_m(x)=U_m(x/2)$ for $m\ge0$, and $\psi_k'=\mathsf{U}_{k-1}$. The first few are $\psi_1(x)=x$, $\psi_2(x)=x^2/2-1$, $\mathsf{U}_0(x)=1$, and $\mathsf{U}_1(x)=x$.

The change of variables $x=2\cos\theta$ gives $\rhosc(x)\,|dx|=(2/\pi)\sin^2\theta\,d\theta$, and hence, for $m,\ell\ge0$,
\begin{equation*}
\int_{-2}^2\mathsf{U}_m(x)\mathsf{U}_\ell(x)\rhosc(x)\,dx
=\frac{2}{\pi}\int_0^\pi\sin((m+1)\theta)\sin((\ell+1)\theta)\,d\theta
=\begin{cases}1,&m=\ell,\\0,&m\ne\ell.\end{cases}
\end{equation*}
Since $\mathsf{U}_m$ has degree $m$ and polynomials are dense in $L^2(\SC_1)$, the family $\{\mathsf{U}_m:m\ge0\}$ is an orthonormal basis of this space. The same trigonometric identities give the elementary bounds
\begin{equation}
\sup_{|x|\le2}|\psi_k(x)|\le\frac{2}{k},
\qquad
\sup_{|x|\le2}|\psi_k'(x)|
=\sup_{|x|\le2}|\mathsf{U}_{k-1}(x)|\le k,
\qquad k\ge1.
\label{eq:Chebyshev-sup-bounds}
\end{equation}
For $k\ge1$, the same identities also give $U_{k-1}(-z)=(-1)^{k-1}U_{k-1}(z)$ and
\begin{equation}
|U_{k-1}(\pm\cosh u)|=\frac{\sinh(ku)}{\sinh u}
=\sum_{j=0}^{k-1}e^{(k-1-2j)u}
\le k e^{(k-1)u},\qquad u>0.
\label{eq:Chebyshev-outside-support}
\end{equation}
We use $U_k$ and $\mathsf U_k$ for scalar Chebyshev polynomials, $\mathcal U_t$ for the matrix OU semigroup, and $\mathcal T_t$ for its Markov operator on test functions.

Using the rescaled Chebyshev polynomials from \eqref{eq:Chebyshev-coordinates}, we define, for $\mathbf{x}\in\mathcal W_n$, the centered linear statistics
\begin{equation}
\Phi_{k,n}(\mathbf{x})
:=\sum_{i=1}^n\psi_k(x_i)
-\E_{W\sim Q_n}\Tr\psi_k(W).
\label{eq:Chebyshev-statistic}
\end{equation}
The statistic $\Phi_{k,n}(\spec(B))$ is a real polynomial of degree $k$ in the matrix entries, centered using the GOE expectation at dimension $n$.

The continuum operator corresponding to \eqref{eq:synchronous-difference} is
\begin{equation*}
(\mathcal L f)(x)
:=\operatorname{p.v.}\int_{-2}^2
\frac{f(y)-f(x)}{(x-y)^2}\rhosc(y)\,dy.
\end{equation*}
The Chebyshev recurrence gives $\mathcal L \mathsf{U}_m=-(m/2)\mathsf{U}_m$. Indeed, $\mathcal L1=0$, $\mathcal Lx=-x/2$, and $\mathcal L(xf)=x\mathcal L f-xf/2+\int_{-2}^2(f(y)-f(x))(y-x)^{-1}\rhosc(y)\,dy$ for every polynomial $f$. For $f=\mathsf{U}_m$ with $m\ge1$, the last integral equals $\mathsf{U}_{m-1}(x)$ (writing $I_m(x)$ for this integral, we have $I_0=0$, $I_1=1=\mathsf U_0$, and the recurrence $\mathsf U_{m+1}=x\mathsf U_m-\mathsf U_{m-1}$ together with $\int_{-2}^2\mathsf U_m\rhosc=0$ for $m\ge1$ gives $I_{m+1}=xI_m-I_{m-1}$), and induction using $\mathsf{U}_{m+1}=x\mathsf{U}_m-\mathsf{U}_{m-1}$ proves the claim. Thus $p_t$ in \eqref{eq:Dyson-heat-kernel} is the integral kernel of $e^{t\mathcal L}$.

The factor $e^{-t/2}$ in the particle displacement, together with the eigenvalue $-(k-1)/2$, gives decay $e^{-kt/2}$ for the $k$th Chebyshev mode. In the following lemma, we carry out this Parseval decomposition on the rigidity event, bounding the discrete heat-kernel energy by the first $K$ Chebyshev modes plus an explicit tail error.

\begin{lemma}%
\label{lem:spectral-Parseval}
Fix $\tau>0$, $b>0$, and $0<\varepsilon<1/10$. Let
$\mathbf{x},\mathbf{y}\in\mathcal R_{n,\varepsilon}$, $\mathbf{d}:=\mathbf{x}-\mathbf{y}$ with $d_i=x_i-y_i$, and $K\le b\log n$. Then
\begin{align}
n\sum_{i=1}^n\left|
\frac{e^{-\tau/2}}n\sum_{j=1}^np_\tau(\gamma_i,\gamma_j)d_j
\right|^2
&\le
\sum_{k=1}^K e^{-k\tau}
|\Phi_{k,n}(\mathbf{x})-\Phi_{k,n}(\mathbf{y})|^2
+c_\tau K^2e^{-K\tau}n^{2\varepsilon}+o(1).
\label{eq:spectral-Parseval}
\end{align}
The $o(1)$ is uniform over these arrays and cutoffs.
\end{lemma}

\begin{proof}

We expand the kernel in the basis $\{\mathsf U_m\}$. This turns the left-hand side into a weighted sum of the squared coefficients $b_k(\mathbf d):=\sum_i\mathsf U_{k-1}(\gamma_i)d_i$, up to a quadrature error. We then identify $b_k(\mathbf d)$ with the difference of the $k$th linear statistics by a Taylor expansion at the classical locations.

The rigidity bounds $|x_i-\gamma_i|,|y_i-\gamma_i|\le n^\varepsilon\ell_i$ and the sums of $\ell_i$ and $\ell_i^2$ give
\begin{equation}
\sum_i|d_i|\le c n^\varepsilon,
\qquad
\sum_i\bigl(|x_i-\gamma_i|^2+|y_i-\gamma_i|^2\bigr)\le c n^{-1+2\varepsilon}.
\label{eq:rigidity-displacement-sums}
\end{equation}
With \eqref{eq:Chebyshev-sup-bounds}, the first bound gives $|b_k(\mathbf d)|\le ckn^\varepsilon$. Inserting \eqref{eq:Dyson-heat-kernel} and exchanging the finite sum over $j$ with the absolutely convergent sum over $m$, we get
\begin{equation*}
\frac{e^{-\tau/2}}n\sum_{j=1}^np_\tau(\gamma_i,\gamma_j)d_j=\frac{g(\gamma_i)}n,
\qquad
g:=\sum_{m\ge0}e^{-(m+1)\tau/2}\,b_{m+1}(\mathbf d)\,\mathsf U_m.
\end{equation*}
Since $|\mathsf U_m|\le m+1$ and $|\mathsf U_m'|\le(m+1)^3$ on $[-2,2]$, the bound on $b_k(\mathbf d)$ gives $\|g\|_\infty+\|g'\|_\infty\le c_\tau n^\varepsilon$. Set $\gamma_0=-2$ and $I_i=(\gamma_{i-1},\gamma_i]$, so that $\SC_1(I_i)=1/n$ by \eqref{eq:classical-location} and $\sum_i|I_i|=4$. By orthonormality, $\int g^2\,d\SC_1=\sum_{k\ge1}e^{-k\tau}b_k(\mathbf d)^2$. The mean value theorem therefore gives
\begin{equation}
\left|n\sum_{i=1}^n\Bigl(\frac{g(\gamma_i)}n\Bigr)^2-\sum_{k\ge1}e^{-k\tau}b_k(\mathbf d)^2\right|
=\left|\sum_i\int_{I_i}\bigl[g(\gamma_i)^2-g(x)^2\bigr]\,\SC_1(dx)\right|
\le\frac{4\|(g^2)'\|_\infty}n
\le c_\tau n^{-1+2\varepsilon}.
\label{eq:discrete-continuum-energy}
\end{equation}
The terms with $k>K$ contribute at most $cn^{2\varepsilon}\sum_{k>K}k^2e^{-k\tau}\le c_\tau K^2e^{-K\tau}n^{2\varepsilon}$.

Finally, we compare $b_k(\mathbf d)$ with the Chebyshev mode difference. Rigidity places both spectra in $[-2-n^{-2/3+\varepsilon},2+n^{-2/3+\varepsilon}]$. Since $K n^{-1/3+\varepsilon/2}=o(1)$, the Chebyshev derivative bounds give $|\psi_k''|\le cK^3$ throughout this interval, uniformly for $k\le K$. Indeed, differentiating $U_{k-1}(y)=2\sum_{r}T_{k-1-2r}(y)+c_k$ over $0\le r\le(k-2)/2$ (with $c_k\in\{0,1\}$) gives $U_{k-1}'(y)=2\sum_{r}(k-1-2r)U_{k-2-2r}(y)$, and applying \eqref{eq:Chebyshev-outside-support} with $|x|/2=\cosh u$ and $u\le cn^{-1/3+\varepsilon/2}$ yields $|\psi_k''(x)|=\frac12|U_{k-1}'(x/2)|\le ck^3e^{ku}\le cK^3$. Taylor's theorem at each $\gamma_i$ therefore gives
\begin{equation*}
\begin{aligned}
\Phi_{k,n}(\mathbf{x})-\Phi_{k,n}(\mathbf{y})
&=\sum_i\psi_k'(\gamma_i)d_i\\
&\quad+O\left(K^3\sum_i\bigl[|x_i-\gamma_i|^2+|y_i-\gamma_i|^2\bigr]\right)\\
&=b_k(\mathbf d)+O(K^3n^{-1+2\varepsilon}).
\end{aligned}
\end{equation*}
The last equality uses $\psi_k'=\mathsf U_{k-1}$ and the second bound in \eqref{eq:rigidity-displacement-sums}.
Using $|a^2-b^2|\le2|b||a-b|+|a-b|^2$, the Taylor remainder and $|b_k(\mathbf d)|\le ck n^\varepsilon$ give
\begin{equation*}
\sum_{k=1}^K e^{-k\tau}
\left|\bigl(\Phi_{k,n}(\mathbf x)-\Phi_{k,n}(\mathbf y)\bigr)^2-b_k(\mathbf d)^2\right|
\le c_\tau\bigl(K^3n^{-1+3\varepsilon}+K^6n^{-2+4\varepsilon}\bigr)=o(1).
\end{equation*}
Combining these estimates with \eqref{eq:discrete-continuum-energy} and the bound on the terms with $k>K$ proves \eqref{eq:spectral-Parseval}.
\end{proof}

\subsection{Choosing the initial coupling}
\label{subsec:Chebyshev-coupling}

We now choose the coupling of the initial spectra so that the low modes nearly agree. For a fixed number $L$ of modes, the low-degree assumption makes every mixed moment of $(\Phi_{1,n},\dots,\Phi_{L,n})$ asymptotically the same under $\sigma_n$ and $\pi_n$, and the GOE central limit theorem identifies the common Gaussian limit.  We then lift an optimal coupling of the mode vectors to the two spectral laws. For the modes $L<k\le b\log n$ we need only second moments, which low-degree moment comparison at degree $2k$ controls. We begin with the following moment bounds under the null.

\begin{lemma}%
\label{lem:Chebyshev-null-moments}
For every fixed $b>0$ and $p\ge1$, there exists a constant $c_{b,p}$ such that
\begin{equation}
\E_{\pi_n}|\Phi_{k,n}|^p\le c_{b,p}k^p,
\qquad 1\le k\le b\log n,
\label{eq:Chebyshev-null-moments}
\end{equation}
for all sufficiently large $n$. For each fixed $L$, the vector
$(\Phi_{1,n},\ldots,\Phi_{L,n})$ under $\pi_n$ converges in distribution and in all mixed moments to a centered Gaussian vector.
\end{lemma}

\begin{proof}
We truncate $\psi_k$ outside a small neighborhood of $[-2,2]$ to obtain a Lipschitz function, apply Gaussian concentration to its trace, and check that the truncation matters only on an event of very small probability.
By \eqref{eq:null-norm-shrinking} in \cref{lem:norm-confinement}, the spectrum of $W$ lies in $[-2-n^{-1/2},2+n^{-1/2}]$ with probability at least $1-n^{-M}$ for every fixed $M>0$ and sufficiently large $n$. Define
\begin{equation*}
f_{k,n}(x):=
\begin{cases}
\psi_k(-2-n^{-1/2}),&x<-2-n^{-1/2},\\
\psi_k(x),&|x|\le2+n^{-1/2},\\
\psi_k(2+n^{-1/2}),&x>2+n^{-1/2}.
\end{cases}
\end{equation*}
For $2<|x|\le2+n^{-1/2}$, write $|x|/2=\cosh u$, where $0<u\le c n^{-1/4}$. By \eqref{eq:Chebyshev-outside-support}, we have $|U_{k-1}(x/2)|\le k e^{(k-1)u}$. Combining this with the bound \eqref{eq:Chebyshev-sup-bounds} on $[-2,2]$ and using $k n^{-1/4}=o(1)$ uniformly for $k\le b\log n$, we obtain
\begin{equation*}
\sup_{|x|\le2+n^{-1/2}}|\psi_k'(x)|
=\sup_{|x|\le2+n^{-1/2}}|U_{k-1}(x/2)|\le c_b k,
\qquad k\le b\log n.
\end{equation*}
Thus $f_{k,n}$ is globally $c_bk$-Lipschitz.
By \eqref{eq:Lipschitz-spectral-trace}, the map $B\mapsto\Tr f_{k,n}(B)$ is $c_bk\sqrt n$-Lipschitz in the Frobenius norm. \cref{lem:gaussian-concentration} therefore gives
\begin{equation*}
\bigl\|\Tr f_{k,n}(W)-\E_{Q_n}\Tr f_{k,n}(W)\bigr\|_{L^p(Q_n)}
\le c_{b,p}k.
\end{equation*}

On the event $\{\|W\|_{\op}\le2+n^{-1/2}\}$, we have $\Tr f_{k,n}(W)=\Tr\psi_k(W)$.
On the complementary event $\{\|W\|_{\op}>2+n^{-1/2}\}$, use $|\psi_k(x)|\le2(1+|x|)^k$ and the null norm moments in \cref{lem:norm-confinement}. For fixed $b,p$, the $L^{2p}(Q_n)$ norm of $\Tr f_{k,n}(W)-\Tr\psi_k(W)$ grows at most polynomially in $n$, uniformly for $k\le b\log n$.
Choosing $M$ sufficiently large in \eqref{eq:null-norm-shrinking} and applying H\"older's inequality gives $\|\Tr f_{k,n}(W)-\Tr\psi_k(W)\|_{L^p(Q_n)}=o(1)$. The difference of the corresponding expectations is also $o(1)$.
This proves \eqref{eq:Chebyshev-null-moments}.

For the fixed-dimensional limit, we use the GOE central limit theorem of \cite[Theorem~2.2 and Remark~2.1(2)]{LytovaPastur2009}, with variance parameter $w=1$. It applies to every fixed linear combination of the polynomials $\psi_1,\ldots,\psi_L$. The Cram\'er--Wold theorem \cite{CramerWold1936} gives joint Gaussian convergence. The bounds at arbitrarily large fixed orders give uniform integrability of every fixed mixed monomial and hence convergence of all mixed moments.
\end{proof}

By \eqref{eq:OU-low-degree-contraction}, the low-degree advantage of $\mathcal U_{t_1}P_n$ is at most $\adv_{\le D_n}(Q_n,P_n)$. Combining this bound with the null moment estimates in \cref{lem:Chebyshev-null-moments}, we obtain the following coupling of the Chebyshev modes.

\begin{lemma}%
\label{lem:Chebyshev-mode-coupling}
Suppose $\adv_{\le D_n}(Q_n,P_n)=o(1)$ and $D_n\to\infty$. For each fixed $L$, there is a coupling $\mathbf{x}\sim\sigma_n$, $\mathbf{y}\sim\pi_n$ such that
\begin{equation}
\E\sum_{k=1}^L
|\Phi_{k,n}(\mathbf{x})-\Phi_{k,n}(\mathbf{y})|^2=o(1).
\label{eq:finite-mode-coupling}
\end{equation}
Moreover, for every fixed $b>0$, whenever $k\le b\log n$ and $2k\le D_n$, every coupling of $\sigma_n$ and $\pi_n$  satisfies
\begin{equation}
\E|\Phi_{k,n}(\mathbf{x})-\Phi_{k,n}(\mathbf{y})|^2
\le c_b k^2(1+\adv_{\le D_n}(Q_n,P_n)).
\label{eq:medium-mode-bound}
\end{equation}
\end{lemma}

\begin{proof}
To prove \eqref{eq:finite-mode-coupling}, we first show that the first $L$ Chebyshev modes under $\sigma_n$ and $\pi_n$ converge in $2$-Wasserstein distance to the same Gaussian limit, and then lift an optimal coupling of these mode vectors back to the full spectra via regular conditional distributions. Fix $L$ and a real polynomial $R$ on $\R^L$.
The statistic $R(\Phi_{1,n}(\spec(B)),\ldots,\Phi_{L,n}(\spec(B)))$ has degree bounded independently of $n$, even though its null-centering constants depend on $n$.
\cref{lem:Chebyshev-null-moments} bounds the null variance of this polynomial statistic uniformly in $n$. Applying \eqref{eq:low-degree-transfer} and \eqref{eq:OU-low-degree-contraction}, we obtain
\begin{equation*}
\left|\E_{\sigma_n}R(\Phi_{1,n},\ldots,\Phi_{L,n})
-\E_{\pi_n}R(\Phi_{1,n},\ldots,\Phi_{L,n})\right|=o(1).
\end{equation*}
Define the mode map and its induced laws by
\begin{equation*}
F_{L,n}(\mathbf{x}):=(\Phi_{1,n}(\mathbf{x}),\ldots,\Phi_{L,n}(\mathbf{x})),
\qquad
\mu_{L,n}:=(F_{L,n})_\#\sigma_n,
\qquad
\nu_{L,n}:=(F_{L,n})_\#\pi_n.
\end{equation*}
Let $\mathrm N(0,\Sigma_L)$ denote the centered Gaussian limit in \cref{lem:Chebyshev-null-moments}. Applying the preceding estimate to every monomial gives, for each $\alpha\in\mathbb Z_{\ge0}^L$,
\begin{equation*}
\int_{\R^L}z^\alpha\,\mu_{L,n}(dz)
=\int_{\R^L}z^\alpha\,\nu_{L,n}(dz)+o(1)
\longrightarrow\int_{\R^L}z^\alpha\,\mathrm N(0,\Sigma_L)(dz),
\qquad z^\alpha:=\prod_{k=1}^Lz_k^{\alpha_k}.
\end{equation*}
The second moments give tightness, and the higher even moments give uniform integrability of every fixed monomial. Since the Gaussian law is moment determinate, the method of moments yields $\mu_{L,n}\Rightarrow\mathrm N(0,\Sigma_L)$ and $\nu_{L,n}\Rightarrow\mathrm N(0,\Sigma_L)$. The second moments $\int\|z\|_2^2\,d\mu_{L,n}$ and $\int\|z\|_2^2\,d\nu_{L,n}$ also converge to that of $\mathrm N(0,\Sigma_L)$. Since weak convergence together with convergence of second moments is equivalent to convergence in $W_2$, the triangle inequality gives
\begin{equation*}
W_2(\mu_{L,n},\nu_{L,n})
\le W_2(\mu_{L,n},\mathrm N(0,\Sigma_L))
+W_2(\nu_{L,n},\mathrm N(0,\Sigma_L))=o(1).
\end{equation*}

Let $\Lambda_n$ be an optimal coupling of $\mu_{L,n}$ and $\nu_{L,n}$, and let $\kappa_n^\sigma(z,d\mathbf{x})$ and $\kappa_n^\pi(w,d\mathbf{y})$ be regular conditional distributions of the spectra given their modes; they exist because $\mathcal W_n$ is a closed subset of $\R^n$ and $F_{L,n}$ is continuous. The coupling
\begin{equation*}
\Gamma_n(d\mathbf{x},d\mathbf{y})
:=\int_{\R^L\times\R^L}
\kappa_n^\sigma(z,d\mathbf{x})\,\kappa_n^\pi(w,d\mathbf{y})\,\Lambda_n(dz,dw)
\end{equation*}
has marginals $\sigma_n$ and $\pi_n$, and under $\Gamma_n$ the pair $(F_{L,n}(\mathbf x),F_{L,n}(\mathbf y))$ has law $\Lambda_n$, because the conditional distributions are supported on the fibers of $F_{L,n}$. Hence the left-hand side of \eqref{eq:finite-mode-coupling} under $\Gamma_n$ equals $W_2^2(\mu_{L,n},\nu_{L,n})=o(1)$. This proves \eqref{eq:finite-mode-coupling}.

For the second assertion, apply \eqref{eq:low-degree-transfer} to the nonnegative degree-$2k$ polynomial $\Phi_{k,n}(\spec(B))^2$. The second and fourth null moment bounds give
\begin{equation*}
\E_{\sigma_n}\Phi_{k,n}^2
\le\E_{\pi_n}\Phi_{k,n}^2
+\adv_{\le D_n}(Q_n,P_n)\sqrt{\Var_{\pi_n}(\Phi_{k,n}^2)}
\le c_b k^2(1+\adv_{\le D_n}(Q_n,P_n)).
\end{equation*}
For any coupling $\mathbf{x}\sim\sigma_n$, $\mathbf{y}\sim\pi_n$, the bound $\E|\Phi_{k,n}(\mathbf{x})-\Phi_{k,n}(\mathbf{y})|^2\le2\E_{\sigma_n}\Phi_{k,n}^2+2\E_{\pi_n}\Phi_{k,n}^2$ and the second null moment bound prove \eqref{eq:medium-mode-bound}, after enlarging $c_b$ if necessary.
\end{proof}

\subsection{Proof of the transport estimate}
\label{subsec:microscopic-transportation}

We now assemble the estimates from Sections~\ref{subsec:homogenization-input}--\ref{subsec:Chebyshev-coupling} to prove \cref{lem:Homogenization}. Combining the transport reduction (\cref{lem:homogenization-transport-reduction}) with the Chebyshev decomposition (\cref{lem:spectral-Parseval}), we split $nW_2^2$ into three ranges of modes: the first $L$ low modes matched by the initial coupling, the middle modes up to $K=\lfloor b\log n\rfloor$ damped by $e^{-k\tau}$, and the tail modes beyond $K$ suppressed by choosing $b\tau>2\varepsilon$.

\begin{proof}[Proof of \cref{lem:Homogenization}]

Fix $t_1,t_2>0$ and put $\tau:=\min\{t_2,1/2\}$. Set $\varepsilon=1/100$, choose $b>0$ so that $b\tau>2\varepsilon$, and let $K=\lfloor b\log n\rfloor$. Let $C_2=C_2(t_1,t_2)$ be at least $2b$ and at least the constant $C(t_1,\tau,\varepsilon)$ of \cref{lem:homogenization-transport-reduction}. This constant is a fixed number, because the proof of \cref{prop:regular-data-homogenization} fixes the control parameter $\delta$, and with it every exponent passed to \cref{section:rigidity}, before the degree threshold is chosen. For large $n$, the assumptions of \cref{lem:Homogenization} give (LD$_{C_2}$) and $2K\le D_n$. The fixed mode cutoff $L$ below does not enter the choice of $C_2$.

Fix $L$ independently of $n$, and use \cref{lem:Chebyshev-mode-coupling} to couple $\mathbf{x}(0)\sim\sigma_n$ and $\mathbf{y}(0)\sim\pi_n$ so that \eqref{eq:finite-mode-coupling} holds. Run the synchronous systems for time $\tau$. Their marginals at that time are $\spec(\mathcal U_{t_1+\tau}P_n)$ and $\pi_n$. On the event $\Omega_n$ of \cref{lem:homogenization-transport-reduction}, both initial spectra lie in $\mathcal R_{n,\varepsilon}$, so \cref{lem:spectral-Parseval} applies there. Dropping the indicator of $\Omega_n$ in front of the nonnegative terms, we obtain

\begin{equation*}
nW_2^2\bigl(\spec(\mathcal U_{t_1+\tau}P_n),\pi_n\bigr)
\le
2\sum_{k=1}^K e^{-k\tau}
\E|\Phi_{k,n}(\mathbf{x}(0))-\Phi_{k,n}(\mathbf{y}(0))|^2+o(1).
\end{equation*}
The tail term is $O(K^2n^{2\varepsilon}e^{-K\tau})=o(1)$ by the choice of $b$. The first $L$ terms tend to zero by \eqref{eq:finite-mode-coupling}; the terms $L<k\le K$ are at most
\begin{equation*}
c_b(1+\adv_{\le D_n}(Q_n,P_n))\sum_{k>L}e^{-k\tau}k^2
\end{equation*}
by \eqref{eq:medium-mode-bound}. This last expression tends to zero as $L\to\infty$, uniformly for sufficiently large $n$.

Taking $n\to\infty$ first and then $L\to\infty$ proves
$nW_2^2(\spec(\mathcal U_{t_1+\tau}P_n),\pi_n)=o(1)$. If $t_2>\tau$, the contraction in \eqref{eq:synchronous-contraction} for the remaining time $t_2-\tau$, together with stationarity, proves \eqref{eq:homogenization-transport}.
\end{proof}

\subsection{Regularity of the interpolating flow}
\label{subsec:interpolating-regularity}

To carry out the homogenization argument, we need rigidity of $\mathbf x^{(\theta)}(t)$ simultaneously for all $\theta,t\in[0,1]$. We separate the randomness of the initial pair from that of the Brownian motions. The initial pair is rigid with probability $1-n^{-M}$ by \cref{section:rigidity}. The event that the initial pair is rigid but rigidity fails along some interpolating flow has probability at most $n^{-D}$ for every fixed $D$. Away from the edges, this follows from the quantile comparison of Aggarwal and Huang, which compares each flow with a free convolution, and from the contraction of free convolution in L\'evy distance. Near the edges, we compare each interpolating flow with the GOE flow, which is stationary. Thus the polynomial failure probability of \cref{section:rigidity} enters only once, through the initial pair. For $\alpha,\varepsilon>0$, let
\begin{equation}
\mathcal H_n(\alpha):=\{\mathbf x(0),\mathbf y(0)\in\mathcal R_{n,\alpha}\},
\qquad
\mathcal G_n(\varepsilon):=\{\mathbf x^{(\theta)}(t)\in\mathcal R_{n,\varepsilon}\text{ for all }\theta,t\in[0,1]\}.
\label{eq:regularity-events}
\end{equation}
The event $\mathcal H_n(\alpha)$ depends only on the initial pair.

\begin{proposition}
\label{prop:interpolating-dynamic-rigidity}
Fix $0<\varepsilon<1/3$ and $D>0$. Let $\mathbf x(0)$ be an almost surely strictly ordered random element of $\mathcal W_n$, coupled in any way with $\mathbf y(0)\sim\pi_n$, and run the flows \eqref{eq:interpolating-DBM}. Then, for all sufficiently large $n$, uniformly over such initial pairs,
\begin{equation}
\Pp\bigl(\mathcal H_n(\varepsilon/4)\setminus\mathcal G_n(\varepsilon)\bigr)\le n^{-D}.
\label{eq:interpolating-uniform-rigidity}
\end{equation}
\end{proposition}

The law of $\mathbf x(0)$ enters only through $\mathcal H_n(\varepsilon/4)$. We prove the proposition at the end of this subsection, after three lemmas. The first one is rigidity of the GOE flow at all times, by time discretization.

\begin{lemma}
\label{lem:GOE-path-rigidity}
Fix $\alpha>0$ and $D>0$. If $\mathbf y(0)\sim\pi_n$ and $\mathbf y$ solves \eqref{eq:Dyson-OU}, then, for all sufficiently large $n$,
\begin{equation*}
\Pp\{\mathbf y(t)\in\mathcal R_{n,\alpha}\text{ for all }t\in[0,1]\}\ge1-n^{-D}.
\end{equation*}
\end{lemma}

\begin{proof}
Since $\pi_n$ is stationary, $\mathbf y(t)\sim\pi_n$ for every $t$, and GOE rigidity \cite{ERDOS20121435} gives $\Pp\{\mathbf y(t)\notin\mathcal R_{n,\alpha/2}\}\le n^{-D-5}$ for each fixed $t$ and large $n$. Let $t_j:=jn^{-4}$ for $0\le j\le n^4$. A union bound shows that $\mathbf y(t_j)\in\mathcal R_{n,\alpha/2}$ for all $j$, except on an event of probability at most $2n^{-D-1}$.

Between grid times we use the matrix process. Since $K_s^{\mathrm{spec}}$ is the transition kernel of \eqref{eq:Dyson-OU} and both processes have continuous paths, $\mathbf y$ has the law of the spectrum of the stationary matrix OU process
\begin{equation*}
dH_t=\frac1{\sqrt n}\,d\Xi_t-\frac12H_t\,dt,
\qquad H_0\sim Q_n,
\end{equation*}
where $\Xi$ is a symmetric matrix Brownian motion whose off-diagonal entries are independent standard Brownian motions and whose diagonal entries have variance rate $2$. The reflection principle and a union bound over at most $2n^6$ entry--interval pairs give $|\Xi_{ab}(t)-\Xi_{ab}(t_j)|\le n^{-2}\log n$ for all $a,b,j$ and $t\in[t_j,t_{j+1}]$, except on an event of probability at most $8n^6\exp(-(\log n)^2/4)$. The same argument over $[0,1]$, the bound $\|H_0\|_{\op}\le3$ from \cref{lem:norm-confinement}, and Gronwall's inequality (using $\sup_{t\le1}|\Xi_{ab}(t)|\le\log n$, so that $\|H_t\|_{\op}\le3+\sqrt n\log n+\frac12\int_0^t\|H_s\|_{\op}\,ds$) give $\sup_{t\le1}\|H_t\|_{\op}\le n$ outside an event of probability at most $n^{-D-1}$. On these events, the integral form of the equation and $\|A\|_{\op}\le n\max_{a,b}|A_{ab}|$ give
\begin{equation*}
\max_j\sup_{t_j\le t\le t_{j+1}}\|H_t-H_{t_j}\|_{\op}
\le n^{-3/2}\log n+\tfrac12n^{-3}=o(n^{-1}).
\end{equation*}
By Weyl's inequality, every eigenvalue moves by $o(n^{-1})$ between consecutive grid times. Since $\ell_i\ge n^{-1}$, this gives $\mathbf y(t)\in\mathcal R_{n,\alpha}$ for all $t\in[0,1]$, for large $n$, outside an event of probability at most $n^{-D}$.
\end{proof}

To control the interpolating flows away from the edges, we use the following quantile comparison for driftless DBM, based on \cite[Corollary~3.2]{HuangLandon2016}. We state it with $\beta=1$, $D=3$, and particles in increasing order.

\begin{lemma}[{\cite[Proposition~2.9]{AggarwalHuang2024}}]%
\label{lem:general-data-quantile-comparison}
Let $\mathbf a\in\mathcal W_n$ be deterministic with $|a_i|\le n^3$ for every $i$, and let $\lambda_1(u)\le\cdots\le\lambda_n(u)$ solve
\begin{equation*}
d\lambda_i(u)=\sqrt{\frac2n}\,dB_i(u)
+\frac1n\sum_{j\ne i}\frac{du}{\lambda_i(u)-\lambda_j(u)},
\qquad \lambda_i(0)=a_i,
\end{equation*}
where $B_1,\ldots,B_n$ are independent standard Brownian motions. Set
\begin{equation*}
\mu_0:=\frac1n\sum_{i=1}^n\delta_{a_i},
\qquad \mu_u:=\mu_0\boxplus\SC_u,\quad u>0,
\end{equation*}
and define the midpoint quantiles, for $u\ge0$, by
\begin{equation*}
\zeta_i(u):=\inf\left\{E\in\R:
\mu_u((-\infty,E])\ge\frac{i-1/2}{n}\right\},
\qquad 1\le i\le n.
\end{equation*}
For indices outside this range, set $\zeta_j(u)=-\infty$ if $j<1$ and $\zeta_j(u)=+\infty$ if $j>n$. There is a constant $c>0$ such that, for all sufficiently large $n$, with probability at least $1-c\exp[-(\log n)^2]$,
\begin{equation*}
\zeta_{i-\lfloor(\log n)^5\rfloor}(u)-n^{-3}
\le\lambda_i(u)
\le\zeta_{i+\lfloor(\log n)^5\rfloor}(u)+n^{-3}
\end{equation*}
simultaneously for every $1\le i\le n$ and $0\le u\le n^3$. The constant and the lower bound on $n$ are uniform over the allowed initial spectra $\mathbf a$.
\end{lemma}

Using \cref{lem:GOE-path-rigidity,lem:general-data-quantile-comparison}, we next prove rigidity along the entire time interval $[0,1]$ for a single fixed interpolation parameter $\theta\in[0,1]$.

\begin{lemma}%
\label{lem:fixed-parameter-rigidity}

Fix $0<\varepsilon<1/3$ and put $\alpha:=\varepsilon/4$. Let the initial pair be as in \cref{prop:interpolating-dynamic-rigidity}, write $\mathcal H_n:=\mathcal H_n(\alpha)$, and let
\begin{equation*}
\mathcal Y_n:=\{\mathbf y(t)\in\mathcal R_{n,\alpha}\text{ for all }t\in[0,1]\}.
\end{equation*}
There is $c>0$ such that, for all sufficiently large $n$ and every $\theta\in[0,1]$, there is an event $\mathcal A_n^{(\theta)}$ with
\begin{equation}
\Pp\bigl(\mathcal H_n\setminus\mathcal A_n^{(\theta)}\bigr)\le c\exp(-(\log n)^2),
\qquad
\mathcal H_n\cap\mathcal Y_n\cap\mathcal A_n^{(\theta)}
\subset\{\mathbf x^{(\theta)}(t)\in\mathcal R_{n,\varepsilon/2}\text{ for all }t\in[0,1]\}.
\label{eq:fixed-parameter-rigidity}
\end{equation}
The constant $c$ and the threshold for $n$ do not depend on $\theta$ or on the initial pair.

\end{lemma}

\begin{proof}

We treat indices away from the edges and near the edges separately. Away from the edges, we compare the empirical distribution function of $\mathbf x^{(\theta)}(t)$ with that of a free convolution of the initial spectrum, using \cref{lem:general-data-quantile-comparison}, and compare this free convolution with $\SC_1$ using L\'evy contraction. Inverting the distribution function then gives rigidity. Near the edges, the maximum principle compares $\mathbf x^{(\theta)}$ with the GOE flow. All constants and lower bounds on $n$ below are independent of $\theta$ and of the initial pair.

On $\mathcal H_n$, the definition of the interpolation gives, simultaneously for every $\theta\in[0,1]$ and every $i$,

\begin{equation*}
|x_i^{(\theta)}(0)-\gamma_i|
\le\theta|x_i(0)-\gamma_i|+(1-\theta)|y_i(0)-\gamma_i|
\le n^\alpha n^{-2/3}\hat i^{-1/3}.
\end{equation*}
Thus every interpolated initial spectrum belongs to $\mathcal R_{n,\alpha}$, and all its particles lie in $[-3,3]$ for large $n$.

Using the distribution-function notation from Preliminaries, let
\begin{equation*}
\mu_t^{(\theta)}:=\widehat{\nu}_{\diag(\mathbf{x}^{(\theta)}(t))}.
\end{equation*}
For a rigid initial spectrum, the semicircle density bound $|F_{\mathrm{sc}}(\gamma_i\pm h)-F_{\mathrm{sc}}(\gamma_i)|\le c(h(\hat i/n)^{1/3}+h^{3/2})$ (applied with $h=n^\alpha\ell_i$) gives
\begin{equation*}
|F_{\mathrm{sc}}(x_i^{(\theta)}(0))-i/n|
\le c\left(n^\alpha n^{-2/3}\hat i^{-1/3}(\hat i/n)^{1/3}
+(n^\alpha n^{-2/3}\hat i^{-1/3})^{3/2}\right)
\le c n^{-1+3\alpha/2}.
\end{equation*}
By monotonicity of the distribution functions, the estimate extends from the particle locations to the whole real line with an additional error of at most $1/n$.
Thus, on $\mathcal H_n$, we have $\sup_E|F_{\mu_0^{(\theta)}}(E)-F_{\mathrm{sc}}(E)|\le c n^{-1+3\alpha/2}$ for all $\theta$, and hence $d_{\mathrm L}(\mu_0^{(\theta)},\SC_1)\le c n^{-1+3\alpha/2}$, where
\begin{equation*}
d_{\mathrm L}(\mu,\nu)
:=\inf\{h>0:F_\mu(E-h)-h\le F_\nu(E)\le F_\mu(E+h)+h
\text{ for all }E\}
\end{equation*}
is the L\'evy distance.

For a fixed deterministic $\theta\in[0,1]$ and given initial spectra $\mathbf x(0)$ and $\mathbf y(0)$, define
\begin{equation*}
\nu_t^{(\theta)}
:=\widehat{\nu}_{e^{-t/2}\diag(\mathbf{x}^{(\theta)}(0))}
\boxplus\SC_{1-e^{-t}},\qquad t>0,
\end{equation*}
and set $\nu_0^{(\theta)}=\mu_0^{(\theta)}$. By \cite[Proposition~4.13]{BercoviciVoiculescu1993}, free convolution satisfies the L\'evy contraction inequality
\begin{equation*}
d_{\mathrm L}(\mu\boxplus\omega,\nu\boxplus\omega)
\le d_{\mathrm L}(\mu,\nu).
\end{equation*}
Dilation by a factor in $(0,1]$ also contracts this distance. Applying these two facts and the identity $\SC_{e^{-t}}\boxplus\SC_{1-e^{-t}}=\SC_1$ to the bound for the initial spectra yields, on $\mathcal H_n$,
\begin{equation*}
\sup_{\theta\in[0,1]}\sup_{0\le t\le1}
d_{\mathrm L}(\nu_t^{(\theta)},\SC_1)
\le c n^{-1+3\alpha/2}.
\end{equation*}
This is a deterministic statement once the initial pair is fixed.

To compare $\mu_t^{(\theta)}$ with $\nu_t^{(\theta)}$, we apply \cref{lem:general-data-quantile-comparison} after removing the OU drift.
Condition on any initial pair in $\mathcal H_n$. The driving Brownian motions retain their original law, since they are independent of the initial pair. The change of variables $u=e^t-1$ and $\widetilde x_i(u)=e^{t/2}x_i^{(\theta)}(t)$ removes the linear drift in \eqref{eq:Dyson-OU}. By It\^o's formula, $\widetilde{\mathbf x}$ solves the driftless equation of \cref{lem:general-data-quantile-comparison}, with initial value $\mathbf x^{(\theta)}(0)$ and the Brownian motions $\widetilde B_i(u):=\int_0^{\log(1+u)}e^{v/2}\,dB_i(v)$. These are independent standard Brownian motions in $u$, because they are continuous Gaussian martingales with quadratic variation $\int_0^{\log(1+u)}e^v\,dv=u$. The initial particles lie in $[-3,3]$ and $u\le e-1<n^3$, so the initial-data and time conditions of \cref{lem:general-data-quantile-comparison} hold uniformly in $\theta$ and in the initial pair under consideration.

The free-convolution measure for the transformed flow is $\widetilde\nu_u^{(\theta)}:=\mu_0^{(\theta)}\boxplus\SC_u$, with $\widetilde\nu_0^{(\theta)}=\mu_0^{(\theta)}$. Its image under $E\mapsto e^{-t/2}E$ is $\nu_t^{(\theta)}$, because $e^{-t}u=1-e^{-t}$. Let $\zeta_i(u)$ be its midpoint quantiles, with the infinite-value convention of \cref{lem:general-data-quantile-comparison}, and put $k_n=\lfloor(\log n)^5\rfloor$. That lemma gives, on an event of conditional probability at least $1-c\exp[-(\log n)^2]$,
\begin{equation*}
\zeta_{i-k_n}(u)-n^{-3}
\le\widetilde x_i(u)
\le\zeta_{i+k_n}(u)+n^{-3}
\end{equation*}
for all $i$ and $0\le u\le e-1$.
To pass from these bounds to distribution functions, note that
\begin{equation*}
\left|\frac1n\#\{i:\zeta_i(u)\le E\}
-F_{\widetilde\nu_u^{(\theta)}}(E)\right|
\le\frac1n.
\end{equation*}
The upper bound on $\widetilde x_i(u)$ shows that each quantile at or below $E-n^{-3}$ gives a particle at or below $E$, apart from at most $k_n$ indices. Conversely, the lower bound shows that the number of particles at or below $E$ exceeds the number of quantiles at or below $E+n^{-3}$ by at most $k_n$. Consequently,
\begin{equation*}
F_{\widetilde\nu_u^{(\theta)}}(E-n^{-3})-\frac{k_n+1}{n}
\le\frac1n\#\{i:\widetilde x_i(u)\le E\}
\le F_{\widetilde\nu_u^{(\theta)}}(E+n^{-3})+\frac{k_n+1}{n}.
\end{equation*}
This also covers the indices for which a shifted quantile is infinite, since omitting those indices changes the cumulative mass by at most $k_n/n$.
Rescaling by $e^{-t/2}\le1$ and using the definition of the L\'evy distance now gives
\begin{equation*}
\sup_{0\le t\le1}d_{\mathrm L}(\mu_t^{(\theta)},\nu_t^{(\theta)})
\le\frac{\lfloor(\log n)^5\rfloor+2}{n}+n^{-3}.
\end{equation*}
Let $\mathcal A_n^{(\theta)}$ be the event on which the preceding L\'evy distance bound holds.
By \cref{lem:general-data-quantile-comparison}, for every initial pair in $\mathcal H_n$,
\begin{equation*}
\Pp\bigl((\mathcal A_n^{(\theta)})^c
\mid\mathbf{x}(0),\mathbf{y}(0)\bigr)
\le c\exp(- (\log n)^2).
\end{equation*}
Integrating over the initial pair proves the first assertion of \eqref{eq:fixed-parameter-rigidity}. Only the initial pair was conditioned on.

Since $\rhosc\le1/\pi$, the definition of the L\'evy distance implies
\begin{equation*}
\sup_E|F_\mu(E)-F_{\mathrm{sc}}(E)|
\le(1+1/\pi)d_{\mathrm L}(\mu,\SC_1).
\end{equation*}
Combining this inequality with the preceding two comparisons, and using $(\log n)^5=o(n^{3\alpha/2})$, gives on $\mathcal H_n\cap\mathcal A_n^{(\theta)}$
\begin{equation}
\sup_{0\le t\le1}\sup_E
\left|\frac1n\#\{i:x_i^{(\theta)}(t)\le E\}-F_{\mathrm{sc}}(E)\right|
\le c n^{-1+3\alpha/2}.
\label{eq:interpolating-CDF-bound}
\end{equation}

Choose $q_n=\lceil c_0 n^{3\alpha/2}\rceil$, with $c_0$ large enough that $q_n$ exceeds $n$ times the right-hand side of \eqref{eq:interpolating-CDF-bound} by at least two. For indices satisfying $\hat i\ge2q_n+2$, inversion of the CDF (at $E=\gamma_{i+q_n}$ and $E=\gamma_{i-q_n}$, \eqref{eq:interpolating-CDF-bound} keeps the empirical count strictly above $i$ and strictly below $i$, respectively) gives
\begin{equation*}
\gamma_{i-q_n}\le x_i^{(\theta)}(t)\le\gamma_{i+q_n},
\qquad 0\le t\le1.
\end{equation*}
For every index $j$ between $i-q_n$ and $i+q_n$, the edge index $\hat j$ is comparable to $\hat i$. Summing the semicircle quantile spacings therefore yields
\begin{equation*}
|x_i^{(\theta)}(t)-\gamma_i|
\le c q_n n^{-2/3}\hat i^{-1/3}\le c n^{3\alpha/2}n^{-2/3}\hat i^{-1/3}.
\end{equation*}

For the remaining indices, work on $\mathcal H_n\cap\mathcal Y_n$ and use \eqref{eq:interpolation-Lipschitz} with $\theta'=0$. The initial rigidity bounds and the rigidity of the GOE flow give
\begin{align*}
|x_i^{(\theta)}(t)-\gamma_i|
&\le\|\mathbf{x}^{(\theta)}(t)-\mathbf{y}(t)\|_\infty
+|y_i(t)-\gamma_i|\\
&\le\|\mathbf{x}^{(\theta)}(0)-\mathbf{y}(0)\|_\infty
+n^\alpha n^{-2/3}\hat i^{-1/3}
\le3n^{-2/3+\alpha}.
\end{align*}
Here $\hat i\le2q_n+1$, so
\begin{equation*}
3n^{-2/3+\alpha}
=3n^\alpha\hat i^{1/3}n^{-2/3}\hat i^{-1/3}
\le c n^\alpha q_n^{1/3}n^{-2/3}\hat i^{-1/3}
\le c n^{3\alpha/2}n^{-2/3}\hat i^{-1/3}.
\end{equation*}
Both index ranges are therefore controlled on $\mathcal H_n\cap\mathcal Y_n\cap\mathcal A_n^{(\theta)}$, simultaneously for $t\in[0,1]$. Since $3\alpha/2=3\varepsilon/8<\varepsilon/2$, this proves the inclusion in \eqref{eq:fixed-parameter-rigidity} for large $n$.
\end{proof}

We now upgrade the fixed-$\theta$ bound of \cref{lem:fixed-parameter-rigidity} to all $\theta\in[0,1]$ simultaneously by taking a union bound over a fine mesh of interpolation parameters and using the Lipschitz bound \eqref{eq:interpolation-Lipschitz}.

\begin{proof}[Proof of \cref{prop:interpolating-dynamic-rigidity}]

Put $\alpha:=\varepsilon/4$ and $\theta_k:=kn^{-2}$ for $0\le k\le n^2$. By \cref{lem:GOE-path-rigidity}, $\Pp(\mathcal Y_n^c)\le n^{-D-1}$. By \cref{lem:fixed-parameter-rigidity} and a union bound over the grid, the event
\begin{equation*}
\mathcal E:=\mathcal H_n(\alpha)\cap\mathcal Y_n\cap\bigcap_{k=0}^{n^2}\mathcal A_n^{(\theta_k)}
\end{equation*}
satisfies $\Pp(\mathcal H_n(\alpha)\setminus\mathcal E)\le n^{-D-1}+(n^2+1)c\exp(-(\log n)^2)\le n^{-D}$, and on $\mathcal E$ we have $\mathbf x^{(\theta_k)}(t)\in\mathcal R_{n,\varepsilon/2}$ for all $k$ and $t$. For $\theta\in[\theta_k,\theta_{k+1}]$, \eqref{eq:interpolation-Lipschitz} and the initial rigidity give
\begin{equation*}
\sup_{0\le t\le1}\|\mathbf x^{(\theta)}(t)-\mathbf x^{(\theta_k)}(t)\|_\infty\le n^{-2}\|\mathbf x(0)-\mathbf y(0)\|_\infty\le2n^{-8/3+\alpha}.
\end{equation*}
Since $\ell_i\ge n^{-1}$, this gives $\mathbf x^{(\theta)}(t)\in\mathcal R_{n,\varepsilon}$ for all $\theta,t\in[0,1]$ on $\mathcal E$, for large $n$. Thus $\mathcal E\subset\mathcal G_n(\varepsilon)$, which proves \eqref{eq:interpolating-uniform-rigidity}.

\end{proof}

Rigidity implies an averaged local semicircle law by a deterministic argument; see \cite[proof of Theorem~2.2]{LandonXian2025}, which refers to \cite[Theorem~10.3]{BenaychGeorgesKnowles2018}. We record it with the exponents needed below.

\begin{lemma}%
\label{lem:rigidity-averaged-local-law}
Fix $0<\varepsilon<1/3$. For all sufficiently large $n$, every $\xi\in\mathcal R_{n,\varepsilon/4}$ satisfies
\begin{equation}
\left|\frac1n\sum_{i=1}^n\frac1{\xi_i-z}-m_{\mathrm{sc}}(z)\right|
\le\frac{n^\varepsilon}
{n\bigl(\eta+\mathbf1_{\{|E|>2+n^{-2/3+\varepsilon}\}}\kappa(E)\bigr)},
\qquad z=E+\mathrm{i}\eta\in\C^+,\quad |z|\le10,
\label{eq:rigidity-averaged-local-law}
\end{equation}
where $\kappa(E):=\dist(E,\{-2,2\})$.
\end{lemma}

\begin{proof}
Put $r_n:=n^{\varepsilon/4}$. Write
$F_\xi(E)=n^{-1}\#\{i:\xi_i\le E\}$.
The semicircle density gives, for $u\in[-2,2]$ and $h>0$,
\begin{equation*}
|F_{\mathrm{sc}}(u\pm h)-F_{\mathrm{sc}}(u)|
\le c\bigl(h\sqrt{\dist(u,\{-2,2\})}+h^{3/2}\bigr).
\end{equation*}
Since $\sqrt{\dist(\gamma_i,\{-2,2\})}\le c(\hat i/n)^{1/3}$, taking $h=r_n\ell_i$ shows that
$|F_{\mathrm{sc}}(\xi_i)-i/n|\le c r_n^{3/2}/n$.
Monotonicity between consecutive particles, including outside the extreme particles, yields
\begin{equation*}
\sup_E|F_\xi(E)-F_{\mathrm{sc}}(E)|\le c r_n^{3/2}/n.
\end{equation*}
Integration by parts against $(u-z)^{-1}$ therefore bounds the left-hand side of \eqref{eq:rigidity-averaged-local-law} by
\begin{equation*}
\frac{c r_n^{3/2}}n\int_\R\frac{du}{(u-E)^2+\eta^2}
\le\frac{c r_n^{3/2}}{n\eta}.
\end{equation*}

For the exterior estimate, suppose $|E|>2+2r_n n^{-2/3}$ and set $\gamma_0=-2$ and $I_i=[\gamma_{i-1},\gamma_i]$.
Each cell has semicircle mass $1/n$ and length comparable to $\ell_i$.
For $u\in I_i$, rigidity gives
$|\xi_i-u|\le c r_n\ell_i$.
The distance from $E$ to $[-2,2]$ is at least twice the largest allowed particle displacement, so the resolvent denominators at $\xi_i$, $\gamma_i$, and $u$ are comparable. Consequently,
\begin{align*}
\left|\frac1n\sum_i\frac1{\xi_i-z}-m_{\mathrm{sc}}(z)\right|
&\le \sum_i\int_{I_i}
\frac{|\xi_i-u|\rhosc(u)\,du}{|\xi_i-z|\,|u-z|}
\le\frac{c r_n}n\sum_i\frac{\ell_i}{|\gamma_i-z|^2}\\
&\le\frac{c r_n}n\int_{-2}^2\frac{du}{(u-E)^2+\eta^2}
\le\frac{c r_n}{n(\kappa(E)+\eta)}.
\end{align*}
For large $n$, $n^{-2/3+\varepsilon}>2r_n n^{-2/3}$, and both $c r_n$ and $c r_n^{3/2}$ are at most $n^\varepsilon$. Combining the two bounds proves the lemma.
\end{proof}

\subsection{Proof of homogenization}
\label{subsec:homogenization-proof}

We prove \cref{prop:regular-data-homogenization} by transferring one induction step of Landon and Xian \cite{LandonXian2025}, and of Bourgade \cite{Bourgade2022ExtremeGaps} in the bulk, to our initial data. We recall the statements we use, explain the two changes our setting requires, and then prove the proposition. We cite \cite{LandonXian2025} by the numbering of arXiv:2509.14192v1, and \cite{Bourgade2022ExtremeGaps} by the numbering of arXiv:1812.10376v3, which is the numbering used in \cite{LandonXian2025}.

We begin by expressing the displacement $\mathbf x(t)-\mathbf y(t)$ as the integral of the $\theta$-derivative flow $\mathbf u^{(\theta)}(t)$ and identifying its homogenized profile with the Chebyshev heat kernel $p_t$. Fix an initial pair as in \cref{prop:interpolating-dynamic-rigidity}, and let $\mathbf d:=\mathbf x(0)-\mathbf y(0)$. For every strictly ordered initial condition, \eqref{eq:Dyson-OU} has a unique strong solution that stays strictly ordered for all times; this is \cite[Lemma~4.3.3]{AGZ2010} after the change of variables $u=e^t-1$ used in the proof of \cref{lem:fixed-parameter-rigidity}. Hence, for each fixed $\theta$, the flow $\mathbf x^{(\theta)}$ is almost surely this solution, and by Fubini's theorem, almost surely this holds for Lebesgue-almost every $\theta\in[0,1]$. For such $\theta$, following \cite[(1.7)--(1.10)]{Bourgade2022ExtremeGaps} and \cite[(2.8)--(2.9)]{LandonXian2025}, let $\mathbf u^{(\theta)}$ solve the linear system
\begin{equation}
\partial_tu_i^{(\theta)}(t)=\frac1n\sum_{j\ne i}\frac{u_j^{(\theta)}(t)-u_i^{(\theta)}(t)}{\bigl(x_i^{(\theta)}(t)-x_j^{(\theta)}(t)\bigr)^2},
\qquad \mathbf u^{(\theta)}(0)=\mathbf d,
\label{eq:derivative-flow}
\end{equation}
whose coefficients are continuous on $[0,1]$, and put $\mathbf u^{(\theta)}:=0$ for the remaining null set of $\theta$. By the maximum principle, $\|\mathbf u^{(\theta)}(t)\|_\infty\le\|\mathbf d\|_\infty$. For $\theta'$ in the countable dense set used to construct the flows, \eqref{eq:synchronous-difference} shows that $e^{t/2}(\mathbf x^{(\theta')}(t)-\mathbf x^{(\theta)}(t))/(\theta'-\theta)$ solves the same system with coefficients $[n(x_i^{(\theta')}-x_j^{(\theta')})(x_i^{(\theta)}-x_j^{(\theta)})]^{-1}$ and initial value $\mathbf d$. As $\theta'\to\theta$, these coefficients converge uniformly on $[0,1]$ by \eqref{eq:interpolation-Lipschitz}, so Gronwall's inequality shows that the difference quotients converge to $e^{-t/2}\mathbf u^{(\theta)}(t)$. Since $\theta\mapsto\mathbf x^{(\theta)}(t)$ is Lipschitz, its derivative exists almost everywhere and equals $e^{-t/2}\mathbf u^{(\theta)}(t)$ there. As $\mathbf x^{(1)}=\mathbf x$ and $\mathbf x^{(0)}=\mathbf y$, this gives
\begin{equation}
x_i(t)-y_i(t)=e^{-t/2}\int_0^1u_i^{(\theta)}(t)\,d\theta.
\label{eq:theta-integration}
\end{equation}
Let $\widetilde\gamma_i$ be the midpoint quantiles, $\int_{-2}^{\widetilde\gamma_i}\rhosc(x)\,dx=(i-1/2)/n$, and put $\varrho_i:=(4-\widetilde\gamma_i^2)^{1/2}\asymp(\hat i/n)^{1/3}$. Then $|\widetilde\gamma_i-\gamma_i|\le c\ell_i$ and $\varrho_i\ge cn^{-1/3}$. With $\widetilde\gamma_i^t:=\widetilde\gamma_i\cosh(t/2)+\mathrm i\varrho_i\sinh(t/2)$, the homogenized profile is
\begin{equation}
\overline u_i(t):=\frac1{n\Im m_{\mathrm{sc}}(\widetilde\gamma_i^t)}\Im\sum_{j=1}^n\frac{d_j}{\widetilde\gamma_j-\widetilde\gamma_i^t}.
\label{eq:homogenized-characteristic}
\end{equation}
After reversing the indices, these definitions are \cite[(1.11)--(1.12) and (1.15)--(1.16)]{LandonXian2025}, where particles are labeled in decreasing order. The profile does not depend on $\theta$, and it is the heat kernel \eqref{eq:Dyson-heat-kernel} at the midpoint quantiles:
\begin{equation}
\overline u_i(t)=\frac1n\sum_{j=1}^np_t(\widetilde\gamma_i,\widetilde\gamma_j)\,d_j.
\label{eq:characteristic-kernel-identification}
\end{equation}
To see this, write $x=2\cos\alpha$ with $\alpha\in(0,\pi)$, and put $w:=e^{t/2+\mathrm i\alpha}$. Then
\begin{equation*}
x\cosh(t/2)+\mathrm i\sqrt{4-x^2}\sinh(t/2)=w+w^{-1},
\qquad
m_{\mathrm{sc}}(w+w^{-1})=-w^{-1},
\end{equation*}
the second identity because $|w|>1$. For $y\in[-2,2]$, the generating function of the Chebyshev polynomials $U_m$ \cite[(18.12.10)]{DLMF}, at the point $y/2$ and with variable $w^{-1}$, gives
\begin{equation*}
\frac1{y-w-w^{-1}}=-\frac{w^{-1}}{1-yw^{-1}+w^{-2}}=-\sum_{m\ge0}\mathsf U_m(y)\,w^{-m-1}.
\end{equation*}
Since $\Im(-w^{-m-1})=e^{-(m+1)t/2}\sin((m+1)\alpha)$, $\Im m_{\mathrm{sc}}(w+w^{-1})=e^{-t/2}\sin\alpha$, and $\mathsf U_m(x)=\sin((m+1)\alpha)/\sin\alpha$, we get
\begin{equation*}
\frac{\Im(y-w-w^{-1})^{-1}}{\Im m_{\mathrm{sc}}(w+w^{-1})}=\sum_{m\ge0}e^{-mt/2}\mathsf U_m(y)\mathsf U_m(x)=p_t(x,y).
\end{equation*}
With $x=\widetilde\gamma_i$ and $y=\widetilde\gamma_j$, this proves \eqref{eq:characteristic-kernel-identification}.

Next, we recall the scale quantities and the homogenization property $\mathcal P_a$ from \cite{LandonXian2025} and summarize the three components of a single induction step. Following \cite[(4.1)]{LandonXian2025}, put $\mathsf B(t,i):=nt\varrho_i(t+\varrho_i)$ and $\mathsf Q_a(t,i):=n(t+\varrho_i)\,\mathsf B(t,i)^{1-a}$ for $a\in[0,1]$. For a control parameter $\phi>1$, $\theta\in[0,1]$, $a\in(0,1]$, and $h>0$, let $\Lambda_a^{(\theta)}(h)$ be the event that
\begin{equation}
|u_i^{(\theta)}(t)-\overline u_i(t)|\le\frac{\phi^h}{\mathsf Q_a(t,i)}
\quad\text{for all }t\in(0,1]\text{ and all }i\text{ with }\mathsf B(t,i)\ge\phi^h.
\label{eq:LX-property}
\end{equation}
Property $\mathcal P_a$ of \cite[Definition~4.1]{LandonXian2025} states that, for every $D>0$, there are $h$ and $n_0$ with $\Pp(\Lambda_a^{(\theta)}(h))\ge1-n^{-D}$ for $n\ge n_0$; there $\theta$ is fixed and suppressed. \cite[Lemma~4.3]{LandonXian2025} proves $\mathcal P_1$, and \cite[Theorem~4.2]{LandonXian2025} proves that $\mathcal P_a$ implies $\mathcal P_{ca}$ for some $c\in(0,1)$. We use one step, from $a=1$, which consists of three parts.
\begin{enumerate}[label=(\roman*),leftmargin=*]
\item \emph{Initial step.} \cite[Lemma~4.3]{LandonXian2025} deduces $\mathcal P_1$ from $|u_i^{(\theta)}(t)|\le\phi^h/(n(t+\varrho_i))$, which is \cite[(4.3)]{LandonXian2025} and \cite[(2.18)]{Bourgade2022ExtremeGaps}, and from $|\overline u_i(t)|\le c\phi^{1/100}/(n(t+\varrho_i))$, which is \cite[(3.10)]{LandonXian2025}. The second bound is deterministic given rigidity of the initial data. The first is stochastic: \cite[Corollary~2.7 and proof of Theorem~2.8]{Bourgade2022ExtremeGaps} derive it from \cite[Proposition~2.6]{Bourgade2022ExtremeGaps}, applied to the flow started from $|\mathbf d|$, and that proposition is proved for processes stopped at the first time rigidity fails \cite[(2.8)]{Bourgade2022ExtremeGaps}.
\item \emph{Indices near the edges.} \cite[Section~4.1]{LandonXian2025} proves \cite[Theorem~4.2]{LandonXian2025} for $i<n/200$ and, by symmetry, for $i>n-n/200$. For these indices, one step turns $\mathcal P_a$ into $\mathcal P_{(1-10^{-5})a}$ \cite[(4.63)]{LandonXian2025}.
\item \emph{Bulk indices.} For $\alpha n\le i\le(1-\alpha)n$ with fixed $\alpha>0$, \cite[Section~4.1]{LandonXian2025} refers to \cite{Bourgade2022ExtremeGaps}. There $\varrho_i\asymp1$ and $\phi^h/\mathsf Q_a(t,i)\asymp\phi^h(nt)^a/(n^2t)$, the error in property $(\mathrm P_a)$ of \cite[(3.6)]{Bourgade2022ExtremeGaps}, and \cite[Proposition~3.6]{Bourgade2022ExtremeGaps} proves that $(\mathrm P_a)$ on $[\alpha n,(1-\alpha)n]$ implies $(\mathrm P_{3a/4})$ on $[3\alpha n,(1-3\alpha)n]$. We take $\alpha=1/800$, so that (ii) and (iii) cover all indices. Since $3/4<1-10^{-5}$ and $\mathsf B(t,i)\ge1$ in \eqref{eq:LX-property}, the bulk bound implies the bound with exponent $1-10^{-5}$. Bourgade's profile \cite[(3.1)]{Bourgade2022ExtremeGaps} uses his classical locations \cite[(1.4)]{Bourgade2022ExtremeGaps}, which are our $\gamma_i$, in place of $\widetilde\gamma_i$. In the bulk, $|\gamma_j-\widetilde\gamma_j|\le c\ell_j$, $|\gamma_i^t-\widetilde\gamma_i^t|\le c/n$ with $\gamma_i^t:=\gamma_i\cosh(t/2)+\mathrm i(4-\gamma_i^2)^{1/2}\sinh(t/2)$, and $\Im\widetilde\gamma_i^t\asymp t$. Together with $|d_j|\le2\phi\ell_j$, a computation as in \cite[(3.2)--(3.5)]{Bourgade2022ExtremeGaps} shows that the two profiles differ by $O(\phi^c/(n^2t))$. Since $nt\ge1$, this is below the error $\phi^h(nt)^a/(n^2t)$ for every $a>0$, so the conversion only increases $h$. The input $\mathcal P_1$ does not depend on the convention, since both profiles are $O(\phi^c/n)$ by \cite[(3.10)]{LandonXian2025} and \cite[(3.5)]{Bourgade2022ExtremeGaps}.
\end{enumerate}
We do not use \cite[Theorem~1.2]{LandonXian2025} or \cite[Theorem~3.1]{Bourgade2022ExtremeGaps}, which iterate the step until $a$ is small. \cite{LandonXian2025} assume generalized Wigner initial data \cite[Theorem~1.2]{LandonXian2025}, $\phi=\exp((\log\log N)^3)$ \cite[(2.1)]{LandonXian2025}, and overwhelming probability \cite[Definition~2.1]{LandonXian2025}; \cite{Bourgade2022ExtremeGaps} assumes generalized Wigner and GOE data \cite[Theorem~3.1]{Bourgade2022ExtremeGaps} and $\phi=\exp(C_0(\log\log N)^2)$ \cite[(1.5)]{Bourgade2022ExtremeGaps}. In (i)--(iii), the initial data enter only through $\mathbf d$, $\overline{\mathbf u}$, and rigidity along all interpolating flows, \cite[Theorem~2.2]{LandonXian2025} and \cite[(2.7)]{Bourgade2022ExtremeGaps}; compare \cite[Remarks~2.9 and~3.2]{Bourgade2022ExtremeGaps}. Our setting differs in two ways.

The first change is the control parameter. \cite{LandonXian2025} and \cite{Bourgade2022ExtremeGaps} use rigidity at the scales $\phi^{1/100}\ell_i$ and $\phi^{1/2}\ell_i$ with a subpolynomial $\phi$. Our rigidity comes from low-degree moment bounds in \cref{section:rigidity} and holds only at the scale $n^\varepsilon\ell_i$, so we take $\phi:=n^\delta$ with a small fixed $\delta>0$. This change is forced by the rigidity scale, not by the failure probability. Across the statements used in one step from $a=1$, the assumptions on $\phi$ fall into four types: (a) lower bounds $\phi\ge c$ and $\phi\ge(\log n)^c$, as in \cite[(5.19), (5.26)]{LandonXian2025}; (b) scale relations between $\phi$, $\mathsf B(t,i)$, and the auxiliary parameters, which hold once $\mathsf B(t,i)\ge\phi^{A_1}$ \cite[(4.29)]{LandonXian2025} for a fixed $A_1\ge10^{20}$ \cite[Lemma~4.9]{LandonXian2025}; (c) finitely many upper bounds $\phi^b\le n^r$ with fixed $b,r>0$, such as $\ell\le\phi^{-100}n$ for the interaction range $\ell$ \cite[(4.7)]{LandonXian2025}; and (d) error terms such as $e^{-c\phi}$ \cite[Theorem~4.5]{LandonXian2025}, which must decay faster than every power of $n$. With $\phi=n^\delta$, conditions (a) and (d) hold for every $\delta>0$; condition (b) is built into the restriction $\mathsf B(t,i)\ge\phi^h$ in \eqref{eq:LX-property} once $h\ge A_1$, as in \cite{LandonXian2025}, and is checked at time $\tau$ below; and condition (c) holds whenever $\delta$ lies below a small threshold $\delta_0>0$. The exponents of $\phi$ in the conclusions do not depend on $\delta$, coming only from numerical constants such as $A_1$, $\mathfrak c_b=1/3000$ and $\mathfrak c_e=1/10$ before \cite[(4.63)]{LandonXian2025}, and the windows \cite[(3.15)]{Bourgade2022ExtremeGaps}. The probability bounds likewise cost no power of $\phi$, coming from Gaussian martingale tails \cite[Lemma~B.1]{LandonXian2025} and moment bounds with logarithmic factors \cite[(5.19)]{LandonXian2025}.

The second change concerns probabilities. For us, rigidity along all interpolating flows holds on the event $\mathcal G_n(\cdot)$ of \cref{prop:interpolating-dynamic-rigidity}, whose complement is only polynomially small. Each stochastic step in the one-step transfer is proved for processes stopped at the first of three times, namely the first failure of rigidity, the first failure of an earlier bound (\cite[(2.18)]{Bourgade2022ExtremeGaps} or $\mathcal P_1$), and a self-referential threshold for the quantity being estimated. Before this stopping time the integrands are controlled deterministically, so the martingale bound produces an event $\mathcal M$ with $\Pp(\mathcal M^c)\le n^{-D}$ on which the self-referential threshold is not reached first. These arguments use only that the initial pair is $\mathcal F_0$-measurable and $(B_i)$ is an $(\mathcal F_t)$-Brownian motion; they never condition on $\mathcal G_n(\cdot)$, which depends on the whole path. Hence, if $\mathcal E_{\mathrm{old}}$ is the event that the earlier bounds hold for all $t\le1$, the new bound holds on $\mathcal G_n(\cdot)\cap\mathcal E_{\mathrm{old}}\cap\mathcal M$, and
\begin{equation}
\Pp\bigl(\mathcal G_n(\cdot)\setminus\mathcal E_{\mathrm{new}}\bigr)\le\Pp\bigl(\mathcal G_n(\cdot)\setminus\mathcal E_{\mathrm{old}}\bigr)+n^{-D}.
\label{eq:chain-accounting}
\end{equation}
Deterministic steps add no exceptional event, so summing \eqref{eq:chain-accounting} over the finitely many stochastic steps in parts (i)--(iii) yields the following one-step transfer lemma.

\begin{lemma}
\label{lem:one-step-transfer}
For every $D>0$, there are $\delta_0\in(0,10^{-3}]$ and $h_0,h_1>0$, depending only on $D$, such that the following holds for every $\delta\in(0,\delta_0]$. Let $\phi:=n^\delta$ and $\mathcal G_n:=\mathcal G_n(\delta/400)$. Then there is $n_0$, depending only on $D$ and $\delta$, such that for every initial pair as in \cref{prop:interpolating-dynamic-rigidity}, every $\theta\in[0,1]$, and every $n\ge n_0$,
\begin{equation}
\Pp\bigl(\mathcal G_n\setminus\Lambda_1^{(\theta)}(h_0)\bigr)\le n^{-D},
\qquad
\Pp\bigl(\mathcal G_n\setminus\Lambda_{1-10^{-5}}^{(\theta)}(h_1)\bigr)\le n^{-D}.
\label{eq:one-step-transfer}
\end{equation}
\end{lemma}

\begin{proof}
On $\mathcal G_n$, every $\mathbf x^{(\theta)}(t)$ lies in $\mathcal R_{n,\delta/400}$. Since $|\gamma_i-\widetilde\gamma_i|\le c\ell_i$ and $n^{\delta/400}+c\le n^{\delta/100}=\phi^{1/100}$ for large $n$, the rigidity estimate \cite[(2.11)]{LandonXian2025} holds on $\mathcal G_n$ for all $\theta$, $t$, and $i$, after reversing the indices. By \cref{lem:rigidity-averaged-local-law} with $\varepsilon=\delta/100$, the averaged local law \cite[(2.12)]{LandonXian2025} holds on $\mathcal G_n$ with the same factor $\phi^{1/100}$. The rigidity event \cite[(2.7)]{Bourgade2022ExtremeGaps} uses our classical locations $\gamma_i$ and the larger scale $\phi^{1/2}\ell_i$, so it contains $\mathcal G_n$. At $t=0$ we also get $|d_i|\le2\phi^{1/100}\ell_i$.

Let $\delta_0$ be the threshold in condition (c). We now verify \eqref{eq:one-step-transfer} by tracing \eqref{eq:chain-accounting} across parts (i)--(iii). For the initial step (i), \cite[Proposition~2.6 and Corollary~2.7]{Bourgade2022ExtremeGaps} yield \cite[(2.18)]{Bourgade2022ExtremeGaps} (which is \cite[(4.3)]{LandonXian2025}) by stopping at the failure of rigidity \cite[(2.7)]{Bourgade2022ExtremeGaps} and at the self-referential threshold \cite[(2.8)]{Bourgade2022ExtremeGaps}, using only condition (a). Combining this with initial rigidity via the deterministic \cite[Lemmas~3.3 and~4.3]{LandonXian2025} (which require only $\mathsf B(t,i)\ge1$) and applying \eqref{eq:chain-accounting} with no earlier bound establishes the first bound in \eqref{eq:one-step-transfer}.

For the second bound, we start from the first bound (with $D$ replaced by a larger fixed exponent) and track the edge and bulk steps separately. Near the edges (part (ii)), \cite[Propositions~5.1--5.2 and Theorem~4.5]{LandonXian2025} stop at rigidity failure and at the threshold \cite[(5.6)]{LandonXian2025}, using \cite[(5.19)]{LandonXian2025} together with conditions (a), (b), and (d); from $\mathcal P_1$ and \cite[Theorem~4.5]{LandonXian2025}, \cite[Lemmas~5.3 and~4.6]{LandonXian2025} (under condition (b) via \cite[Lemma~4.6 and (5.43)]{LandonXian2025}) and \cite[Proposition~4.8]{LandonXian2025} (via the maximum principle and condition (b) in \cite[(4.23)--(4.25)]{LandonXian2025}) follow deterministically. We then apply \cite[Theorem~6.2, Lemmas~6.3--6.7, and Lemma~4.11]{LandonXian2025}, which stop at the failure of rigidity or \cite[(4.3)]{LandonXian2025} (see \cite[(6.31)]{LandonXian2025}) and of $\mathcal P_1$ (see \cite[(6.58)]{LandonXian2025}) using \cite[Lemma~B.1]{LandonXian2025} and condition (b) in \cite[(4.43), (6.21), (6.75)]{LandonXian2025}; the deterministic assembly in \cite[Section~4.1.2, (4.47)--(4.63)]{LandonXian2025} combines these estimates under conditions (b) (\cite[(4.29), (4.48), (4.57)--(4.58)]{LandonXian2025}) and (c) (\cite[(4.7)]{LandonXian2025}). In the bulk (part (iii)), \cite[Proposition~2.11]{Bourgade2022ExtremeGaps} stops at the failure of rigidity or \cite[(2.18)]{Bourgade2022ExtremeGaps} and at the threshold \cite[(2.20)]{Bourgade2022ExtremeGaps} under condition (a). Finite speed of propagation \cite[Lemma~3.8]{Bourgade2022ExtremeGaps}, valid for $\ell\ge n|u-v|$, holds outside an exceptional event; its proof stops at the first failure of rigidity, which does not occur on $\mathcal G_n$, so it is one more stochastic step in \eqref{eq:chain-accounting}. On its event \cite[(3.9)]{Bourgade2022ExtremeGaps}, \cite[Lemmas~3.4 and~3.7]{Bourgade2022ExtremeGaps} (the latter for $\ell>\phi$) combine rigidity, \cite[(2.18)]{Bourgade2022ExtremeGaps}, and $(\mathrm P_1)$ deterministically, and \cite[Lemma~3.9 and Proposition~3.6]{Bourgade2022ExtremeGaps} complete the bulk step via the maximum principle under condition (b) (\cite[(3.10), (3.15)]{Bourgade2022ExtremeGaps}), with \cite[Proposition~2.11]{Bourgade2022ExtremeGaps} replacing \cite[Lemma~3.10]{Bourgade2022ExtremeGaps} when $a=1$ (see the remark after \cite[(3.16)]{Bourgade2022ExtremeGaps}). Summing \eqref{eq:chain-accounting} over these finitely many stochastic steps yields the second bound in \eqref{eq:one-step-transfer}.
\end{proof}

With the one-step transfer established for each fixed $\theta$, we now integrate over $\theta\in[0,1]$ to complete the proof of \cref{prop:regular-data-homogenization}.

\begin{proof}[Proof of \cref{prop:regular-data-homogenization}]
We choose the parameters in the following order. Put $D:=M+5$ and $\kappa_1:=10^{-5}$, and let $\delta_0$, $h_0$, $h_1$ be as in \cref{lem:one-step-transfer} for this $D$. Then set
\begin{equation*}
\delta:=\min\Bigl\{\delta_0,\ \varepsilon,\ \frac{\kappa_1}{6h_1}\Bigr\},
\qquad\phi:=n^\delta,
\end{equation*}
and let $C$ be the degree threshold of \cref{prop:burnin-rigidity} for the burn-in time $t_1$, the rigidity exponent $\delta/1600$, and the failure exponent $M+1$. Thus $C$ depends only on $t_1$, $\varepsilon$, and $M$.

Let $\mathcal H_n:=\mathcal H_n(\delta/1600)$ and $\mathcal G_n:=\mathcal G_n(\delta/400)$, and let $A_n$ be the random set of $\theta\in[0,1]$ for which $\Lambda_{1-\kappa_1}^{(\theta)}(h_1)$ fails. The map $(\theta,\omega)\mapsto\mathbf1\{\theta\in A_n\}$ is jointly measurable, since $(\theta,t)\mapsto\mathbf x^{(\theta)}(t)$ is continuous and $\mathbf u^{(\theta)}(t)$ is a limit of its difference quotients for almost every $\theta$. We now combine the exceptional events. By \cref{prop:burnin-rigidity} (using (LD$_C$) for $\mathbf x(0)\sim\sigma_n$) and GOE rigidity \cite{ERDOS20121435}, we have $\Pp(\mathcal H_n^c)\le n^{-M-1}+n^{-D}$; by \cref{prop:interpolating-dynamic-rigidity} with $\varepsilon=\delta/400$, we have $\Pp(\mathcal H_n\setminus\mathcal G_n)\le n^{-D}$; and by \cref{lem:one-step-transfer}, Fubini's theorem, and Markov's inequality, we obtain $\E[\mathbf1_{\mathcal G_n}|A_n|]\le n^{-D}$ and hence $\Pp(\mathcal G_n\cap\{|A_n|>n^{-3}\})\le n^{3-D}$. Setting $\Omega_n:=\mathcal G_n\cap\{|A_n|\le n^{-3}\}$, a union bound yields $\Pp(\Omega_n^c)\le n^{-M-1}+2n^{-D}+n^{3-D}\le n^{-M}$ for all sufficiently large $n$.

We now work on $\Omega_n$. Taking $\theta\in\{0,1\}$ and $t=0$ in $\mathcal G_n$ gives $\mathbf x(0),\mathbf y(0)\in\mathcal R_{n,\delta/400}\subset\mathcal R_{n,\varepsilon}$, so $|d_j|\le2n^{\delta/400}\ell_j$. At the time $\tau$, $\tau+\varrho_i\ge\tau$ and $\mathsf B(\tau,i)\ge n\tau^2\varrho_i\ge c_\tau n^{2/3}$. Since $\delta h_1\le\kappa_1/6<2/3$, this gives $\mathsf B(\tau,i)\ge\phi^{h_1}$ for all $i$ and large $n$. Hence, for $\theta\notin A_n$,
\begin{equation*}
|u_i^{(\theta)}(\tau)-\overline u_i(\tau)|
\le\frac{\phi^{h_1}}{n(\tau+\varrho_i)\mathsf B(\tau,i)^{\kappa_1}}
\le c_\tau n^{\delta h_1-1-2\kappa_1/3}
\le c_\tau n^{-1-\kappa_1/2}.
\end{equation*}
For $\theta\in A_n$, we use $|u_i^{(\theta)}(\tau)|\le\|\mathbf d\|_\infty$ and $|\overline u_i(\tau)|\le\|p_\tau\|_\infty\|\mathbf d\|_\infty$, from \eqref{eq:characteristic-kernel-identification}; both are at most $c_\tau n^{-1/2}$ (since $\|\mathbf d\|_\infty\le 2n^{-2/3+\delta/400}\le n^{-1/2}$). By \eqref{eq:theta-integration},
\begin{equation*}
\bigl|x_i(\tau)-y_i(\tau)-e^{-\tau/2}\overline u_i(\tau)\bigr|
\le\int_0^1|u_i^{(\theta)}(\tau)-\overline u_i(\tau)|\,d\theta
\le c_\tau n^{-1-\kappa_1/2}+c_\tau n^{-7/2}.
\end{equation*}
Finally, we replace the midpoint quantiles in \eqref{eq:characteristic-kernel-identification} by $\gamma_i$. Bounded first derivatives of $p_\tau$, $|\widetilde\gamma_i-\gamma_i|\le c\ell_i$, $\sum_j\ell_j\le c$, and $\sum_j\ell_j^2\le c/n$ give
\begin{equation*}
\frac{1}n\sum_j\bigl|p_\tau(\widetilde\gamma_i,\widetilde\gamma_j)-p_\tau(\gamma_i,\gamma_j)\bigr||d_j|
\le c_\tau n^{-1+\delta/400}\Bigl(\ell_i\sum_j\ell_j+\sum_j\ell_j^2\Bigr)
\le c_\tau n^{-5/3+\delta/400}.
\end{equation*}
Since $\chi=10^{-6}<\kappa_1/2$ and $\delta\le10^{-3}$, the last two displays prove \eqref{eq:regular-data-homogenization} on $\Omega_n$ for large $n$.
\end{proof}

\section{From transport to total variation}
\label{sec:spectral-smoothing}

We complete the proof of \cref{thm:main} by converting the transportation estimate into total-variation convergence after an independent Gaussian increment. The Gaussian relative-entropy formula \cite[Lemma~A.5]{BoursierChafaiLabbe2023} and Pinsker's inequality give the required comparison at the matrix level. We pass to spectral laws by data processing and average over a coupling of the input spectra. This yields a bound proportional to $\sqrt n\,W_2$, which is sufficient by \cref{lem:Homogenization}.

For probability measures $\mu,\nu$, let
$\operatorname{Ent}(\mu\mid\nu):=\int\log(d\mu/d\nu)\,d\mu$ if $\mu\ll\nu$, and set it equal to $+\infty$ otherwise. We use
$d_{\mathrm{TV}}(\mu,\nu)=\sup_A|\mu(A)-\nu(A)|$.

\begin{lemma}%
\label{lem:spectral-Gaussian-smoothing}
For every $s>0$ and probability laws $\mu,\nu$ on $\mathcal W_n$ with finite second moments,
\begin{equation}
\operatorname{Ent}\left(\mu K_s^{\mathrm{spec}}\mid
\nu K_s^{\mathrm{spec}}\right)
\le\frac{ne^{-s}}{4(1-e^{-s})}W_2^2(\mu,\nu),
\label{eq:spectral-entropy-smoothing}
\end{equation}
and consequently
\begin{equation}
d_{\mathrm{TV}}\left(\mu K_s^{\mathrm{spec}},
\nu K_s^{\mathrm{spec}}\right)
\le\left(\frac{ne^{-s}}{8(1-e^{-s})}\right)^{1/2}W_2(\mu,\nu).
\label{eq:spectral-TV-smoothing}
\end{equation}
\end{lemma}

\begin{proof}
For $\mathbf{x},\mathbf{y}\in\mathcal W_n$, set $A=\diag(\mathbf{x})$ and $B=\diag(\mathbf{y})$. In coordinates orthonormal for the Frobenius inner product, a normalized GOE matrix has covariance $(2/n)I$. 
The laws $\mathcal U_s\delta_A$ and $\mathcal U_s\delta_B$, obtained by applying the OU noise for time $s$ to $A$ and $B$, are therefore Gaussian with respective means $e^{-s/2}A$ and $e^{-s/2}B$ and common covariance $2(1-e^{-s})I/n$. 
The equal-covariance Gaussian entropy formula gives
\begin{equation}
\operatorname{Ent}(\mathcal U_s\delta_A\mid \mathcal U_s\delta_B)
=\frac{ne^{-s}}{4(1-e^{-s})}\|A-B\|_F^2
=\frac{ne^{-s}}{4(1-e^{-s})}\|\mathbf{x}-\mathbf{y}\|_2^2.
\label{eq:matrix-Gaussian-entropy}
\end{equation}
Applying data processing to the ordered eigenvalue map and using \eqref{eq:spectral-kernel}, we obtain the upper bound in \eqref{eq:matrix-Gaussian-entropy} for
$\operatorname{Ent}(K_s^{\mathrm{spec}}(\mathbf{x},\cdot)\mid
K_s^{\mathrm{spec}}(\mathbf{y},\cdot))$.

For any coupling $\Gamma\in\Pi(\mu,\nu)$, the two output laws are mixtures of these kernels with the same mixing law $\Gamma$. Joint convexity of relative entropy yields
\begin{equation*}
\operatorname{Ent}(\mu K_s^{\mathrm{spec}}\mid\nu K_s^{\mathrm{spec}})
\le
\int\operatorname{Ent}\left(K_s^{\mathrm{spec}}(\mathbf{x},\cdot)
\mid K_s^{\mathrm{spec}}(\mathbf{y},\cdot)\right)\Gamma(d\mathbf{x},d\mathbf{y}).
\end{equation*}
Bounding the spectral-kernel entropy by \eqref{eq:matrix-Gaussian-entropy} and taking the infimum over $\Gamma\in\Pi(\mu,\nu)$ gives \eqref{eq:spectral-entropy-smoothing}. Pinsker's inequality,
$d_{\mathrm{TV}}^2\le\operatorname{Ent}/2$, gives \eqref{eq:spectral-TV-smoothing}.
\end{proof}

Combining the three increments of the OU noise, we now complete the proof of \cref{thm:main}.

\begin{proof}[Proof of \cref{thm:main}]
Fix $T>0$ and choose $t_1=t_2=t_3=T/3$. Let the fixed degree parameter $C>0$ be at least $C_2(t_1,t_2)$, with $C_2$ chosen in the proof of \cref{lem:Homogenization}. The assumption on the low-degree advantage then yields
\begin{equation*}
nW_2^2\left(\spec(\mathcal U_{t_1+t_2}P_n),\pi_n\right)=o(1)
\end{equation*}
by \cref{lem:Homogenization}. The finite-moment requirement in \cref{lem:spectral-Gaussian-smoothing} follows from \eqref{eq:spectral-fourth-moment}. Apply \eqref{eq:spectral-TV-smoothing} for the independent final increment of duration $t_3$. By \eqref{eq:spectral-semigroup},
\begin{equation*}
d_{\mathrm{TV}}\left(\spec(\mathcal U_TP_n),\spec(Q_n)\right)
\le
\left(\frac{ne^{-t_3}}{8(1-e^{-t_3})}\right)^{1/2}
W_2\left(\spec(\mathcal U_{t_1+t_2}P_n),\pi_n\right)
=o(1).
\end{equation*}
This proves \cref{thm:main}. The bound \eqref{eq:spectral-entropy-smoothing} is symmetric in $\mu$ and $\nu$, and $\pi_nK_{t_3}^{\mathrm{spec}}=\pi_n$. Applying it in both orders shows that $\operatorname{Ent}(\spec(\mathcal U_TP_n)\mid\spec(Q_n))$ and $\operatorname{Ent}(\spec(Q_n)\mid\spec(\mathcal U_TP_n))$ both tend to zero.
\end{proof}

\input{eigenvector_examples}

\section*{Use of AI tools}
We used AI models extensively in this project, in particular GPT-5.6, GPT-6, and GPT-6.2, with Pro and Astra modes. The project started from the question, raised in \cite{HsiehKaneKothariLiMohantyTiegel2026}, of whether low-degree indistinguishability rules out spectral methods. During our meetings in August 2026, we engaged in extensive conversations with GPT-5.6 on this question, searching for counterexamples and testing which spectral statistics we could control. We first controlled the largest eigenvalue, and then linear and pairwise eigenvalue statistics. From these cases we came to expect that the whole spectrum could be handled at once, by combining low-degree control of the global shape of the spectrum with the local relaxation of Dyson Brownian motion, and we designed the three-step strategy of \cref{section:overview}. Through repeated discussions with the models, we then worked out the technical estimates, in particular the stability of the free convolution in \cref{subsec:pastur-stability} and the adaptation of the homogenization argument of \cite{LandonXian2025} in \cref{sec:spectral-coupling}, and we checked each of them ourselves. We found the eigenvector examples in the same way: through conversations with GPT-6 in September 2026 we found the one-eigenvector example of \cref{ex:eigenvector}(a), and working with GPT-6.2 in Codex in November 2026 we constructed the polynomial-time example of \cref{ex:eigenvector}(b). We also used Claude Opus 5.5 and Gemini Neon to polish the writing, check the proofs, and prepare the figures. We take full responsibility for the mathematical content and the correctness of this paper.

\bibliographystyle{alpha}
    \bibliography{ref}
\end{document}

%% file: eigenvector_examples.tex
\section{Proofs for the eigenvector examples}
\label{sec:eigenvector-examples}

In this section, we prove \cref{ex:eigenvector}, treating part~(a) in \cref{subsec:one-eigenvector} and part~(b) in \cref{subsec:poly-eigenvectors}. Throughout, we fix $T>0$, set $r=e^{-T/2}$ and $s=1-e^{-T}$, and write $H=rX+\sqrt s\,W\sim\mathcal U_TP_n$ with independent $X\sim P_n$ and $W\sim Q_n$.

Because an eigenvector is determined only up to sign and an eigenspace admits many orthonormal bases, we formulate our tests in terms of orthogonal projectors: if the columns of $V\in\R^{n\times k}$ are orthonormal, then $VV^{\mathsf T}$ is invariant under replacing $V$ by $VO$ for orthogonal $O$ (and reduces to $vv^{\mathsf T}$ when $k=1$). For a real symmetric matrix $B$ with simple spectrum and $I\subseteq[n]$, let $\Pi_I(B)$ be the orthogonal projector onto the span of the eigenvectors of $\lambda_i(B)$, $i\in I$, and let $\Pi_k(B)=\Pi_{\{n-k+1,\dots,n\}}(B)$. A uniformly random projector of rank $k$ has the law of $OE_kO^{\mathsf T}$, where $O$ is Haar distributed on the orthogonal group and $E_k$ is a fixed rank-$k$ projector. Under $Q_n$, the eigenvectors of $H$ form a Haar orthogonal matrix independent of $\spec(H)$, so $\Pi_I(H)$ is uniformly random of rank $|I|$. For a real symmetric matrix $A$ and a rank-$k$ orthogonal projector $\Pi$, we define the residual of $A$ in the subspace represented by $\Pi$ by
\begin{equation*}
L(A,\Pi)=\|(I_n-\Pi)A\Pi\|_F,\qquad m_k=\frac{k(n-k)}{n}.
\end{equation*}
Thus $L(A,\Pi)$ measures how far $A$ maps the range of $\Pi$ out of itself, and satisfies $L(A+cI_n,\Pi)=L(A,\Pi)=L(A,I_n-\Pi)$.
\subsection{One eigenvector}
\label{subsec:one-eigenvector}
We prove part~(a) of \cref{ex:eigenvector} by rotating a GOE spectrum using one of a finite deterministic list of orthogonal matrices, and testing whether any candidate matrix associated with this list has an unusually small residual in the projector $\Pi_I(H)$ of the noisy matrix. We first compute the exact residual of $X$ in $\Pi_I(H)$ (\cref{lem:noisy-eigenbasis}) and then estimate the residual of a fixed matrix in a uniformly random subspace (\cref{lem:random-subspace-residual}).
\begin{lemma}
\label{lem:noisy-eigenbasis}
Suppose that $\spec(X)$ has the law $\spec(Q_n)$. Let $V$ be the orthogonal matrix of eigenvectors of $H$, ordered by eigenvalue and with independent uniformly random signs, and let $\Lambda=\diag(\spec(H))$. Then $(\Lambda,V^{\mathsf T}XV)$ has the law of $(\Lambda,r\Lambda+\sqrt s\,W')$, where $W'\sim Q_n$ is independent of $\Lambda$ and $\Lambda$ has the law of $\diag(\spec(Q_n))$. In particular, for every $I\subseteq[n]$ with $|I|=k$,
\begin{equation}
L\bigl(X,\Pi_I(H)\bigr)^2\overset{d}{=}\frac sn\,\chi^2_{k(n-k)}.
\label{eq:residual-law}
\end{equation}
\end{lemma}

\begin{proof}
Let first $X\sim Q_n$ and $Z=\sqrt s\,X-rW$. Since $r^2+s=1$, the map $(X,W)\mapsto(H,Z)$ acts on each pair of corresponding entries by the same orthogonal matrix, so $H$ and $Z$ are independent GOE matrices and $X=rH+\sqrt s\,Z$. Hence $V^{\mathsf T}XV=r\Lambda+\sqrt s\,V^{\mathsf T}ZV$, and $V^{\mathsf T}ZV\sim Q_n$ independently of $\Lambda$ by orthogonal invariance. For general $X$, write $X=OMO^{\mathsf T}$ with $M=\diag(\spec(X))$ and $O$ orthogonal. Then $\widetilde W=O^{\mathsf T}WO\sim Q_n$ is independent of $(O,M)$, and $H=O\widetilde HO^{\mathsf T}$ with $\widetilde H=rM+\sqrt s\,\widetilde W$. If $\widetilde V$ is a uniformly signed eigenvector matrix of $\widetilde H$, then $O\widetilde V$ is one of $H$, so $(\Lambda,V^{\mathsf T}XV)$ has the law of $(\diag(\spec(\widetilde H)),\widetilde V^{\mathsf T}M\widetilde V)$. This law depends only on the law of $M$, so the first case applies. Finally, $L(X,\Pi_I(H))^2$ is the sum of the squares of the entries $(V^{\mathsf T}XV)_{ji}$ with $i\in I$ and $j\notin I$, which have the law of $k(n-k)$ independent $N(0,s/n)$ variables.
\end{proof}

The proof uses only the law of $\spec(X)$, even when the eigenvectors of $X$ depend on its eigenvalues. So $\spec(\mathcal U_tP_n)=\spec(Q_n)$ for every $t\ge0$ in both constructions below.

For $1\le k\le n-1$, \eqref{eq:residual-law} shows that $L(X,\Pi_I(H))/\sqrt{m_k}\to\sqrt s$ in probability. We next compute the second moment and tail concentration of the residual of a fixed matrix $A$ in a uniformly random rank-$k$ subspace; when $A$ has eigenvalues $\gamma_1,\dots,\gamma_n$, this yields $L(A,\Pi)/\sqrt{m_k}\to1$ in probability.

\begin{lemma}
\label{lem:random-subspace-residual}
Let $n\ge3$, let $A$ be real symmetric with $\|A\|_{\op}\le3$, let $\mathring A=A-n^{-1}(\Tr A)I_n$, and let $\Pi$ be a uniformly random projector of rank $k\in[1,n-1]$. Then
\begin{equation*}
\E L(A,\Pi)^2=\frac{k(n-k)}{(n-1)(n+2)}\Tr\mathring A^2,
\qquad
\Pp\bigl\{|L(A,\Pi)-\E L(A,\Pi)|\ge x\bigr\}\le2e^{-cnx^2}\quad(x>0),
\end{equation*}
where $c>0$ is an absolute constant.
\end{lemma}

\begin{proof}
Assume $\Tr A=0$, and write $\Pi=UU^{\mathsf T}$ where the columns $u_1,\dots,u_k$ of $U$ are the first $k$ columns of a Haar orthogonal matrix. Then $L(A,\Pi)^2=\Tr(U^{\mathsf T}A^2U)-\|U^{\mathsf T}AU\|_F^2$. For a uniform unit vector $u$, $\E\,u^{\mathsf T}A^2u=\Tr A^2/n$ and $\E(u^{\mathsf T}Au)^2=2\Tr A^2/(n(n+2))$, and summing $(u_1^{\mathsf T}Au_j)^2$ over an orthonormal basis gives $\E(u_1^{\mathsf T}Au_2)^2=\Tr A^2/((n-1)(n+2))$. The formula for $\E L(A,\Pi)^2$ follows. Writing $\Pi=OE_kO^{\mathsf T}$ with $O$ Haar on the special orthogonal group, the map $O\mapsto L(A,\Pi)$ is $12$-Lipschitz for the Frobenius distance, and the tail bound follows from concentration of measure on this group, which holds for $n\ge3$ \cite{Ledoux2001}.
\end{proof}

Combining these two estimates, we build a test that searches a finite deterministic list of candidate matrices with eigenvalues $\gamma_1,\dots,\gamma_n$ and declares the planted law if some candidate has residual at most $\frac12(1+\sqrt s)\sqrt{m_k}$ in $\Pi_I(H)$. Under $\mathcal U_TP_n$, the candidate matching the planted basis approximates $X$ closely enough to fall below this threshold, whereas under $Q_n$ we apply \cref{lem:random-subspace-residual} and a union bound over the list to show that every candidate stays near $\sqrt{m_k}$. We now construct such a list with vanishing low-degree advantage, completing part~(a) of \cref{ex:eigenvector}.

\begin{proposition}
\label{prop:one-eigenvector}
Let $D_n\log n=o(n)$. There are laws $P_n$ such that $\spec(X)$ has the law $\spec(Q_n)$ for $X\sim P_n$, $\adv_{\le D_n}(Q_n,P_n)\le n^{-1/2}$, and for every $I\subseteq[n]$ with $1\le|I|\le n-1$ some test that uses only $\Pi_I(H)$ has both errors $o(1)$. For the average of $P_n$ over simultaneous permutations of rows and columns, the same holds when $\min(|I|,n-|I|)\ge C_T\log n$, with $C_T$ depending only on $T$.
\end{proposition}

\begin{proof}
As in \cref{rem:necessity}, let $N_n$ be the dimension of the polynomials of degree at most $D_n$ with mean zero under $Q_n$, and let $M_n=nN_n$. For orthogonal $O_1,\dots,O_{M_n}$, let $P_n$ be the law of $O_\iota\diag(\spec(Y))O_\iota^{\mathsf T}$, with $\iota$ uniform on $[M_n]$ and $Y\sim Q_n$ independent of $\iota$. If the $O_j$ are independent and Haar distributed, then for an orthonormal basis $(f_l)$ of the polynomials above,
\begin{equation*}
\E\,\adv_{\le D_n}(Q_n,P_n)^2=\sum_l\E\Bigl(\frac1{M_n}\sum_j\phi_l(O_j)\Bigr)^2\le\frac{N_n}{M_n}=\frac1n,
\end{equation*}
where $\phi_l(O)=\E_Y f_l(O\diag(\spec(Y))O^{\mathsf T})$. Here we used that $O\diag(\spec(Y))O^{\mathsf T}\sim Q_n$ for Haar $O$, so that $\E\phi_l(O)=0$ and $\E\phi_l(O)^2\le1$. We fix $O_1,\dots,O_{M_n}$ with $\adv_{\le D_n}(Q_n,P_n)\le n^{-1/2}$. Note that $\log M_n\le3D_n\log n=o(n)$.

Let $\Gamma=\diag(\gamma_1,\dots,\gamma_n)$ with the classical locations \eqref{eq:classical-location}, and $A_j=O_j\Gamma O_j^{\mathsf T}$. Fix $I$ with $|I|=k$, write $\Pi=\Pi_I(H)$, and assume $k\le n/2$, so that $m_k\ge\max\{k/2,1/2\}$. The test declares the planted law when $\min_jL(A_j,\Pi)\le\frac12(1+\sqrt s)\sqrt{m_k}$. Under $\mathcal U_TP_n$, GOE rigidity \cite{ERDOS20121435} gives $\|X-A_\iota\|_{\op}=\max_i|\lambda_i(Y)-\gamma_i|\le n^{-1/2}$ with probability $1-o(1)$, hence $L(A_\iota,\Pi)\le L(X,\Pi)+\sqrt{2m_k/n}$, and $L(X,\Pi)/\sqrt{m_k}\to\sqrt s$ in probability by \eqref{eq:residual-law}. Under $Q_n$, $\Pi$ is uniformly random. Since $n^{-1}\Tr\mathring\Gamma^2\to1$, \cref{lem:random-subspace-residual} gives $\E L(A_j,\Pi)^2=(1+o(1))m_k$, and its tail bound gives $\Var L(A_j,\Pi)\le2/(cn)$. As $m_k\ge1/2$, it follows that $\E L(A_j,\Pi)\ge(1-o(1))\sqrt{m_k}$ for every $j$. The tail bound and a union bound over $j$ bound the error by
\begin{equation}
2M_n\exp\bigl(-c(1-\sqrt s)^2nm_k/16\bigr)=o(1).
\label{eq:one-eigenvector-null}
\end{equation}
For the permutation average, the spectrum is unchanged and the advantage does not increase, since relabeling preserves the degree and the null variance of a polynomial. The test then minimizes over the $n!\,M_n\le e^{2n\log n}$ matrices $SA_jS^{\mathsf T}$, and \eqref{eq:one-eigenvector-null} with $M_n$ replaced by $n!\,M_n$ still tends to zero when $m_k\ge C_T\log n/2$ with $C_T=128/(c(1-\sqrt s)^2)$.
\end{proof}

\subsection{A fixed fraction of eigenvectors in polynomial time}
\label{subsec:poly-eigenvectors}

\Cref{prop:one-eigenvector} hides a signal in a projector while the spectral law agrees exactly with that of the GOE, but its test searches a list of $n^{O(D_n)}$ rotations, or $n!\,n^{O(D_n)}$ after relabeling. For a polynomial-time test, we hide instead a code that is known to be detectable in polynomial time. Its low-degree moments must agree with those of independent bits, and its structure must stay readable by an efficient algorithm after noise. We insert it by orthogonal conjugation. If $Y\sim Q_n$ and $X_0=OYO^{\mathsf T}$ for an orthogonal $O$ that may depend on $Y$, then $\spec(X_0)=\spec(Y)$ and $\Pi_k(X_0)=O\,\Pi_k(Y)O^{\mathsf T}$, so we can design the eigenvectors without changing the spectrum.

\begin{proposition}
\label{prop:poly-eigenvectors}
Fix $\alpha\in(0,1)$ and $q>0$, and let $k=\lfloor\alpha n\rfloor$ and $D_n=\lceil(\log n)^q\rceil$. There are laws $P_n$, invariant under simultaneous permutations of rows and columns, a constant $T_0>0$, and a deterministic polynomial-time test with the following properties. For $X\sim P_n$, $\spec(X)$ has the law $\spec(Q_n)$, and $\adv_{\le D_n}(Q_n,P_n)\le n^{-10D_n}$ for large $n$. If $T\in(0,T_0]$ and the test receives a matrix within $n^{-4}$ of $\Pi_k(H)$ entrywise, its errors under $\mathcal U_TP_n$ and under $Q_n$ tend to zero.
\end{proposition}

\paragraph{Computing the approximate projector.}
The precision requirement in \cref{prop:poly-eigenvectors} gives the test a finite rational input approximating $\Pi_k(H)$ (whose entries may be irrational) to entrywise error $n^{-4}$. With probability $1-o(1)$ under both hypotheses, we can compute such an approximation in polynomial time from the rational matrix $\widehat H$ obtained by rounding each entry of $H$ to a multiple of $n^{-10}$. Indeed, by \cref{lem:noisy-eigenbasis}, $\spec(H)$ has the law $\spec(Q_n)$ under both hypotheses, so with probability $1-o(1)$ we have $\|H\|_{\op}\le3$ and $\lambda_{n-k+1}(H)-\lambda_{n-k}(H)\ge n^{-2}$, since the smallest gap between consecutive GOE eigenvalues is of order $n^{-3/2}$ \cite{FengTianWei2019}. Then $\|H-\widehat H\|_{\op}\le n^{-9}$, and $\|\Pi_k(H)-\Pi_k(\widehat H)\|_{\op}\le2n^{-7}$ by the Davis--Kahan theorem. Using exact eigenvalue counts via Sylvester's law of inertia and bisection, we find a rational $\mu$ at distance at least $n^{-2}/4$ from $\spec(\widehat H)$ with exactly $k$ eigenvalues of $\widehat H$ above it. Setting $\Delta=(\widehat H-\mu I_n)/8$, every eigenvalue $x$ of $\Delta$ satisfies $n^{-2}/32\le|x|<1$. For $N_1=2^{\lceil3\log_2n\rceil}$, the function $((1+x)^{N_1}-(1-x)^{N_1})/((1+x)^{N_1}+(1-x)^{N_1})=\tanh(N_1\operatorname{artanh}x)$ is within $2e^{-n/16}$ of $\operatorname{sign}(x)$, so $\frac12(I_n+B_+^{-1}B_-)$ with $B_\pm=(I_n+\Delta)^{N_1}\pm(I_n-\Delta)^{N_1}$ lies within $e^{-n/16}$ of $\Pi_k(\widehat H)$, hence within $n^{-4}$ of $\Pi_k(H)$. Repeated squaring and Gaussian elimination over $\mathbb Q$ compute this rational matrix exactly using polynomially many bits.
\Cref{prop:poly-eigenvectors} makes no claim about sampling $P_n$ efficiently, and even Mao's Boolean law is sampled with advice \cite[Theorem~1.2]{Mao2026}.

To construct a low-degree-hard signal that remains efficiently detectable, we use the Boolean code of Mao \cite{Mao2026}. Collections of its bits of a prescribed polylogarithmic size are independent and uniform, defeating low-degree tests, while globally the bits are evaluations of low-degree polynomials over $\mathbb F_2$, which we can detect by an $\mathbb F_2$-rank test. For $y\in\mathbb F_2^\ell$ and an integer $d$, we define the monomial lift $\Psi_d(y)=(\prod_{a\in S}y_a)_S$, indexed by $S\subseteq[\ell]$ with $|S|\le d$. Whenever the coordinates of $y$ are polynomials of degree at most $u$ in a variable $\xi$, the coordinates of $\Psi_d(y)$ have degree at most $ud$ in $\xi$, so as $\xi$ varies the lifted vectors span a subspace of dimension at most the number of degree-$ud$ monomials in $\xi$.

\begin{lemma}[\cite{Mao2026}]
\label{lem:mao-code}
For every integer $u\ge3$ there are $c_u,A_u>0$ and integers $m,\ell,d$ depending on $n$, with $\ell\le A_u\log n$ and $\ell\to\infty$, such that for large $n$ there are random bits $b_{ij}=b_{ji}\in\mathbb F_2$, $i\ne j$, and random points $\xi_1,\dots,\xi_n\in\mathbb F_2^m$ with the following properties.
\begin{enumerate}[label=(\roman*)]
\item Any $\lfloor c_u(\log n)^{u-1}\rfloor$ of the bits $b_{ij}$, $i<j$, are independent and uniform.
\item For each $i$ there is a polynomial $g_i$ on $\mathbb F_2^m$ of degree at most $u$ with $b_{ij}=g_i(\xi_j)$ for all $j\ne i$.
\item If $y_1,\dots,y_{N_0}$, $N_0=\lfloor\sqrt n\rfloor$, are independent and uniform in $\mathbb F_2^\ell$, then $\Psi_d(y_1),\dots,\Psi_d(y_{N_0})$ are linearly independent with probability $1-o(1)$.
\item For any $i_1,\dots,i_\ell$, the vectors $\Psi_d(g_{i_1}(\xi_j),\dots,g_{i_\ell}(\xi_j))$, $j\in[n]$, span a space of dimension at most $n^{1/4}$.
\end{enumerate}
\end{lemma}

\begin{proof}
Take the construction of \cite[Sections~2--3]{Mao2026} with Reed--Muller degree $u$ (called $r$ there), and let $\beta,a,\sigma$ be the constants called $b,a,\sigma$ in \cite[Section~2.1]{Mao2026} and in the proof of \cite[Lemma~3.5]{Mao2026}. They satisfy $u\sigma/\beta<1/2$ and $\beta h_2(u\sigma/\beta)<1/4$, where $h_2$ is the binary entropy. Each vertex $i$ carries a point $\xi_i\in\mathbb F_2^m$, $m=\lceil\beta\lceil\log_2n\rceil\rceil$, and $b_{ij}=\omega(\phi(\xi_i),\phi(\xi_j))$ for a uniformly random alternating form $\omega$ and a map $\phi$ whose coordinates have degree at most $u$. Then (i) is \cite[Proposition~2.6]{Mao2026}, and (ii) holds with $g_i=\omega(\phi(\xi_i),\phi(\cdot))$, as in the proof of \cite[Lemma~3.4]{Mao2026}. With $\ell=\lfloor a\lceil\log_2n\rceil\rfloor$ and $d=\lfloor\sigma\lceil\log_2n\rceil\rfloor$, (iii) is \cite[Lemma~3.3]{Mao2026}, whose hypotheses hold by \cite[Lemma~3.5]{Mao2026}. For (iv), each coordinate of these vectors is a polynomial of degree at most $ud$ in $\xi_j$, and $ud/m\le u\sigma/\beta<1/2$. So they span a space of dimension at most $\sum_{l\le ud}\binom ml\le2^{m\,h_2(u\sigma/\beta)}\le2^{\beta+1}n^{\beta h_2(u\sigma/\beta)}$, which is at most $n^{1/4}$ for large $n$.
\end{proof}

\paragraph{The polynomial-time rank test.}
In \cref{lem:mao-code}, property~(i) supplies the local independence behind the low-degree bound, (ii) encodes the polynomial structure, and (iii)--(iv) create a rank gap: for $\ell$ rows and $N_0=\lfloor\sqrt n\rfloor$ columns, the lifted columns $\Psi_d(y_j)$ are linearly independent under uniform random bits, whereas for Mao's code they span a space of dimension at most $n^{1/4}\ll N_0$. Because each corrupted column increases the rank by at most one, this gap persists as long as many columns are read correctly.

To apply this test to the observed eigenvectors, we fix a constant $\varepsilon\in(0,\min\{\alpha,1-\alpha\}/2]$ (chosen in the proof of \cref{prop:poly-eigenvectors}) and set $n_1=\lceil16/\varepsilon\rceil\ell$ and $\mathcal C_0=\{n_1+1,\dots,n_1+N_0\}$. Given a symmetric entrywise $n^{-4}$ approximation $\widetilde\Pi$ to $\Pi_k(H)$, we extract a bit from the sign of each entry: for each $\ell$-element row subset $I=\{i_1<\dots<i_\ell\}\subseteq[n_1]$, we form the column vectors $y^I_j\in\mathbb F_2^\ell$ with entries $\mathbf 1\{\widetilde\Pi_{i_aj}<0\}$ ($a\le\ell$) and compute the rank deficit $\operatorname{def}(I)=N_0-\operatorname{rank}\{\Psi_d(y^I_j):j\in\mathcal C_0\}$. Our test searches over all $I\subseteq[n_1]$ of size $\ell$ (needed because we hide the coded rows by a random relabeling) and declares the planted law if $\operatorname{def}(I)\ge n^{1/3}$ for some $I$. Since there are at most $2^{n_1}=n^{O(1)}$ subsets $I$ and each rank is computed by Gaussian elimination over $\mathbb F_2$ on a matrix with at most $2^\ell=n^{O(1)}$ rows, the test runs in polynomial time.

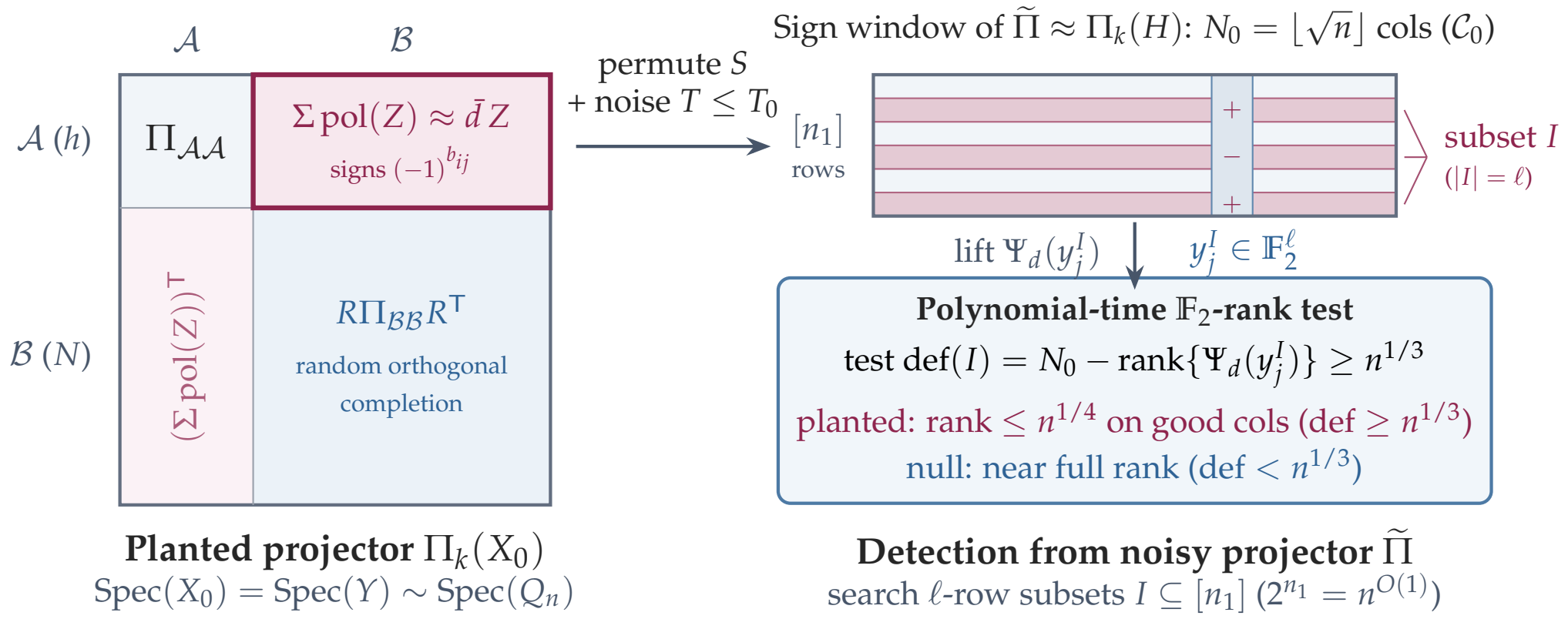
\begin{figure}[H]
\centering
\definecolor{figblue}{HTML}{2E6295}%
\definecolor{figbluebg}{HTML}{EBF2F8}%
\definecolor{figplum}{HTML}{8E244D}%
\definecolor{figplumbg}{HTML}{F7E8EE}%
\definecolor{figslate}{HTML}{475569}%
\definecolor{figgray}{HTML}{F1F5F9}%
\begin{tikzpicture}[
  arr/.style={-{Stealth[length=2.1mm]}, thick, figslate},
  scale=0.88]

\begin{scope}[shift={(0,0)}]
  \def\ha{1.30}
  \def\nb{2.90}
  \def\tot{4.20}
  \fill[figgray] (0,\nb) rectangle (\ha,\tot);
  \fill[figplumbg] (\ha,\nb) rectangle (\tot,\tot);
  \fill[figplumbg!55] (0,0) rectangle (\ha,\nb);
  \fill[figbluebg] (\ha,0) rectangle (\tot,\nb);
  \draw[figslate!85, thick] (0,0) rectangle (\tot,\tot);
  \draw[figslate!55, thin] (\ha,0) -- (\ha,\tot) (0,\nb) -- (\tot,\nb);
  \draw[figplum, very thick] (\ha,\nb) rectangle (\tot,\tot);
  \node[font=\footnotesize, text=black!85] at (\ha/2, \nb+\ha/2) {$\Pi_{\mathcal A\mathcal A}$};
  \node[font=\scriptsize, align=center, text=figplum] at (\ha+\nb/2, \nb+\ha/2)
    {$\Sigma\operatorname{pol}(Z)\approx\bar d\,Z$\\[2pt]
     {\tiny signs $(-1)^{b_{ij}}$}};
  \node[font=\scriptsize, text=figplum!75, rotate=90] at (\ha/2, \nb/2)
    {$(\Sigma\operatorname{pol}(Z))^{\mathsf T}$};
  \node[font=\scriptsize, align=center, text=figblue] at (\ha+\nb/2, \nb/2)
    {$R\Pi_{\mathcal B\mathcal B}R^{\mathsf T}$\\[2pt]
     {\tiny random orthogonal}\\[-1pt]{\tiny completion}};
  \node[font=\scriptsize, text=figslate, left=3pt] at (0, \nb+\ha/2) {$\mathcal A$ ($h$)};
  \node[font=\scriptsize, text=figslate, left=3pt] at (0, \nb/2) {$\mathcal B$ ($N$)};
  \node[font=\scriptsize, text=figslate, above=2pt] at (\ha/2, \tot) {$\mathcal A$};
  \node[font=\scriptsize, text=figslate, above=2pt] at (\ha+\nb/2, \tot) {$\mathcal B$};
  \node[font=\footnotesize\bfseries, text=black!85] at (\tot/2, -0.45)
    {Planted projector $\Pi_k(X_0)$};
  \node[font=\scriptsize, text=figslate] at (\tot/2, -0.86)
    {$\spec(X_0)=\spec(Y)\sim\spec(Q_n)$};
\end{scope}

\draw[arr] (4.45, 3.50) -- node[above=3pt, font=\scriptsize, align=center, text=black!85]
  {permute $S$\\[-1pt]+ noise $T\le T_0$} (6.35, 3.50);

\begin{scope}[shift={(7.35,0)}]
  \def\wx{0}
  \def\wy{2.82}
  \def\ww{5.10}
  \def\wh{1.38}
  \fill[figgray] (\wx,\wy) rectangle (\wx+\ww,\wy+\wh);
  \foreach \yy in {3.05, 3.28, 3.51, 3.74, 3.97} {
    \draw[figslate!20, very thin] (\wx,\yy) -- (\wx+\ww,\yy);
  }
  \fill[figplum!24] (\wx, 3.74) rectangle (\wx+\ww, 3.97);
  \fill[figplum!24] (\wx, 3.28) rectangle (\wx+\ww, 3.51);
  \fill[figplum!24] (\wx, 2.82) rectangle (\wx+\ww, 3.05);
  \draw[figplum!70, thin] (\wx, 3.74) -- (\wx+\ww, 3.74) (\wx, 3.97) -- (\wx+\ww, 3.97);
  \draw[figplum!70, thin] (\wx, 3.28) -- (\wx+\ww, 3.28) (\wx, 3.51) -- (\wx+\ww, 3.51);
  \draw[figplum!70, thin] (\wx, 3.05) -- (\wx+\ww, 3.05);
  \fill[figblue!16] (3.30, \wy) rectangle (3.70, \wy+\wh);
  \draw[figblue!75, thin] (3.30, \wy) rectangle (3.70, \wy+\wh);
  \draw[figslate!85, thick] (\wx,\wy) rectangle (\wx+\ww,\wy+\wh);
  \node[font=\tiny\bfseries, text=figplum] at (3.50, 3.855) {$+$};
  \node[font=\tiny\bfseries, text=figplum] at (3.50, 3.395) {$-$};
  \node[font=\tiny\bfseries, text=figplum] at (3.50, 2.935) {$+$};
  \node[font=\scriptsize, text=black!85, above=3pt] at (\wx+\ww/2, \wy+\wh)
    {Sign window of $\widetilde\Pi\approx\Pi_k(H)$: $N_0=\lfloor\sqrt n\rfloor$ cols ($\mathcal C_0$)};
  \node[font=\scriptsize, text=figslate, left=3pt, align=right] at (\wx, \wy+\wh/2)
    {$[n_1]$\\[-1pt]{\tiny rows}};
  \draw[thin, figplum!80] (\wx+\ww+0.08, 3.855) -- (\wx+\ww+0.30, 3.395);
  \draw[thin, figplum!80] (\wx+\ww+0.08, 3.395) -- (\wx+\ww+0.30, 3.395);
  \draw[thin, figplum!80] (\wx+\ww+0.08, 2.935) -- (\wx+\ww+0.30, 3.395);
  \node[font=\scriptsize, text=figplum, right=1pt, align=left] at (\wx+\ww+0.28, 3.395)
    {subset $I$\\[-1pt]{\tiny ($|I|=\ell$)}};

  \node[anchor=south, draw=figblue!85, thick, rounded corners=3pt, fill=figbluebg!70,
        align=center, font=\scriptsize, inner sep=4.5pt, minimum width=5.10cm]
    (testbox) at (\wx+\ww/2, 0.00)
    {\textbf{\color{black!85}Polynomial-time $\mathbb F_2$-rank test}\\[2pt]
     test $\operatorname{def}(I)=N_0-\operatorname{rank}\{\Psi_d(y_j^I)\}\ge n^{1/3}$\\[2.5pt]
     {\color{figplum}planted: rank $\le n^{1/4}$ on good cols ($\operatorname{def}\ge n^{1/3}$)}\\[1pt]
     {\color{figblue}null: near full rank ($\operatorname{def}<n^{1/3}$)}};

  \draw[arr] (2.55, \wy-0.06) -- node[left=4pt, font=\scriptsize, text=figslate]
    {lift $\Psi_d(y_j^I)$} (2.55, 2.10);
  \node[font=\scriptsize, text=figblue] at (3.62, 2.46) {$y_j^I\in\mathbb F_2^\ell$};

  \node[font=\footnotesize\bfseries, text=black!85] at (\wx+\ww/2, -0.45)
    {Detection from noisy projector $\widetilde\Pi$};
  \node[font=\scriptsize, text=figslate] at (\wx+\ww/2, -0.86)
    {search $\ell$-row subsets $I\subseteq[n_1]$ ($2^{n_1}=n^{O(1)}$)};
\end{scope}
\end{tikzpicture}
\caption{Construction and detection in \cref{prop:poly-eigenvectors}. Left: conjugating $Y\sim Q_n$ by $I_h\oplus R$ preserves both the spectrum and $\Pi_{\mathcal A\mathcal A}$, while making the $\mathcal A\times\mathcal B$ block of $\Pi_k(X_0)$ equal to $\Sigma\operatorname{pol}(Z)\approx\bar d\,Z$, where $\bar d=(\alpha(1-\alpha)/N)^{1/2}$; most signs of this block agree with Mao's code signs $(-1)^{b_{ij}}$. Right: after random relabeling $S$ and small OU noise $T\le T_0$, our polynomial-time test inspects the $[n_1]\times\mathcal C_0$ sign window of $\widetilde\Pi\approx\Pi_k(H)$, lifts the sign vectors $y_j^I\in\mathbb F_2^\ell$ for each $\ell$-row subset $I\subseteq[n_1]$ via $\Psi_d$, and tests for a rank deficit over $\mathbb F_2$.}
\label{fig:coded-projector}
\end{figure}

\paragraph{Embedding the code into the projector.}
We now embed Mao's code into the $\mathcal A\times\mathcal B$ block of the orthogonal  projector $\Pi_k(X_0)$ while preserving the exact GOE spectral law (\cref{fig:coded-projector}). Because $\Pi^2=\Pi$ forces the off-diagonal block $\Pi_{\mathcal A\mathcal B}$ to have row Gram matrix $\Pi_{\mathcal A\mathcal A}-\Pi_{\mathcal A\mathcal A}^2$, we factor $\Pi_{\mathcal A\mathcal B}$ into the square root of its Gram matrix times a matrix $\Theta$ with orthonormal rows. Applying polar normalization to the half-normal matrix $Z$ carrying the code produces another matrix $U$ with orthonormal rows, and conjugating by a block orthogonal matrix $I_h\oplus R$ replaces $\Theta$ by $U$ while keeping both the spectrum and $\Pi_{\mathcal A\mathcal A}$ unchanged.

\begin{enumerate}[label=(\arabic*)]
\item
With $\varepsilon$ as fixed above, let $u=\lceil3q\rceil+3$ and take the bits of \cref{lem:mao-code}. Set $h=\lfloor\varepsilon n\rfloor$, $N=n-h$, $\mathcal A=[h]$ and $\mathcal B=[n]\setminus[h]$, and index $h\times N$ matrices by $\mathcal A\times\mathcal B$. For an $h\times N$ matrix $F$ of rank $h$, write $\operatorname{pol}(F)=(FF^{\mathsf T})^{-1/2}F$ for the matrix with orthonormal rows closest to $F$ in Frobenius norm.
\item Let $Y\sim Q_n$ and $\Pi=\Pi_k(Y)$. Since $\Pi^2=\Pi$, the block $\Pi_{\mathcal A\mathcal B}$ has the Gram matrix $\Pi_{\mathcal A\mathcal A}-\Pi_{\mathcal A\mathcal A}^2$, so we cannot put a sign matrix there directly. Since $h\le k\le n-h$, we have $0<\Pi_{\mathcal A\mathcal A}<I_h$ almost surely, and $\Pi_{\mathcal A\mathcal B}=\Sigma\Theta$ with $\Sigma=(\Pi_{\mathcal A\mathcal A}-\Pi_{\mathcal A\mathcal A}^2)^{1/2}$ and $\Theta=\Sigma^{-1}\Pi_{\mathcal A\mathcal B}$, which has orthonormal rows. Only the orientation $\Theta$ of the rows is free.
\item Independently, let $Z_{ij}=(-1)^{b_{ij}}|\zeta_{ij}|$ for $(i,j)\in\mathcal A\times\mathcal B$, with independent standard Gaussian $\zeta_{ij}$, and let $U=\operatorname{pol}(Z)$. This is the half-normal embedding of Klivans and Kothari \cite{KlivansKothari2014}, which Mao uses to carry his construction over to Gaussian data with Ornstein--Uhlenbeck noise \cite[Section~4]{Mao2026}. The matrix $Z$ is independent of $Y$ and has rank $h$ almost surely. Since $ZZ^{\mathsf T}/N$ is close to $I_h$ when $h$ is a small fraction of $N$, $U$ is close to $Z/\sqrt N$.
\item Given $(Y,Z)$, let $R$ be uniform on the orthogonal matrices with $R\Theta^{\mathsf T}=U^{\mathsf T}$, and $X_0=(I_h\oplus R)Y(I_h\oplus R)^{\mathsf T}$. Conjugation by $I_h\oplus R$ keeps $\Pi_{\mathcal A\mathcal A}$ and the spectrum and replaces $\Theta$ by $\Theta R^{\mathsf T}=U$. The condition fixes $R$ only on the row span of $\Theta$, and the uniform choice of the rest is what makes the low-degree analysis work.
\item Let $P_n$ be the law of $SX_0S^{\mathsf T}$ for an independent uniformly random permutation matrix $S$.
\end{enumerate}
Then $\spec(X_0)=\spec(Y)$, and $\Pi_k(X_0)=(I_h\oplus R)\Pi(I_h\oplus R)^{\mathsf T}$ has the block
\begin{equation}
\Pi_k(X_0)_{\mathcal A\mathcal B}=\Sigma\Theta R^{\mathsf T}=\Sigma\operatorname{pol}(Z).
\label{eq:coded-block}
\end{equation}
Thus we embed the code carried by $Z$ into the $\mathcal A\times\mathcal B$ block of $\Pi_k(X_0)$ through $\Sigma\operatorname{pol}(Z)$ while keeping the spectrum exactly GOE. We next verify that the resulting law has vanishing low-degree advantage (\cref{lem:polar-moments}) and that enough code signs survive a small fixed amount of OU noise (\cref{lem:coded-block-noise}).

To bound the low-degree advantage of $X_0$, notice that by \cref{lem:mao-code}(i) every polynomial in $Z$ of degree at most $c_u(\log n)^{u-1}$ has the same expectation as for an $h\times N$ matrix $G$ of independent standard Gaussians (since a uniform sign times an independent half-normal variable is standard Gaussian). Because $X_0$ depends on $Z$ through the nonpolynomial polar factor $\operatorname{pol}(Z)$, we average over the random completion $R$ to write $\E[f(X_0)\mid Z]=p(\operatorname{pol}(Z))$ for a degree-$2D$ polynomial $p$, and truncate the Taylor series of $\operatorname{pol}$ at order $L=O(D^2\log n)$ to obtain polynomial approximations $U_L$ and $U_{L,G}$. In \cref{lem:polar-moments} below, we establish the chain
\begin{equation*}
\E f(X_0)=\E p(\operatorname{pol}(Z))\approx\E p(U_L)=\E p(U_{L,G})\approx\E p(\operatorname{pol}(G))=0,
\end{equation*}
where the middle equality uses moment matching up to degree $O(D^3\log n)$ and the last equality holds because replacing $Z$ by $G$ makes $X_0$ a GOE matrix.

\begin{lemma}
\label{lem:polar-moments}
There are absolute constants $\varepsilon_0,c_0>0$ and $K_0<\infty$ with the following property. Let $h\le\varepsilon_0N$ and $D^3\log n\le c_0N$. Let $\Theta$ be any measurable function of $Y$ with orthonormal rows, and let $Z$ be a random $h\times N$ matrix, independent of $Y$ and of rank $h$ almost surely, such that every polynomial of degree at most $K_0D^3\log n$ in the entries of $Z$ has the same expectation as for independent standard Gaussian entries. Define $U$, $R$ and $X_0$ as in steps (3) and (4). Then $\adv_{\le D}(Q_n,\mathcal L(X_0))\le n^{-10D}$ for large $n$.
\end{lemma}

\begin{proof}
Fix a polynomial $f$ of degree at most $D$ with $\E_{Q_n}f=0$ and $\Var_{Q_n}f=1$. For a polynomial $p$, let $\|p\|_1$ be the sum of the absolute values of its coefficients, so that $\|p_1p_2\|_1\le\|p_1\|_1\|p_2\|_1$ and $|p(V)|\le\|p\|_1\max(1,\max_{ij}|V_{ij}|)^{\deg p}$. Constants $C_1,C_2,\dots$ and $c_1$ are absolute. We assume $n\ge4$, and note that $1\le D\le n^{1/3}$ since $D^3\le c_0N$.

\emph{Step 1: averaging over the completion.} For an $h\times N$ matrix $V$ with orthonormal rows, let $\E^V$ be the expectation over a uniformly random orthonormal basis $\omega_1,\dots,\omega_{N-h}$ of the orthogonal complement of the row span of $V$, and let $\Omega_V\in O(N)$ have the columns $V^{\mathsf T},\omega_1,\dots,\omega_{N-h}$. Let $R_\Theta$ be an orthogonal matrix, measurable in $Y$, whose first $h$ columns are $\Theta^{\mathsf T}$, and $M=(I_h\oplus R_\Theta)^{\mathsf T}Y(I_h\oplus R_\Theta)$. The map $R\mapsto RR_\Theta$ carries the uniform law on the $R$ with $R\Theta^{\mathsf T}=U^{\mathsf T}$ to the uniform law on the orthogonal matrices with first $h$ columns $U^{\mathsf T}$. So, given $(Y,Z)$, $\Omega=RR_\Theta$ has the law of $\Omega_U$, and $X_0=(I_h\oplus\Omega)M(I_h\oplus\Omega)^{\mathsf T}$. As $M$ is a function of $Y$, it is independent of $Z$. With $g_M(\Omega)=f((I_h\oplus\Omega)M(I_h\oplus\Omega)^{\mathsf T})$ and $\E_M$ the expectation over an independent copy of $M$,
\begin{equation*}
\E[f(X_0)\mid Z]=\E\bigl[\E[g_M(\Omega)\mid Y,Z]\bigm|Z\bigr]=\E_M\,\E^Ug_M(\Omega_U)=:p(U).
\end{equation*}
We show that $p$ is a polynomial of degree at most $2D$ and bound $\|p\|_1$. First, $\|f\|_1\le n^{4D}$. Indeed, in the coordinates $x_{ab}=\sqrt n\,X_{ab}$, $a<b$, and $x_{aa}=\sqrt{n/2}\,X_{aa}$, which are independent standard Gaussians under $Q_n$, $f$ is a combination of at most $\binom{n(n+1)/2+D}{D}\le n^{2D}$ normalized Hermite products $\prod_c\mathrm{He}_{\beta_c}(x_c)/\sqrt{\beta_c!}$, with coefficients whose squares sum to $\E_{Q_n}f^2=1$. Since $\|\mathrm{He}_m\|_1\le m^m$, each product has $\|\cdot\|_1\le D^Dn^{D/2}\le n^{2D}$ in the entries $X_{ab}$. Second, each entry of $(I_h\oplus\Omega)M(I_h\oplus\Omega)^{\mathsf T}$ is a polynomial of degree at most two in $\Omega$ with $\|\cdot\|_1\le n^2\|M\|_{\op}$, so $g_M$ has degree at most $2D$ in $\Omega$ and $\|g_M\|_1\le n^{6D}(1+\|M\|_{\op})^D$.

Third, we integrate the columns $\omega_t$. A monomial of $g_M$ contains at most $2D$ distinct columns, which by exchangeability we may take to be $\omega_1,\dots,\omega_\tau$, $\tau\le2D$. Given $\omega_1,\dots,\omega_{t-1}$, the column $\omega_t$ is uniform on the unit sphere of a subspace of dimension $e_t=N-h-t+1\ge N-h-2D+1\ge1$, with projector $P_t=I_N-U^{\mathsf T}U-\sum_{s<t}\omega_s\omega_s^{\mathsf T}$. If $x$ is uniform on the unit sphere of a subspace of dimension $e$ with projector $P$, then
\begin{equation*}
\E\prod_{r=1}^{2a}x_{i_r}=\frac{1}{e(e+2)\cdots(e+2a-2)}\sum_\pi\prod_{\{r,r'\}\in\pi}P_{i_ri_{r'}},
\end{equation*}
summed over the $(2a-1)!!$ pairings $\pi$ of $[2a]$, and odd products have mean zero. This follows from Wick's formula for a Gaussian vector $y$ with covariance $P$, since $y=\|y\|_2x$ with $\|y\|_2$ independent of $x$ and $\E\|y\|_2^{2a}=e(e+2)\cdots(e+2a-2)$. We integrate $\omega_\tau$ first, then $\omega_{\tau-1}$, and so on. Each step replaces $2a\le2D$ coordinates of $\omega_t$ by a sum of $(2a-1)!!\le(2D)^D$ products of $a$ entries of $P_t$, which have degree two in $(U,\omega_1,\dots,\omega_{t-1})$ and $\|\cdot\|_1\le1+h+2D\le2n$. So the degree does not increase, and $\|\cdot\|_1$ grows by a factor at most $(4Dn)^D$ per step. After at most $2D$ steps, $\E^Ug_M(\Omega_U)=q_M(U)$ with $\deg q_M\le2D$ and $\|q_M\|_1\le n^{6D}(1+\|M\|_{\op})^D(4Dn)^{2D^2}$. Finally $p=\E_Mq_M$ coefficientwise, $\|M\|_{\op}=\|Y\|_{\op}$, and $\E(1+\|Y\|_{\op})^D\le5^D$ for large $n$, by Minkowski's inequality and \eqref{eq:null-norm-moments}, since $D\le n^{1/3}$. As $(4Dn)^{2D^2}\le n^{6D^2}$,
\begin{equation}
\E[f(X_0)\mid Z]=p(U),\qquad\deg p\le2D,\qquad\|p\|_1\le n^{C_1D^2}.
\label{eq:frame-polynomial}
\end{equation}
Here $p$ is the fixed polynomial produced by this computation, and below we also evaluate it at matrices whose rows are not orthonormal, such as $U_L$. It depends only on $f$, $n$, $h$ and the law of $M$, so the same $p$ appears when $Z$ is replaced by $G$. Since $GO$ has the law of $G$ and $\operatorname{pol}(GO)=\operatorname{pol}(G)O$ for $O\in O(N)$, $\operatorname{pol}(G)$ is uniform on the $h\times N$ matrices with orthonormal rows. So $\Omega_{\operatorname{pol}(G)}$ is Haar distributed on $O(N)$ and independent of $Y$, and so is $O_G=\Omega_{\operatorname{pol}(G)}R_\Theta^{\mathsf T}$. The matrix built from $G$ is $X_0^G=(I_h\oplus O_G)Y(I_h\oplus O_G)^{\mathsf T}$, which has the law $Q_n$ by the orthogonal invariance of the GOE. Hence $\E p(\operatorname{pol}(G))=\E f(X_0^G)=\E_{Q_n}f=0$.

\emph{Step 2: truncating the polar factor.} Let $\bar Z=Z/\sqrt N$ and $E=\bar Z\bar Z^{\mathsf T}-I_h$, so that $ZZ^{\mathsf T}=N(I_h+E)$ and $U=(I_h+E)^{-1/2}\bar Z$. For $|x|<1$, $(1+x)^{-1/2}=\sum_{l\ge0}\binom{-1/2}lx^l$ with $|\binom{-1/2}l|=\binom{2l}l4^{-l}\le1$. Let $U_L=\sum_{l=0}^L\binom{-1/2}{l}E^l\bar Z$, a polynomial of degree $2L+1$ in $Z$, and define $\bar G,E_G,U_{L,G}$ in the same way from $G$.

We first bound the moments of $E_G$. For a fixed unit vector $x\in\R^h$, $1+x^{\mathsf T}E_Gx=\|G^{\mathsf T}x\|_2^2/N$ has the law of $\chi^2_N/N$, and a standard chi-square bound gives $\Pp(|\chi^2_N/N-1|\ge t)\le2e^{-N\min(t,t^2)/16}$. For a symmetric $B$ and a $1/4$-net $\mathcal N$ of the unit sphere of $\R^h$ with $|\mathcal N|\le9^h$, we have $\|B\|_{\op}\le2\max_{x\in\mathcal N}|x^{\mathsf T}Bx|$. Hence
\begin{equation*}
\Pp(\|E_G\|_{\op}>t)\le\sum_{x\in\mathcal N}\Pp(|x^{\mathsf T}E_Gx|>t/2)\le2\cdot9^he^{-N\min(t,t^2)/64}\qquad(t>0).
\end{equation*}
Let $t_0=(384\varepsilon_0)^{1/2}\le1$. Since $h\le\varepsilon_0N$, for $t\ge t_0$ we have $9^h\le e^{3\varepsilon_0N}\le e^{N\min(t,t^2)/128}$. Integrating the tail, separately for $t\le1$ and $t\ge1$, gives $\E\|E_G\|_{\op}^\kappa\le t_0^\kappa+2\Gamma(\kappa/2+1)(128/N)^{\kappa/2}+2\Gamma(\kappa+1)(128/N)^\kappa$, so $\|\|E_G\|_{\op}\|_{L^\kappa}\le t_0+C_2((\kappa/N)^{1/2}+\kappa/N)$ for $\kappa\ge1$. As $\|\bar G\|_{\op}^2=\|I_h+E_G\|_{\op}\le1+\|E_G\|_{\op}$, for $\varepsilon_0$ and $c_1$ small we get
\begin{equation*}
\bigl\|\|E_G\|_{\op}\bigr\|_{L^\kappa}\le\tfrac14\quad\text{and}\quad\bigl\|\|\bar G\|_{\op}\bigr\|_{L^\kappa}\le2\qquad(1\le\kappa\le c_1N).
\end{equation*}

On the event $\mathcal D=\{\|E\|_{\op}\le1/2\}$, we have $\|\bar Z\|_{\op}^2=\|I_h+E\|_{\op}\le3/2$, so $\|U-U_L\|_{\op}=\|\sum_{l>L}\binom{-1/2}lE^l\bar Z\|_{\op}\le(3/2)^{1/2}\sum_{l>L}2^{-l}\le2^{1-L}$. Then $\|U_L\|_{\op}\le2$, all entries of $U$ and $U_L$ are at most $2$ in absolute value, and a monomial of degree $a\le2D$ changes by at most $a2^{a-1}$ times the largest change of an entry. So $|p(U)-p(U_L)|\le\|p\|_1\,D\,4^D2^{1-L}\le n^{C_3D^2}2^{-L}$ on $\mathcal D$.

For the complement, $\|E\|_{\op}^{2L}\le\Tr E^{2L}$, and $\Tr E^{2L}$ is a polynomial of degree $4L$ in $Z$. Markov's inequality, moment matching, and the moment bound with $\kappa=2L\le c_1N$ give
\begin{equation*}
\Pp(\mathcal D^c)\le4^L\E\Tr E^{2L}=4^L\E\Tr E_G^{2L}\le4^Lh\,\E\|E_G\|_{\op}^{2L}\le h4^{-L},
\end{equation*}
and the same bound holds for $\mathcal D_G=\{\|E_G\|_{\op}\le1/2\}$. On these complements we need second moments. Since $\|U_{L,G}\|_{\op}\le\sum_{l\le L}\|E_G\|_{\op}^l\|\bar G\|_{\op}$ and $|p(V)|\le\|p\|_1(1+\|V\|_{\op})^{2D}$, Minkowski's inequality and $\|AB\|_{L^{4D}}\le\|A\|_{L^{8D}}\|B\|_{L^{8D}}$ give
\begin{equation*}
\bigl\|1+\|U_{L,G}\|_{\op}\bigr\|_{L^{4D}}\le1+\bigl\|\|\bar G\|_{\op}\bigr\|_{L^{4D}}+\sum_{l=1}^L\bigl\|\|E_G\|_{\op}\bigr\|_{L^{8Dl}}^l\bigl\|\|\bar G\|_{\op}\bigr\|_{L^{8D}}\le3+\sum_{l\ge1}2\cdot4^{-l}<4,
\end{equation*}
provided the moment orders $8Dl\le8DL$ and $8D$ are at most $c_1N$, that is, $8D(L+1)\le c_1N$. Hence $\E p(U_{L,G})^2\le\|p\|_1^2\,4^{4D}$, and $\E p(U_L)^2=\E p(U_{L,G})^2$ because $p(U_L)^2$ has degree at most $4D(2L+1)$ in $Z$. Splitting on $\mathcal D$, and using $|p(U)|\le\|p\|_1$ since the entries of $U$ are at most one,
\begin{align*}
|\E p(U)-\E p(U_L)|&\le\E\bigl[|p(U)-p(U_L)|\mathbf 1_{\mathcal D}\bigr]+\|p\|_1\Pp(\mathcal D^c)+\bigl(\E p(U_L)^2\bigr)^{1/2}\Pp(\mathcal D^c)^{1/2}\\
&\le n^{C_3D^2}2^{-L}+n^{C_1D^2}h4^{-L}+n^{C_1D^2}4^{2D}h^{1/2}2^{-L}\le n^{C_4D^2}2^{-L}.
\end{align*}
The same argument with $G$ in place of $Z$ gives $|\E p(\operatorname{pol}(G))-\E p(U_{L,G})|\le n^{C_4D^2}2^{-L}$.

\emph{Step 3: matching.} Let $L=\lceil K_1D^2\log n\rceil$, with $K_1\ln2\ge C_4+11$. Then $2^{-L}\le n^{-K_1D^2\ln2}$, so $n^{C_4D^2}2^{-L}\le n^{-11D^2}\le\frac12n^{-10D}$. The polynomials in $Z$ whose expectations we matched are $\Tr E^{2L}$, $p(U_L)$ and $p(U_L)^2$, of degrees $4L$, at most $2D(2L+1)$, and at most $4D(2L+1)$. Since $L\le K_1D^2\log n+1$ and $D\le D^3\log n$, we have $4D(2L+1)\le8K_1D^3\log n+12D\le K_0D^3\log n$ with $K_0=8K_1+12$. Also $8D(L+1)\le(8K_1+16)D^3\log n\le c_1N$ and $N-h\ge N/2\ge2D$ if $c_0\le\min\{c_1/(8K_1+16),1/4\}$ and $\varepsilon_0\le1/2$. By Step 1 for $Z$ and for $G$, then Step 2 and the moment matching $\E p(U_L)=\E p(U_{L,G})$,
\begin{align*}
|\E f(X_0)|&=|\E p(U)-\E p(\operatorname{pol}(G))|\\
&\le|\E p(U)-\E p(U_L)|+|\E p(U_L)-\E p(U_{L,G})|+|\E p(U_{L,G})-\E p(\operatorname{pol}(G))|\\
&\le\tfrac12n^{-10D}+0+\tfrac12n^{-10D}=n^{-10D}.
\end{align*}
As $f$ was arbitrary, this proves the lemma.
\end{proof}

We next show that the code remains detectable in the noisy projector $\Pi_k(H)$. Individual eigenvectors near the bulk cutoff $k=\lfloor\alpha n\rfloor$ need not survive the noise, because neighboring eigenvalues are spaced by $O(1/n)$ while the perturbation has operator norm $O(\sqrt T)$. Working with the rank-$k$ projector avoids this sensitivity: only eigenvectors with eigenvalues within $O(T^{1/3})$ of the cutoff acquire a non-negligible component across the cut, giving $\|\Pi_k(H)-\Pi_k(X_0)\|_F^2=O(nT^{1/3})$, which is negligible compared with the squared Frobenius norm $\Theta(\varepsilon n)$ of the coded block when $T\le T_0$ is small. In the next lemma, we combine this projector stability with the approximation $\Sigma\operatorname{pol}(Z)\approx\bar d\,Z$ to show that almost all entries of the noisy coded block retain the sign $(-1)^{b_{ij}}$ with magnitude well above the $n^{-4}$ rounding precision.

\begin{lemma}
\label{lem:coded-block-noise}
Let $\eta\in(0,1/8)$ and $\tau>0$ with $\Pp(|N(0,1)|\le\tau)\le\eta/16$, and let $\varsigma=(\alpha(1-\alpha))^{1/2}$ and $\bar d=\varsigma/\sqrt N$. If $\varepsilon$ is small enough depending on $\alpha,\eta,\tau$, and $T_0$ is small enough depending also on $\varepsilon$, then the following holds for every $T\in(0,T_0]$. Let $\Pi'=\Pi_k(rX_0+\sqrt s\,W)$ with $W\sim Q_n$ independent of $(Y,Z,R)$. With probability $1-o(1)$, the set $\mathcal E$ of $(i,j)\in\mathcal A\times\mathcal B$ with $|Z_{ij}|\le\tau$ or $|\Pi'_{ij}-\bar dZ_{ij}|>\bar d\tau/2$ has at most $\eta hN/4$ elements, and for $(i,j)\notin\mathcal E$, $\Pi'_{ij}$ has the sign $(-1)^{b_{ij}}$ and absolute value larger than $\bar d\tau/2$.
\end{lemma}

\begin{proof}
Let $C'_1,C'_2,C'_3$ denote constants depending only on $\alpha$, let $c'>0$ be absolute, assume $\varepsilon\le\varepsilon_0/2$, so that $h\le\varepsilon_0N$ because $N=n-h\ge n/2$, and let $\Pi(0)=\Pi_k(X_0)$. We compare the coded block of $\Pi(0)$ with $\bar dZ$, then bound $\Pi'-\Pi(0)$, and finally count entries.

\emph{The block before noise.} Let $P=\Pi_{\mathcal A\mathcal A}$ with $\Pi=\Pi_k(Y)$ uniformly random. For a fixed unit vector $x\in\R^h$, $x^{\mathsf T}Px=\|\Pi x\|_2^2$ has the law of the squared norm of the first $k$ coordinates of a uniform unit vector in $\R^n$, a $2$-Lipschitz function on the sphere with mean $k/n$. So $\Pp(|x^{\mathsf T}Px-k/n|>t)\le2e^{-c'nt^2}$ \cite{Ledoux2001}, and a union bound over a $1/4$-net of the unit sphere of $\R^h$, of size at most $9^h\le e^{3\varepsilon n}$, gives $\Pp(\|P-(k/n)I_h\|_{\op}>2t)\le2e^{3\varepsilon n-c'nt^2}$. With $t=(4\varepsilon/c')^{1/2}$ and $|k/n-\alpha|\le1/n$, we get $\|P-\alpha I_h\|_{\op}\le C'_1\sqrt\varepsilon$ with probability $1-o(1)$. For $\varepsilon$ small, the eigenvalues of $P$ then lie in $[\alpha/2,(1+\alpha)/2]$, where $\phi(x)=(x-x^2)^{1/2}$ is Lipschitz with a constant depending on $\alpha$. As $\Sigma=\phi(P)$ and $\varsigma=\phi(\alpha)$, this gives $\|\Sigma-\varsigma I_h\|_{\op}\le C'_2\sqrt\varepsilon$.

Next we compare $U=\operatorname{pol}(Z)$ with $\bar Z=Z/\sqrt N$, with $E=\bar Z\bar Z^{\mathsf T}-I_h$ as in the proof of \cref{lem:polar-moments}. If $\bar Z=A\diag(s_1,\dots,s_h)B^{\mathsf T}$ is a singular value decomposition, then $U=AB^{\mathsf T}$ and $s_i^2=\lambda_i(I_h+E)$, so, since $|1-s|\le|1-s^2|$ for $s\ge0$, $\|U-\bar Z\|_F^2=\sum_i(1-s_i)^2\le\sum_i(1-s_i^2)^2=\Tr E^2$. The mean and second moment of $\Tr E^2$ are expectations of polynomials of degrees four and eight in $Z$, so they are the Gaussian ones. If $g_1,\dots,g_h$ are the rows of $G$, the entries of $E_G$ are $\|g_i\|_2^2/N-1$ and $g_i^{\mathsf T}g_{i'}/N$, with second moments $2/N$ and $1/N$, so $\E\Tr E_G^2=2h/N+h(h-1)/N=h(h+1)/N\le2\varepsilon h$ for large $n$. The gradient of $G\mapsto\Tr E_G^2$ is $4E_GG/N$ and $\|E_G\|_F^2\le h\|E_G\|_{\op}^2$, so the Gaussian Poincar\'e inequality and the moment bounds in the proof of \cref{lem:polar-moments} give
\begin{equation*}
\Var\Tr E_G^2\le\frac{16}{N^2}\E\|E_GG\|_F^2\le\frac{16h}N\bigl(\E\|E_G\|_{\op}^4\bigr)^{1/2}\bigl(\E\|\bar G\|_{\op}^4\bigr)^{1/2}\le\frac{4h}N\le8\varepsilon.
\end{equation*}
By Chebyshev's inequality, $\Pp(\Tr E^2>3\varepsilon h)\le8\varepsilon/(\varepsilon h)^2=o(1)$. Since $\bar dZ=\varsigma\bar Z$ and $\|U\|_F^2=h$, \eqref{eq:coded-block} gives, with probability $1-o(1)$,
\begin{equation*}
\|\Pi(0)_{\mathcal A\mathcal B}-\bar dZ\|_F\le\|(\Sigma-\varsigma I_h)U\|_F+\varsigma\|U-\bar Z\|_F\le C'_2\sqrt\varepsilon\,h^{1/2}+\varsigma(3\varepsilon h)^{1/2}\le(C'_3\varepsilon h)^{1/2}.
\end{equation*}

\emph{The projector after noise.} With probability $1-o(1)$, $\|X_0\|_{\op}=\|Y\|_{\op}\le3$ and $\|W\|_{\op}\le3$. Since $1-r\le T/2$, $s\le T$ and $T\le T_0\le1/36$, the matrix $H_0=rX_0+\sqrt s\,W$ then satisfies $\|H_0-X_0\|_{\op}\le\frac32T+3\sqrt T\le\delta:=6\sqrt{T_0}\le1$. Put $w=\delta^{2/3}\ge\delta$. Write $X_0=\sum_i\lambda_ix_ix_i^{\mathsf T}$ with $\lambda_1<\dots<\lambda_n$ and orthonormal $x_i$, and let $\mu=\lambda_{n-k+1}$, so that $\Pi(0)=\sum_{i>n-k}x_ix_i^{\mathsf T}$. As $\Pi'$ and $\Pi(0)$ are projectors of rank $k$,
\begin{equation*}
\|\Pi'-\Pi(0)\|_F^2=2k-2\Tr(\Pi'\Pi(0))=2\Tr\bigl(\Pi'(I_n-\Pi(0))\bigr)=2\sum_{i\le n-k}\|\Pi'x_i\|_2^2.
\end{equation*}
We split this sum at $\mu-2w$. By Weyl's inequality $\lambda_{n-k+1}(H_0)\ge\mu-\delta$, so $H_0\ge\mu-\delta$ on the range of $\Pi'$, which is invariant under $H_0$. If $\lambda_i\le\mu-2w$, then $H_0-\lambda_i\ge2w-\delta\ge w$ on this range, and $(H_0-\lambda_i)\Pi'x_i=\Pi'(H_0-\lambda_i)x_i=\Pi'(H_0-X_0)x_i$, so $w\|\Pi'x_i\|_2\le\|(H_0-\lambda_i)\Pi'x_i\|_2=\|\Pi'(H_0-X_0)x_i\|_2\le\delta$ and $\|\Pi'x_i\|_2^2\le\delta^2/w^2=\delta^{2/3}$. The other $i\le n-k$ have $\lambda_i\in(\mu-2w,\mu)$. The spectrum of $X_0$ is that of $Y$, the semicircle law holds uniformly over intervals, and the semicircle density is at most $1/\pi$, so with probability $1-o(1)$ every interval of length $2w$ contains at most $n(2w/\pi+o(1))\le wn$ eigenvalues of $X_0$. For these $i$ we use $\|\Pi'x_i\|_2\le1$. Altogether, $\|\Pi'-\Pi(0)\|_F^2\le2(n\delta^{2/3}+wn)=4n\delta^{2/3}$.

\emph{Counting entries.} Since $\Pi'_{\mathcal A\mathcal B}-\Pi(0)_{\mathcal A\mathcal B}$ is a block of $\Pi'-\Pi(0)$,
\begin{equation*}
\|\Pi'_{\mathcal A\mathcal B}-\bar dZ\|_F^2\le2\|\Pi'-\Pi(0)\|_F^2+2\|\Pi(0)_{\mathcal A\mathcal B}-\bar dZ\|_F^2\le8n\delta^{2/3}+2C'_3\varepsilon h\le\frac{\eta}{32}\varsigma^2\tau^2h,
\end{equation*}
where the last step holds if first $\varepsilon\le\eta\varsigma^2\tau^2/(128C'_3)$ and then $\delta^{2/3}\le\eta\varsigma^2\tau^2\varepsilon/1024$, using $h\ge\varepsilon n/2$. Each entry with $|\Pi'_{ij}-\bar dZ_{ij}|>\bar d\tau/2$ contributes more than $\bar d^2\tau^2/4=\varsigma^2\tau^2/(4N)$ to the left side, so there are fewer than $(\eta/32)\varsigma^2\tau^2h\cdot4N/(\varsigma^2\tau^2)=\eta hN/8$ such entries. The variables $|Z_{ij}|=|\zeta_{ij}|$ are independent, and each is at most $\tau$ with probability at most $\eta/16$. So the number of $(i,j)$ with $|Z_{ij}|\le\tau$ has mean at most $\eta hN/16$ and variance at most $hN$, and by Chebyshev's inequality it is at most $\eta hN/8$ with probability $1-o(1)$. Hence $|\mathcal E|\le\eta hN/4$. Finally, for $(i,j)\notin\mathcal E$ we have $|Z_{ij}|>\tau$ and $|\Pi'_{ij}-\bar dZ_{ij}|\le\bar d\tau/2<\bar d|Z_{ij}|/2$, so $\Pi'_{ij}$ has the sign of $Z_{ij}$, which is $(-1)^{b_{ij}}$, and $|\Pi'_{ij}|\ge\bar d|Z_{ij}|-\bar d\tau/2>\bar d\tau/2$.
\end{proof}

We now combine \cref{lem:polar-moments,lem:coded-block-noise} to complete the proof of \cref{prop:poly-eigenvectors}. After fixing the parameters to obtain the low-degree bound, we show in the planted case that even after random relabeling some $\ell$-row subset $I\subseteq[n_1]$ reads enough uncorrupted code columns to create a rank deficit of at least $n^{1/3}$, and in the null case we couple the observed signs with independent Gaussian signs and apply McDiarmid's inequality with a union bound over all subsets $I$.

\begin{proof}[Proof of \cref{prop:poly-eigenvectors}]
\emph{Constants, spectrum and advantage.} Choose $\eta\in(0,1/8)$ with $A_u\log(1-4\eta)^{-1}\le1/20$, then $\tau>0$ with $\Pp(|N(0,1)|\le\tau)\le\eta/16$, then $\varepsilon$ and $T_0$ as in \cref{lem:coded-block-noise}, with also $\varepsilon\le\varepsilon_0/2$. The spectrum of $SX_0S^{\mathsf T}$ is that of $Y$. Since $N\ge n/2$, $h\le\varepsilon n\le\varepsilon_0N$. Since $D_n\le2(\log n)^q$ and $u-1=\lceil3q\rceil+2>3q+1$, for large $n$ we have $D_n^3\log n\le c_0N$ and $K_0D_n^3\log n\le8K_0(\log n)^{3q+1}\le c_u(\log n)^{u-1}$. So every polynomial in $Z$ of degree at most $K_0D_n^3\log n$, in particular $p(U_L)$, has the Gaussian expectation, and \cref{lem:polar-moments} with $D=D_n$ gives $\adv_{\le D_n}(Q_n,\mathcal L(X_0))\le n^{-10D_n}$. Averaging over $S$ does not increase the advantage, because $f\mapsto f(S\cdot S^{\mathsf T})$ preserves the degree and the mean and variance under $Q_n$. We now run the rank test defined after \cref{lem:mao-code} on a matrix $\widetilde\Pi$ within $n^{-4}$ of $\Pi_k(H)$.

\emph{The planted case: relabeling.} Let $H=rSX_0S^{\mathsf T}+\sqrt s\,W$ and $W'=S^{\mathsf T}WS$. Since $W$ is independent of $S$ and of $(Y,Z,R)$, and $Q_n$ is invariant under permutations, $W'\sim Q_n$ is independent of $S$ and of $(Y,Z,R)$. Then $\Pi_k(H)=S\Pi'S^{\mathsf T}$ with $\Pi'=\Pi_k(rX_0+\sqrt s\,W')$. If $Se_i=e_{\sigma(i)}$, the test reads the entries of $\Pi'$ in the rows $\mathcal P=\sigma^{-1}([n_1])$ and the columns $\mathcal C=\sigma^{-1}(\mathcal C_0)$. We condition on everything except $S$, in the event of \cref{lem:coded-block-noise} applied with $W'$, which has probability $1-o(1)$. Then $\mathcal P$ is a uniformly random $n_1$-subset of $[n]$, and given $\mathcal P$, $\mathcal C$ is a uniformly random $N_0$-subset of $[n]\setminus\mathcal P$.

\emph{Good columns and coded rows.} For $j\in\mathcal B$ let $B_j=\{i\in\mathcal A:(i,j)\in\mathcal E\}$, and call $j$ good if $|B_j|\le\eta h$. Since $|\mathcal E|\le\eta hN/4$, fewer than $N/4$ columns are bad. The number $\nu=|\mathcal P\cap\mathcal A|$ is hypergeometric, and given $\nu$ the set $\mathcal P\cap\mathcal A$ is a uniformly random $\nu$-subset of $\mathcal A$. As $16\ell/\varepsilon\le n_1\le17\ell/\varepsilon$ and $h\ge\varepsilon n-1$, its mean $n_1h/n$ is at least $15\ell$ for large $n$, and Hoeffding's inequality for sampling without replacement gives $\Pp(\nu<2\ell)\le\exp(-2(13\ell)^2/n_1)\le e^{-19\varepsilon\ell}=o(1)$. For a good $j$, the number $\nu_j=|\mathcal P\cap B_j|$ has conditional mean $\nu|B_j|/h\le\eta\nu$ given $\nu$, so on $\{\nu\ge2\ell\}$ Hoeffding's inequality gives $\Pp(\nu_j>2\eta\nu\mid\nu)\le e^{-2\eta^2\nu}\le e^{-4\eta^2\ell}$. The expected number of good $j$ with $\nu_j>2\eta\nu$ is then at most $Ne^{-4\eta^2\ell}$, so by Markov's inequality there are at most $N/8$ of them, except with probability $8e^{-4\eta^2\ell}=o(1)$. Removing also the $n_1\le N/8$ columns in $\mathcal P$, we obtain a set $\mathcal J\subseteq\mathcal B\setminus\mathcal P$ of at least $N/2$ good columns with $\nu_j\le2\eta\nu$.

\emph{A row subset that reads the code.} Let $I$ be a uniformly random $\ell$-subset of $\mathcal P\cap\mathcal A$. For $j\in\mathcal J$, using $\nu\ge2\ell$ and $\ell\le A_u\log n$,
\begin{equation*}
\Pp(I\cap B_j=\emptyset)=\prod_{t=0}^{\ell-1}\Bigl(1-\frac{\nu_j}{\nu-t}\Bigr)\ge\Bigl(1-\frac{2\eta\nu}{\nu-\ell}\Bigr)^\ell\ge(1-4\eta)^\ell\ge n^{-1/20}.
\end{equation*}
So the expected number of $j\in\mathcal J$ with $I\cap B_j=\emptyset$ is at least $|\mathcal J|n^{-1/20}\ge n^{19/20}/4$, and some $\ell$-subset $I_*$, which depends on $\mathcal P$ but not on $\mathcal C$, attains this. Let $\mathcal G=\{j\in\mathcal J:I_*\cap B_j=\emptyset\}$. Given $\mathcal P$, $|\mathcal G\cap\mathcal C|$ is hypergeometric with mean $N_0|\mathcal G|/(n-n_1)\ge n^{9/20}/5$, so Hoeffding's inequality gives $\Pp(|\mathcal G\cap\mathcal C|<n^{9/20}/10)\le\exp(-2(n^{9/20}/10)^2/N_0)\le e^{-n^{2/5}/50}$. For $j\in\mathcal G\cap\mathcal C$ and $i\in I_*$, we have $(i,j)\notin\mathcal E$, so $\Pi'_{ij}$ has the sign $(-1)^{b_{ij}}$ and $|\Pi'_{ij}|>\bar d\tau/2\ge\varsigma\tau/(2\sqrt n)>n^{-4}$ by \cref{lem:coded-block-noise}. The entry $\widetilde\Pi_{\sigma(i)\sigma(j)}$ read by the test is within $n^{-4}$ of $\Pi'_{ij}$, so it has the same sign, and the bit read is $b_{ij}=g_i(\xi_j)$ by \cref{lem:mao-code}(ii).

\emph{The planted deficit.} Let $I=\sigma(I_*)$. For $j\in\mathcal G\cap\mathcal C$, the vector $y^I_{\sigma(j)}$ is $(g_{i_1}(\xi_j),\dots,g_{i_\ell}(\xi_j))$ for the ordering $i_1,\dots,i_\ell$ of $I_*$ induced by $\sigma$. By \cref{lem:mao-code}(iv), the lifts of these columns span a space of dimension at most $n^{1/4}$, and each of the other $N_0-|\mathcal G\cap\mathcal C|$ columns adds at most one to the rank. So, with probability $1-o(1)$,
\begin{equation*}
\operatorname{def}(I)\ge N_0-\bigl(n^{1/4}+N_0-|\mathcal G\cap\mathcal C|\bigr)\ge n^{9/20}/10-n^{1/4}\ge n^{1/3},
\end{equation*}
and the test declares the planted law.

\emph{The null case: Gaussian signs.} Under $Q_n$, $\Pi=\Pi_k(H)$ is uniformly random. Let $\mathcal K=[n_1]$, $n'=n-n_1$, $P=\Pi_{\mathcal K\mathcal K}$, $F=\Pi_{\mathcal K\mathcal K^c}$, $\phi(x)=(x-x^2)^{1/2}$ and $\varsigma_n=\phi(k/n)$. The test reads only $\widetilde\Pi$ in the block of $F$. As in the proof of \cref{lem:coded-block-noise}, now with a net of size $9^{n_1}=n^{O(1)}$, $\Pp(\|P-(k/n)I_{n_1}\|_{\op}>n^{-1/3})\le2\cdot9^{n_1}e^{-c'n^{1/3}/4}\le e^{-n^{1/4}}$. Also $FF^{\mathsf T}=P-P^2$. Conjugation by $I_{n_1}\oplus O$ with $O\in O(n')$ Haar and independent of $\Pi$ preserves the law of $\Pi$ and maps $(P,F)$ to $(P,FO^{\mathsf T})$, and $F=\phi(P)\operatorname{pol}(F)$. So $(P,F)$ has the law of $(P,\phi(P)\operatorname{pol}(G))$, where $G$ is an $n_1\times n'$ matrix of independent standard Gaussian entries, independent of $P$, and we use this representation. Let $E'=GG^{\mathsf T}/n'-I_{n_1}$. The tail bound for $E_G$ in the proof of \cref{lem:polar-moments} and $\Pp(\chi^2_{n_1}>n^{1/10})\le e^{-n^{1/10}/4}$ for each column and large $n$ show that $\|E'\|_{\op}\le n^{-1/3}$ and $\max_j\|G_{\cdot j}\|_2\le n^{1/20}$ with probability at least $1-e^{-n^{1/20}}$ for large $n$. On these events,
\begin{equation*}
\frac{n^{1/2}}{\varsigma_n}F=A'G,\qquad A'=\Bigl(\frac n{n'}\Bigr)^{1/2}\frac{\phi(P)}{\varsigma_n}(I_{n_1}+E')^{-1/2},\qquad\|A'-I_{n_1}\|_{\op}=O(n^{-1/3}),
\end{equation*}
because $n/n'-1=O(n_1/n)$, $\phi$ is Lipschitz near $k/n$, and $\|(I+E')^{-1/2}-I\|_{\op}\le\|E'\|_{\op}$. So every entry of $(n^{1/2}/\varsigma_n)F-G$ is at most $\|A'-I_{n_1}\|_{\op}\max_j\|G_{\cdot j}\|_2=O(n^{-17/60})\le n^{-1/4}$. As $\widetilde\Pi$ is within $n^{-4}$ of $F$ and $n^{1/2-4}/\varsigma_n\le n^{-1/4}$, the sign read by the test at $(i,j)$ is that of $G_{ij}$ unless $|G_{ij}|\le2n^{-1/4}$.

\emph{The null case: few wrong signs.} The number $K'$ of $(i,j)\in\mathcal K\times\mathcal C_0$ with $|G_{ij}|\le2n^{-1/4}$ is binomial with mean at most $2n_1N_0n^{-1/4}\le n^{0.3}/2$, so $\Pp(K'>n^{0.3})\le e^{-n^{0.3}/6}$ by the Chernoff bound. For $I\subseteq[n_1]$ with $|I|=\ell$, let $\operatorname{def}_G(I)$ be the deficit computed from the signs of $G$. At most $K'$ columns differ, and each changes the rank by at most one, so $|\operatorname{def}(I)-\operatorname{def}_G(I)|\le K'$ for all $I$.

\emph{The null case: McDiarmid's inequality.} Fix $I$. The signs of $G$ are independent and uniform, so the $N_0$ columns of signs used by $\operatorname{def}_G(I)$ are independent and uniform in $\mathbb F_2^\ell$, and changing one of them changes $\operatorname{def}_G(I)$ by at most one. With $m_I=\E\operatorname{def}_G(I)$, McDiarmid's inequality gives $\Pp(\pm(\operatorname{def}_G(I)-m_I)\ge x)\le e^{-2x^2/N_0}$ for $x>0$. By \cref{lem:mao-code}(iii), $\Pp(\operatorname{def}_G(I)=0)\ge1/2$ for large $n$. If $m_I>0$, the lower tail with $x=m_I$ gives $1/2\le e^{-2m_I^2/N_0}$, so $m_I\le(N_0\ln2/2)^{1/2}\le n^{1/4}$. The upper tail with $x=n^{0.3}-n^{1/4}\ge n^{0.3}/2$ then gives $\Pp(\operatorname{def}_G(I)\ge n^{0.3})\le e^{-n^{0.6}/(2N_0)}\le e^{-n^{0.1}/2}$.

\emph{The null case: union bound.} There are at most $2^{n_1}=n^{O(1)}$ sets $I$. So with probability at least $1-2^{n_1}e^{-n^{0.1}/2}-e^{-n^{0.3}/6}-e^{-n^{1/20}}-e^{-n^{1/4}}=1-o(1)$, for all $I$ simultaneously, $\operatorname{def}(I)\le\operatorname{def}_G(I)+K'<2n^{0.3}<n^{1/3}$, and the test declares the null law.
\end{proof}